\documentclass[prl,showpacs,twocolumn,floats,10pt,aps,citeautoscript,longbibliography,superscriptaddress]{revtex4-2}

\usepackage[final]{changes}
\definechangesauthor[name=GK, color=red]{GK}
\definechangesauthor[name=AG, color=red]{AG}

\usepackage{microtype} 
\usepackage{graphicx}
\usepackage{enumerate}
\usepackage{bm}
\usepackage{amsmath}
\usepackage{amsfonts}
\usepackage{dsfont}
\usepackage{color}
\usepackage{upgreek}
\usepackage[normalem]{ulem}

\allowdisplaybreaks 

\usepackage{xspace} 

\renewcommand{\vec}[1]{{\mathbf{#1}}} 
\newcommand{\bk}{{\vec{k}}}
\newcommand{\mi}{\mathrm{i}}
\usepackage{bbm}  

\newcommand{\kF}{\vec{k}_{\mr{F}}}
\newcommand{\kL}{\vec{k}_{\mr{L}}}

\newcommand{\mc}[1]{\ensuremath{\mathcal{#1}}}
\newcommand{\mr}[1]{\ensuremath{\mathrm{#1}}}

\newcommand{\re}{\ensuremath{\mathrm{Re}}}
\newcommand{\im}{\ensuremath{\mathrm{Im}}}

\newcommand{\bra}[1]{\ensuremath{\langle #1 |}}
\newcommand{\ket}[1]{\ensuremath{| #1 \rangle}}

\newcommand{\G}[1]{\langle\!\langle#1\rangle\!\rangle}

\newcommand{\Eq}[1]{Eq.~\eqref{#1}}

\definecolor{darkgreen}{rgb}{0,0.5,0}

\newcommand{\pdag}{{\vphantom{dagger}}}  

\RequirePackage[
  hyperindex,colorlinks,bookmarksnumbered,
  plainpages=true,pdfstartview=FitH,hypertexnames=false]{hyperref}
\hypersetup{linkcolor=blue,urlcolor=blue,citecolor=blue} 
\usepackage[hypertexnames=false]{hyperref}
\usepackage[all]{hypcap}

\makeatletter
\def\maketitle{
\@author@finish
\title@column\titleblock@produce
\suppressfloats[t]}
\makeatother

\newcommand{\maintitle}{Pseudogapped Fermi liquids from emergent quasiparticles}

\begin{document} 

\title{\maintitle}
\author{Andreas Gleis}
\email{andreas.gleis@rutgers.edu}
\affiliation{Department of Physics and Astronomy, Rutgers University, Piscataway, NJ 08854, USA}
\author{Gabriel Kotliar}
\email{kotliar@physics.rutgers.edu}
\affiliation{Department of Physics and Astronomy, Rutgers University, Piscataway, NJ 08854, USA}
\affiliation{Condensed Matter Physics and Materials Science Department,\looseness=-1\,  
Brookhaven National Laboratory, Upton, NY 11973, USA}

\date{\today}

\begin{abstract}
We propose an interacting model that is exactly solvable in any spatial dimension and gives rise to a Fermi liquid~(FL) featuring a pseudogapped (PG) single-particle spectral function and a vanishing quasiparticle~(QP) weight at half-filling, without invoking Mott physics.
The PG originates from a purely fermionic mechanism through emergent QPs arising from a correlated hopping interaction.
By employing an appropriate coherent-state basis, we derive a Gaussian path-integral representation of the partition function, which enables systematic treatments of deviations from the Gaussian limit using standard many-body techniques, such as diagrammatic perturbation theory or mean-field theory. We explicitly demonstrate and discuss several properties of the exactly solvable limit on the square lattice, including the mechanism for temperature-dependent PG opening, 
the singular behavior of the self-energy, the violation of the Luttinger sum rule, and the role
of Luttinger and Fermi surfaces. Finally, we explore quantum phase transitions between PG-FLs and Landau FLs.
\end{abstract}

\maketitle

Pseudogapped~(PG) metals represent one of the most longstanding unresolved mysteries in strongly-correlated electron physics. 
While most famously observed in cuprates~\cite{Loeser1996,Ding1996,Norman1998,Timusk1999,Damascelli2003,Kanigel2006,Sobota2021,Armanno2025},
they also appear, for instance, in nickelates~\cite{Uchida2011}, iridates~\cite{Peng2022,Alexanian2025}, and magic-angle twisted bilayer graphene~\cite{Oh2021}.
In contrast to conventional Landau Fermi liquids~(FL) --- where electron spectroscopy reveals a large electronic density of states near the Fermi level ---
PG metals exhibit a depression of the electronic density of states at the Fermi level, which becomes increasingly prominent at lower temperatures.
Furthermore, while quantum oscillations~\cite{Chan2025} in PG metals detect closed Fermi surfaces~(FS), electronic spectroscopy typically only observes disconnected Fermi arcs~\cite{Meng2009,Chang2008,King2011,Freutel2019}.
This dichotomy suggests that quasiparticles~(QP) are ``hidden'' along segments of the FS.
Apart from the previously mentioned bulk systems, local PG metals with hidden QP physics also emerge in certain impurity systems, which are experimentally realizable~\cite{Minamitani2012,Hiraoka2017,Yang2019,Guo2021}
and amenable to highly controlled theoretical analysis~\cite{Jones1987,DeLeo2004,Sakai1992,Zarand2006,Chung2007,Nishikawa2012,Nishikawa2012a,Nishikawa2018,Blesio2018,Zitko2021}. 

Several mechanisms for the depletion of low-energy spectral weight have been proposed.
Those invoke either fluctuating order~\cite{Emery1995,Eberlein2016,Verret2017,Bonetti2020,Bonetti2022,Klett2022,Lihm2026,Forni2026} or Mott physics~\cite{Hubbard1963,Yang2006,Sakai2010,Yamaji2011,Robinson2019,Scheurer2018,Wu2018,Zhang2020,Mascot2022,Zhou2024,Ledwith2025,Ledwith2025a}.
In the latter, the FS is reconstructed through a self-energy pole that has a hidden fermion interpretation~\cite{Zhu2013,Sakai2015,Sakai2016,Imada2019,Ido2020}.
Recently, it has also been argued that Luttinger surfaces~(LS)~\cite{Dzyaloshinskii2003}, which are inherently present in PG metals, can host hidden QPs~\cite{Fabrizio2020,Fabrizio2022,Fabrizio2023}.
Overall, the precise origin of PG metals remains unsettled and may differ depending on the specific physical context.

In this letter, we propose an interacting model that realizes hidden QPs and depletion of low-energy spectral weight 
through an intrinsically fermionic mechanism, without invoking Mott physics.
In our model, which is exactly solvable in arbitrary spatial dimensions, a correlated hopping~(CH) interaction gives rise to new emergent QPs,
and phase space constraints subsequently produce a PG in the electronic spectral function. 
A Gaussian path-integral construction demonstrates the stability of our PG-FL fixed point and provides a framework for studies beyond the exactly solvable limit~\cite{Abrikosov1963,Altland2023}.
Further, our model generically exhibits LSs, whose physical role we examine. Finally, we discuss quantum phase transitions separating PG-FLs from Landau FLs across different dimensions.

\textit{Hamiltonian and solution.---}
We consider spinful fermions on an arbitrary lattice with a CH interaction, 
\begin{subequations}
\label{eq:Hq}
\begin{align}
H_q &= H_0 + H_{\mr{int}} \, , \;\;
H_0 = \sum_{ij\sigma} (t_{ij\sigma}^{q} - \mu \delta_{ij}) c^{\dag}_{i\sigma} c^{\pdag}_{j\sigma}
\\
H_{\mr{int}} &= \sum_{ij\sigma} t_{ij\sigma}^{q} (4 n_{i\bar{\sigma}} n_{j\bar{\sigma}} - 2 n_{i\bar{\sigma}} - 2 n_{j\bar{\sigma}}) c^{\dag}_{i\sigma} c^{\pdag}_{j\sigma} \, ,
\end{align}
\end{subequations}
where $c_{i\sigma}$ annihilates a spin-$\sigma$ electron at site $i$, $\bar{\sigma}$ denotes the spin projection opposite to $\sigma$, and $n_{i\sigma} = c^{\dagger}_{i\sigma} c^{\pdag}_{i\sigma}$.

Correlated hopping was originally introduced into the condensed matter community by Hirsch and Marsiglio~\cite{Hirsch1989,Hirsch1989a} to study superconductivity.
Since then, it has been extensively studied~\cite{Essler1992,Bariev1993,Simon1993,Arrachea1994,Gagliano1995,Schulz1998,Japaridze1999,Arrachea2000,Aligia2000,Aligia2007,Jiang2023,Kovalska2025}
and has recently garnered renewed interest due to its realization in engineered quantum systems~\cite{Greschner2014,Bermudez2015,Barbiero2019,Goerg2019,Schweizer2019,Lienhard2020,Montorsi2022,Jamotte2022,Segura2023,RouraBas2023}.
Because CH terms are usually symmetry-allowed, they can generically arise in effective low-energy models~\cite{Hirsch1991,Schuettler1992,Simon1993a,Feiner1996,Raimondi1996,Simon1997,Jiang2023} derived via downfolding procedures~\cite{Wilson1983,Profe2025}.
Equation~\eqref{eq:Hq} should be understood as such an effective low-energy model; its realization through downfolding of microscopic models is left for future work.

For an exact solution to Eq.~\eqref{eq:Hq}, we define~\cite{Zhu2013,Hubbard_commutator}
\begin{align}
q_{i\sigma} = c_{i\sigma} (2 n_{i\bar{\sigma}} - 1) \, ,
\end{align}
which are canonical fermionic operators~\footnote{The canonical anticommutation relations $\{q_{i\sigma},q_{j\sigma'}\} = 0$  and $\{q^{\dagger}_{i\sigma}, q_{j\sigma'}\} = \delta_{ij} \delta_{\sigma\sigma'}$  follow from $(2 n_{i\bar{\sigma}} - 1)^2 = 1$. The $q$-operators and electronic operators are \textit{not} mutually canonical.}.
Further, \textit{local} density and spin operators are quadratic in $q_{i\sigma}$,
\begin{subequations}
\label{subeq:n_S}
\begin{align}
\label{eq:n_i}
n_{i\sigma} &= c^{\dagger}_{i\sigma} c_{i\sigma} = q^{\dagger}_{i\sigma} q_{i\sigma} \, ,
\\
\vec{S}_{i} &= \tfrac{1}{2}\sum_{ss'} c^{\dagger}_{is} \boldsymbol{\sigma}_{ss'} c_{is'} = \tfrac{1}{2}\sum_{ss'} q^{\dagger}_{is} \boldsymbol{\sigma}_{ss'} q_{is'} \, .
\end{align}
\end{subequations}
Since $H_q$ is quadratic if written in terms of $q^{(\dagger)}_{i\sigma}$~\footnote{This follows from $n_{i\bar\sigma}^2 = n_{i\bar\sigma}$ and $(2 n_{i\bar{\sigma}} - 1)(2 n_{j\bar{\sigma}} - 1) 
= 4 n_{i\bar{\sigma}} n_{j\bar{\sigma}} - 2 n_{i\bar{\sigma}} - 2 n_{j\bar{\sigma}} + 1$.},
\begin{align}
H_q = \sum_{ij\sigma} (t^{q}_{ij\sigma} - \mu \delta_{ij}) q^{\dagger}_{i\sigma} q^{\pdag}_{j\sigma} = \sum_{\vec{k}\sigma} \epsilon_{\vec{k}\sigma} \, q^{\dagger}_{\vec{k}\sigma} q^{\pdag}_{\vec{k}\sigma} \, ,
\end{align}
where $q_{\vec{k}\sigma}$ and $\epsilon_{\vec{k}\sigma}$ is the Fouriertransforms of $q_{i\sigma}$ and $t^{q}_{ij\sigma} - \mu \delta_{ij}$, respectively,
its  eigenstates are 
\begin{align}
\label{eq:q_states}
\ket{\vec{k}_1\sigma_1\cdots\vec{k}_n\sigma_n}_q &= q^{\dagger}_{\vec{k}_1\sigma_1} \cdots q^{\dagger}_{\vec{k}_n\sigma_n} \ket{0} \, ,
\end{align}
where $\ket{0}$ is the empty state. 

$H_q$ therefore describes a Fermi gas of canonical quasiparticles~(QP), created by $q^{\dagger}_{\vec{k}\sigma}$.
Thermodynamic properties such as specific heat, compressibility, or magnetic susceptibility [cf. Eq.~\eqref{subeq:n_S}] are identical to those of a free electron gas. 
Nevertheless, the electrons are strongly interacting, and the eigenstates in Eq.~\eqref{eq:q_states} are generically not superpositions of a few electronic Slater determinants.

Indeed, the single-particle physics of $H_q$ is quite different from a free electron gas.
For the former, the momentum space occupation numbers of the $q$-particles, $q^{\dag}_{\vec{k}\sigma} q_{\vec{k}\sigma}$ are conserved.
These are very different from the electronic momentum-space occupation numbers (in contrast to the \textit{local} occupation numbers, cf.\ Eq.~\eqref{subeq:n_S}), 
since $c_{\vec{k}\sigma}$ is a composite operator when written in terms of $q$-operators, with contributions across all momenta,
\begin{align}
\label{eq:ck_qk}
c_{\vec{k}\sigma} &= 
\frac{2}{N}\sum_{\vec{k}'\vec{q}}  q^{\phantom{\dagger}}_{\vec{k}+\vec{q}\sigma} q_{\vec{k}'\bar{\sigma}}^{\dagger} q^{\phantom{\dagger}}_{\vec{k}'-\vec{q}\bar{\sigma}}  - q^{\phantom{\dagger}}_{\vec{k}\sigma} \, ,
\\
n_{\vec{k}\sigma} &= c^{\dagger}_{\vec{k}\sigma} c^{\pdag}_{\vec{k}\sigma} \neq q^{\dagger}_{\vec{k}\sigma} q^{\pdag}_{\vec{k}\sigma} \, .
\end{align}
Here, $N$ is the number of lattice sites. 
We will show below that the QP weight of the electrons is $Z_{\sigma} = (2n_{\bar{\sigma}} - 1)^2$, where $n_{\sigma}$ is the density of spin-$\sigma$ electrons. 
Thus, when $n_{\bar{\sigma}} = \tfrac{1}{2}$, the QP weight is zero and $\langle n_{\vec{k}\sigma} \rangle$ does not exhibit a discontinuity at $\vec{k}_{\mr{F}}$.

Further, $q$ and $c$ operators are related by the unitary
\begin{align}
\label{eq:cq_unitary}
\mc{U} &= \mc{U}^{\dagger} = \prod_i \mr{e}^{\mr{i}\pi n_{i\uparrow} n_{i\downarrow}} \, , \;
q_{i\sigma} = \mc{U} c_{i\sigma} \mc{U} \, . 
\end{align}
This will be useful below, where we study quantum phase transitions when tuning between $H_q$ and free electrons.
If there is a single transition, Eq.~\eqref{eq:cq_unitary} tells us that it occurs when the Hamiltonian is left invariant by $\mc{U}$.

\textit{Action and stability.---}
The PG-FL described by $H_q$ is only practical if it is stable to perturbations, and if these can be systematically dealt with.
Most important are additional single-electron hopping or two-electron interactions, such that the total Hamiltonian is
\begin{align}
\label{eq:H_generic}
H &= H_{c} + H_2 + H_{q} \; , \quad  H_c = \sum_{ij\sigma}  t^{c}_{i j\sigma} c^{\dagger}_{i\sigma} c_{j\sigma} 
\\ \nonumber
H_2 &= \sum_{ijk\ell} \sum_{\sigma\sigma'} U_{ijk\ell}^{\sigma\sigma'} \, c^{\dag}_{i\sigma} c^{\dag}_{j\sigma'} c^{\pdag}_{k\sigma'} c^{\pdag}_{\ell\sigma} \, .
\end{align}
If $H_c$ is the dominant term, it is well established that  $H$ generically ($d>1$) gives rise to a Landau FL.
It is then standard to write the partition function in terms of a path integral, which facilitates a perturbative expansion around $H_c$, a stability analysis~\cite{Shankar1994} of the fixed point defined by $H_c$, and the study of symmetry-breaking order out of a L-FL.
For that, it is crucial that the action corresponding to $H_c$ is quadratic.

To achieve the same when $H_q$ is dominant, we define the coherent states
\begin{align}
\label{eq:q_coherent}
|\chi \rangle_q &= \prod_{i\sigma} \mr{e}^{-\chi_{i\sigma} q^{\dagger}_{i\sigma}} |0\rangle
\, , \;
q_{j\sigma'} |\chi \rangle_q = \chi_{j\sigma'} |\chi \rangle_q \, ,
\end{align}
with Grassmann fields $\chi_{i\sigma}$, and 
write the partition function of $H$ [Eq.~\eqref{eq:H_generic}] as a path integral,
\begin{align}
\label{eq:Zq_pathIntegral}
Z = \mr{Tr} \, \mr{e}^{-\beta H} = \int \mc{D}\{\chi\} \, \mr{e}^{-S^q[\chi]} \, .
\end{align}
The explicit form of $S^{q}[\chi]$ including single-electron hopping and density interactions 
 is given in the Supplemental Material (SM)~\cite{supplement}.
 \nocite{Rojas1995,Rieger1999,Zlatic2000,Mori1965,Haydock1980a,Viswanath1994,Tiegel2014,julien2008_EROS,Auerbach2018,Auerbach2019,Foley2024_Liouvillian,Pelz2026,Bulla1998,Nozieres1980,Andrei1984,Tsvelick1985,Affleck1991,Ludwig1991,Anders2005}
Crucially, the contribution from $H_q$ is quadratic,
\begin{align}
\label{eq:Sq}
S^{q}_0[\chi] = \int_{0}^{\beta} \sum_{\vec{k}\sigma} \bar{\chi}_{\vec{k}\sigma}(\tau) (\partial_{\tau} + \epsilon_{\vec{k}\sigma}) \chi_{\vec{k}\sigma}(\tau) \, .
\end{align}
Single-electron hopping contributes terms up to $\mc{O}(\chi^{6})$ while $H_2$ generically contributes up to $\mc{O}(\chi^{8})$. 
If $H_2$ describes a density-density interaction, it is quartic in the $\chi$-fields. 
As long as $H_c$ and $H_2$ are not long-ranged, they are not relevant perturbations~\cite{Shankar1994} to the fixed point described by the action~\eqref{eq:Sq}.

Due to Eq.~\eqref{subeq:n_S}, a symmetry-breaking instability in the density or spin channels leads to conventional 
charge or spin density wave order, albeit on top of a PG-FL. 
In the Cooper channel, on the other hand, the order parameter $q^{\dagger}_{\vec{k}\uparrow} q^{\dagger}_{-\vec{k}\downarrow}$ is unconventional.
We leave a detailed investigation for future work. 

\begin{figure}
\includegraphics[width=\linewidth]{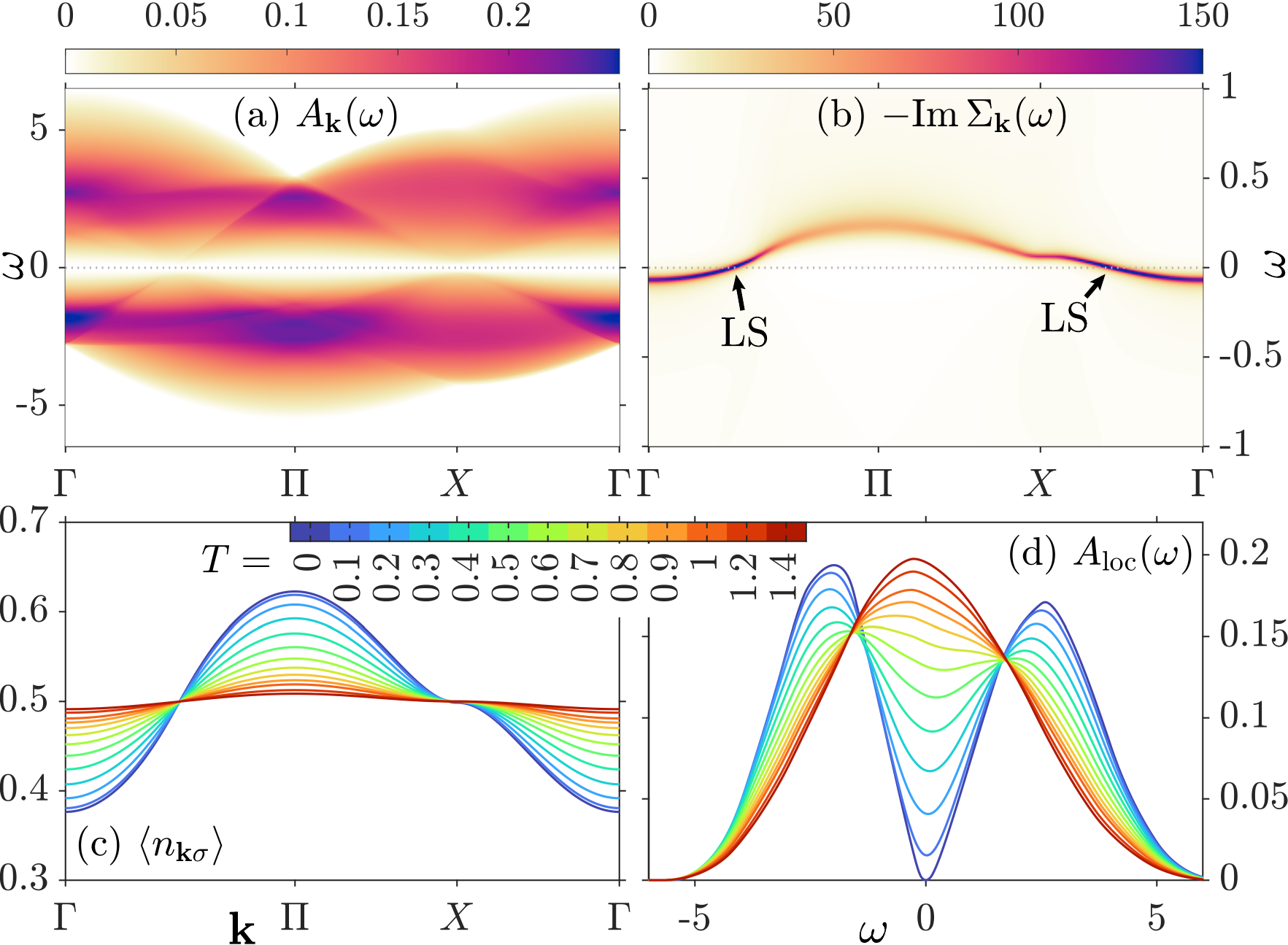}
\caption{
Single-particle properties on the square lattice with $t = 0.5$ and $t' = -0.15 t$ [cf.\ Eq.~\eqref{eq:t_tp_dispersion}] at half-filling:
(a) $T=0$ $\vec{k}$-resolved spectral function [$\Gamma = (0,0)$, $X = (0,\pi)$, $\Pi = (\pi,\pi)$], (b) corresponding self-energy (LS: Luttinger surface), (c) $T$-dependent electronic occupation number
and (d) local spectral function.
We used periodic boundary conditions with $L_x \times L_y =  1024 \times 1024$. }
\label{fig:Hq_ttp_square}
\end{figure}

\textit{Pseudogap.---}
The single-electron Green's function of the PG-FL is non-trivial, and given by~\cite{supplement}
\begin{align}
A^{(3)}_{\vec{k}\vec{k}'\vec{q}\sigma} &= \left[1-f(\epsilon_{\vec{k}'-\vec{q}\bar{\sigma}})\right] \left[1-f(\epsilon_{\vec{k}+\vec{q}\sigma})\right] f(\epsilon_{\vec{k}'\bar{\sigma}})
\\ \nonumber
&+ f(\epsilon_{\vec{k}'-\vec{q}\bar{\sigma}}) f(\epsilon_{\vec{k}+\vec{q}\sigma}) \left[1 - f(\epsilon_{\vec{k}'\bar{\sigma}})\right]
\\
E^{(3)}_{\vec{k}\vec{k}'\vec{q}\sigma} &=  \epsilon_{\vec{k}' - \vec{q}\bar{\sigma}} - \epsilon_{\vec{k}'\bar{\sigma}} + \epsilon_{\vec{k}+\vec{q}\sigma}
\\
\label{eq:Gk}
G_{\vec{k}\sigma}(z) &= 
\underbrace{4\int_{\vec{k}'}\int_{\vec{q}}
\frac{A^{(3)}_{\vec{k}\vec{k}'\vec{q}\sigma}}{z - E^{(3)}_{\vec{k}\vec{k}'\vec{q}\sigma}}}_{\equiv G^{(3)}_{\vec{k}\sigma}(z)}
+ \underbrace{\frac{(2 n_{\bar{\sigma}} - 1)^2}{z - \epsilon_{\vec{k}\sigma}}}_{\equiv G^{\mr{qp}}_{\vec{k}\sigma}(z)} \, , 
\end{align}
where we have used 
\begin{align}
 f(\epsilon_{\vec{k}\sigma}) &= \frac{1}{\mr{e}^{\beta \epsilon_{\vec{k}\sigma}} + 1} \; , \;\; \mr{with } \,\, \beta = 1/T \, , \\
n_{\bar{\sigma}} &= \int_{\vec{k}} f(\epsilon_{\vec{k}\bar{\sigma}}) \, , \quad \int_{\vec{k}} = \int_{\mr{BZ}} \frac{\mr{d}^{d} k}{\mc{V}_{\mr{BZ}}} \, .
\end{align}
The Green's function in Eq.~\eqref{eq:Gk} has two terms, $G^{(3)}_{\vec{k}\sigma}(z)$ and $G^{\mr{qp}}_{\vec{k}\sigma}(z)$. 
The latter is the coherent QP pole
due to the overlap of $c^{\dagger}_{\vec{k}\sigma}$ with $q^{\dagger}_{\vec{k}\sigma}$. This overlap (the QP weight) is $Z_{\sigma} = (2 n_{\bar{\sigma}} - 1)^2$,
ie.\ it depends quadratically on the deviation from half-filling of the opposite-spin electrons. At half-filling, the QP weight is zero, ie.\ we have a \textit{coherent Fermi liquid with zero QP weight}~\cite{Nandkishore2012}.

The first term, $G^{(3)}_{\vec{k}\sigma}(z)$, describes the 
fractionalization~\footnote{The electron is composed of three $q$-particles which are deconfined.}
of the electron 
created by $c^{\dagger}_{\vec{k}\sigma}$ into
three $q$-QPs (two particle and one hole excitation). 
The integral that determines its spectral function $A^{(3)}_{\vec{k}\sigma}(\omega) = -\tfrac{1}{\pi}\mr{Im} \, G^{(3)}_{\vec{k}\sigma}(\omega^{+})$ is well known from the computation 
of the second-order self-energy of L-FLs~\cite{Hodges1971,Galan1993,Daul1997,DasSarma2021} and is governed by phase-space constraints. 
Those generically~\footnote{Exceptions occur in $d=1$ and at van Hove singularities, but $A^{(3)}_{\vec{k}\sigma}(\omega=0,T=0) = 0$ still holds in these cases.} lead to a quadratic low-frequency and temperature dependence,
\begin{align}
\label{eq:A3_lowfrequency}
A^{(3)}_{\vec{k}\sigma}(\omega,T) = C_{\vec{k}\sigma} (\omega^2 + \pi^2 T^2) + \cdots \, ,
\end{align}
resulting in a PG which gradually opens when lowering the temperature.
This PG arises because the phase space for an electron to fractionalize into three $q$-QPs is highly restricted close to the $q$-QP Fermi surface (FS), and it therefore has a purely fermionic origin.
In the vicinity of $n_{\bar{\sigma}} = \tfrac{1}{2}$, the spectral weight of $A^{(3)}_{\vec{k}\sigma}(\omega)$ is $1 - Z_{\sigma} \gg Z_\sigma$, and the PG is the dominant spectral feature.

The $\vec{k}$-space occupation $\langle n_{\vec{k}\sigma} \rangle$ is obtained by integrating $A_{\vec{k}\sigma}(\omega)$ against the Fermi-Dirac distribution.
It has a smooth contribution from $A_{\vec{k}\sigma}^{(3)}(\omega)$ that does not exhibit non-analyticities, while the QP part contributes a step of size $Z_{\sigma}$ at $T=0$ at the FS.

To make our discussion more concrete, we consider the spin-independent $t$-$t'$ square lattice dispersion
\begin{align}
\label{eq:t_tp_dispersion}
\epsilon_{\vec{k}} = -2t \left[ \cos k_x + \cos k_y \right] - 4 t' \cos k_x \cos k_y - \mu \, ,
\end{align}
where we choose $t = \tfrac{1}{2}$, $t' = -0.15 t$ and half-filling.
The $\vec{k}$-dependnet spectral function and self-energy at $T=0$ are shown in Fig.~\ref{fig:Hq_ttp_square}(a,b). 
As expected, the spectral function features a PG around $\omega = 0$, and spectral features are entirely incoherent due to the electron fractionalizing into three QP excitations. 
The self-energy exhibits a dispersive pole, which suppresses the spectral weight around $\omega = 0$. 
Indeed, at $T=0$ and $\omega=0$, the Green's function generically exhibits zeros and therefore 
Luttinger surfaces~(LS), discussed below in more detail. 

Figure~\ref{fig:Hq_ttp_square}(c) shows
 $\langle n_{\vec{k}\sigma} \rangle$. 
Because $Z_{\sigma} = 0$ ($n_{\sigma} = \tfrac{1}{2}$) 
$\langle n_{\vec{k}\sigma} \rangle$ does not exhibit non-anlytic features. 
Further, at $T\to 0$, $\langle n_{\vec{k}\sigma} \rangle$ takes values in the range $0.38 \lesssim \langle n_{\vec{k}\sigma} \rangle \lesssim 0.62$.
This differs
significantly from the idempotent values $0$ or $1$ characteristic of electronic Slater determinants, illustrating
that the ground state is strongly correlated and not close to an electronic Slater determinant. 
A more quantitative discussion in terms of the nonfreeness~\cite{Gottlieb2005,Gottlieb2014,AlivertiPiuri2024} is provided in the SM~\cite{supplement}.

In Fig.~\ref{fig:Hq_ttp_square}(d), we show the temperature dependence of the local spectral function. At low temperatures, it
exhibits a pseudogap which gradually fills as the temperature is increased and eventually disappears at $T>t$. 

\begin{figure}
\includegraphics[width=\linewidth]{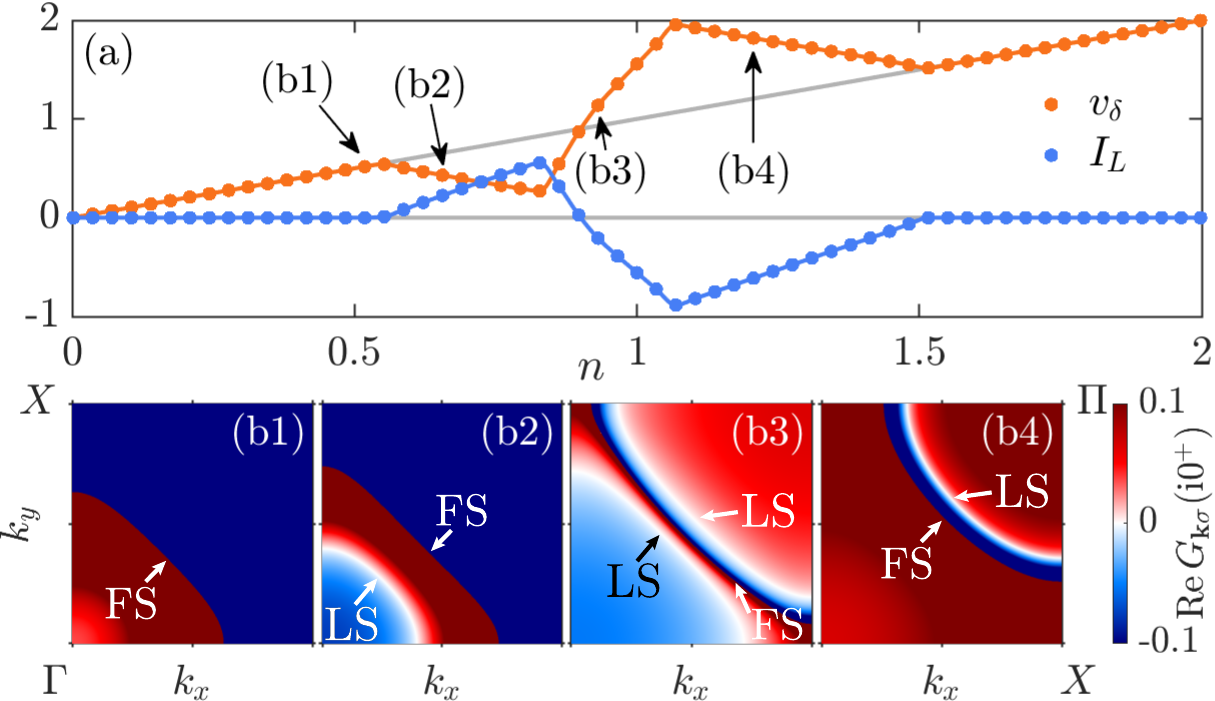}
\caption{(a) Average phase shift $v_{\delta}$ [c.f.\ Eq.~\eqref{eq:Luttinger_sum_rule}] and Luttinger integral $I_L$ versus filling, for a square lattice with $t = 0.5$ and $t' = -0.15 t$. 
(b) Real parts of $G_{\vec{k}\sigma}(\mr{i} 0^{+})$ for selected fillings marked in (a), 
illustrating the location of Luttinger surfaces (LS, white) and Fermi surfaces (FS, abrupt blue to red boundary).}
\label{fig:LuttingerSumRule}
\end{figure}

\textit{Luttinger surfaces.---}
In the vicinity of half-filling, our model generically features a PG-FL with LSs. 
Based on this model, it is therefore possible to clarify some of the ongoing debate~\cite{Dzyaloshinskii2003,Coleman2005,Rosch2007,Farid2007,Stanescu2007,Dave2013,Seki2017,Heath2020,Hazra2021,Fabrizio2020,Fabrizio2022,Skolimowski2022,Fabrizio2023,LaNave2025} on the physical content of LSs and their contribution to the low-energy response of PG-FLs.

The locations of Luttinger and Fermi surfaces are 
\begin{align}
G^{(3)}_{\kL \sigma}(0) = \frac{Z_{\sigma}}{\epsilon_{\kL\sigma}} \, , \quad A_{\kL\sigma}(\omega) \propto \omega^2 \, , 
\quad \epsilon_{\kF\sigma} = 0 \, ,
\end{align}
where $\kL$ and $\kF$ are Luttinger and Fermi wavevectors, respectively.
Coherent $q$-QP excitations occur at $\kF$, which is generically different from $\kL$.
As these are the only elementary excitations in our system, it
shows that there generically are no physical QP excitations at LSs. 
We emphasize that this is also true at half-filling, where $G_{\vec{k}\sigma}(\omega)$ does not exhibit a pole at $\kF$ since $Z_{\sigma} = 0$
and, generically, $G^{(3)}_{\kF \sigma}(0) \neq 0$.

However, LSs may still contribute when the low-energy, long-distance response (e.g.\ the specific heat) of PG-FLs is computed from electronic propagators, as recently argued by Fabrizio~\cite{Fabrizio2020,Fabrizio2022}
under the ($T=0$) condition $A_{\kL \sigma}(\omega) \propto \omega^2$.
This condition is met in our model, since $A_{\kL \sigma}(\omega) = A^{(3)}_{\kL \sigma}(\omega) \propto \omega^2$ [c.f.\ Eq.~\eqref{eq:A3_lowfrequency}], i.e.\ the phenomenology studied in Refs.~\cite{Fabrizio2020,Fabrizio2022} is realized in our microscopic model.
Nevertheless, we find that there is \textit{no} LS contribution to the low-energy specific heat, as explicitly demonstrated in the ``End Matter''.
When $Z_{\sigma} \neq 0$, this follows from the fact that the QP propagator defined in Refs.~\cite{Fabrizio2020,Fabrizio2022} exhibits a pole of weight $1$ at $\kF$, 
thus already accounting for the entire low-energy response of the $q$-QP, i.e.\ there is no room for additional LS contributions. 
A more in-depth analysis of where the compelling arguments in Refs.~\cite{Fabrizio2020,Fabrizio2022} go wrong requires a detailed investigation of the vertex functions of the PG-FL, which we leave for future work.

The appearance of LSs is usually connected to a violation of the Luttinger sum rule,
which connects the average phase shift of $G_{\vec{k}\sigma}(\omega)$, $v_{\delta}$, to the particle number~\cite{Luttinger1960,Kokalj2007,Seki2017},
\begin{subequations}
\label{eq:Luttinger_sum_rule}
\begin{align}
n &= \sum_{\sigma} n_{\sigma} =  \int_{\vec{k}} \sum_{\sigma} \delta_{\vec{k}\sigma} + I_{L} =  v_{\delta} + I_{L} \, ,
\\
\delta_{\vec{k}\sigma} &= 1-\frac{1}{\pi} \mr{Im} \, \ln \, G_{\vec{k}\sigma}^{-1}(\mr{i} 0^{+}) \, ,
\end{align}
\end{subequations}
where $I_L$ is the Luttinger integral. 
In L-FLs, $I_L = 0$ and $v_{\delta}$ is the FS volume~\cite{Luttinger1960}.
However, LSs also contribute to $v_{\delta}$~\cite{Dzyaloshinskii2003}, and it is known that $I_L$ can be non-zero in non-perturbative settings.
Whether $I_L$ is generically constrained is a topic of ongoing debate~\cite{Dzyaloshinskii2003,Coleman2005,Rosch2007,Farid2007,Stanescu2007,Dave2013,Seki2017,Heath2020,Hazra2021,Fabrizio2020,Fabrizio2022,Skolimowski2022,Fabrizio2023,LaNave2025}.

To make progress on this front, we show both $v_{\delta}$ and $I_L$ for the previously discussed square lattice model for different fillings in Fig.~\ref{fig:LuttingerSumRule}(a), 
while Fig.~\ref{fig:LuttingerSumRule}(b1-b4) show $\mr{Re} \, G_{\vec{k}\sigma}(\mr{i} 0^{+})$ at selected fillings. 
$v_{\delta}$ corresponds to the portion of red area ($\mr{Re} \, G_{\vec{k}\sigma}(\mr{i} 0^{+}) > 0$) in Fig.~\ref{fig:LuttingerSumRule}(b1-b4). 
Sharp red-to-blue jumps indicate FSs, while LSs are visible as white lines separating red and blue areas. 
Figure~\ref{fig:LuttingerSumRule} shows that $I_L$ deviates from zero in the vicinity of half-filling, seemingly without special constraints.
This deviation is due to the presence of LSs, whose additional contributions to $v_{\delta}$ exactly cancel with $I_L$.
Thus, the FS volume of this model still coincides with the filling, a non-generic situation, especially in Mott systems.
Nevertheless, this shows that $I_L$ is generically not quantized in terms of special fractions of the BZ.

Figures~\ref{fig:LuttingerSumRule}(b1-b4) illustrate how the locations of the LSs evolve with filling. 
Within the regime where LSs are present, the system either exhibits a single LS [Figs.~\ref{fig:LuttingerSumRule}(b2,b4)], or two LSs in close vicinity to half-filling [Fig.~\ref{fig:LuttingerSumRule}(b3)]. 
In the latter nearly half-filled case, the QP weight approaches zero ($Z_{\sigma} \simeq 0$), splitting the LS condition: the first LS is determined by $G^{(3)}_{\kL\sigma}(0) \simeq 0$, while the second is located at $\epsilon_{\kL\sigma} = Z_{\sigma}/G^{(3)}_{\kL\sigma}(0) \simeq 0$ while $G^{(3)}_{\kL\sigma}(0)$ is significantly non-zero. 
Consequently, this second LS is nearly degenerate with the Fermi surface ($\epsilon_{\kF\sigma} = 0$) and ultimately annihilates with it as $Z_{\sigma} \to 0$.
This process leaves only the single LS defined by $G^{(3)}_{\kL\sigma}(0) = 0$ at half-filling.
Our model therefore provides a mechanism for FS-LS annihilation, and for the near-degeneracy of a LS with (part of) a FS, which is responsible for Fermi arcs in the Hubbard model~\cite{Stanescu2006}.

\begin{figure}
\includegraphics[width=\linewidth]{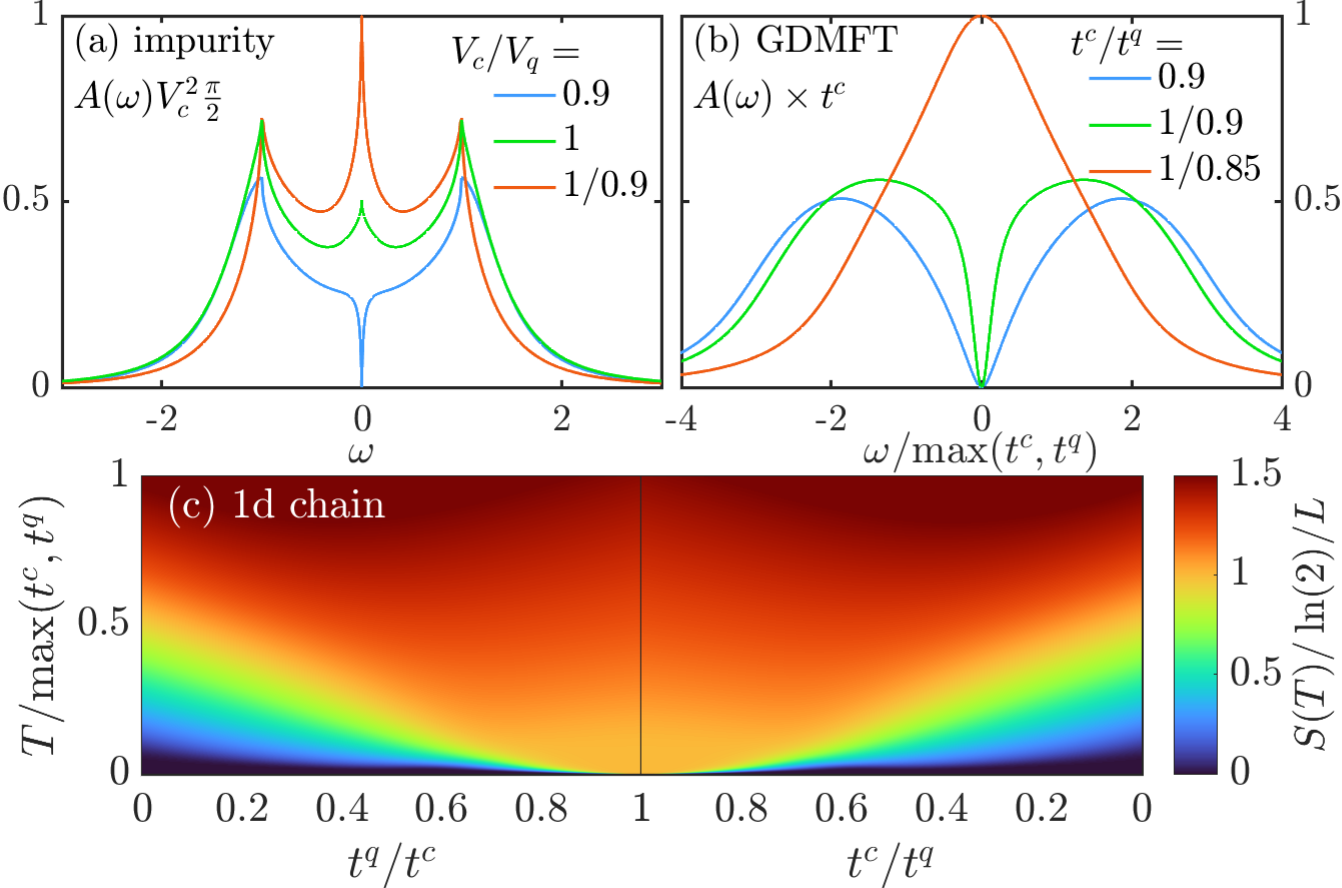}
\caption{(a) Spectral function of an impurity model with hybridization $V_x$ for both $q$ and $c$ [c.f.\ Eq.~\eqref{eq:Himp}]. 
(b) Local spectral function from a GDMFT solution on the $d \to \infty$ Bethe lattice. The GDMFT has been initialized with a PG-FL solution.
(c) Entropy of an $L=8$ 1d chain.}
\label{fig:QC_transition}
\end{figure}

\textit{PG-FL to L-FL transitions.---}%
We now study physics beyond the exactly solvable limit by investigating the interplay between
single-particle hopping with amplitude $t^c$ and CH with amplitude $t^q$ while keeping $H_2 = 0$ in Eq.~\eqref{eq:H_generic}.
This allows us to tune between the limiting cases of free $q$-QP and free electrons, with the goal
to study possible quantum phase transitions~(QPTs) separating them.
Without symmetry constraints,
\begin{align}
H_{\phi} = \sum_{ij\sigma} t_{ij\sigma} \mr{e}^{-\mr{i} \phi (n_{i\bar{\sigma} }- n_{j\bar{\sigma}})} c^{\dagger}_{i\sigma} c_{j\sigma} 
\end{align}
smoothly interpolates between free electrons ($\phi = 0$) and free $q$-QP ($\phi = \pi$) without a phase transition, since $H_{\phi}$ describes a gas of free fermions annihilated by $p_{i\sigma} = \mr{e}^{\mr{i} \phi n_{i\bar{\sigma}}} c_{i\sigma}$.
However, there is no smooth path if the Hamiltonian is real~\footnote{Same-sign $q$-particle and electronic hopping are frustrated close to half-filling.
For a free-electron Hamiltonian with hopping $t^{c}_{ij}$, we have $t^c_{ij\sigma} \langle c^{\dag}_{i\sigma} c^{\pdag}_{j\sigma}\rangle < 0$.
The corresponding $q$-particle bond expectation value is $\langle q^{\dag}_{i\sigma} q^{\pdag}_{j\sigma} \rangle = \langle c^{\dag}_{i\sigma} c^{\pdag}_{j\sigma} \rangle [(2 n_{\bar{\sigma}} - 1)^2 - |\langle c^{\dag}_{i\bar{\sigma}} c^{\pdag}_{j\bar\sigma} \rangle|^2]$.
Close to half-filling, its sign is opposite to $\langle c^{\dag}_{i\sigma} c^{\pdag}_{j\sigma} \rangle$, and a small $q$-particle hopping amplitude $t_{ij}$ with the same sign as $t^{c}_{ij}$ therefore raises the energy, i.e.\ they are frustrated.} 
and particle-hole symmetric~\cite{supplement}, which is what we focus on below.

First, we consider an impurity model with Hamiltonian
\begin{align}
\label{eq:Himp}
H_{\mr{imp}} &= \sum_{\sigma} \left[V_{c}  c^{\dagger}_{\sigma} f_{0,c,\sigma} + V_{q} q^{\dagger}_{\sigma} f_{0,q,\sigma} + \mr{h.c.}\right]
\\ \nonumber
&+ \tfrac{1}{2} \sum_{\ell \geq 0} \sum_{\sigma = \uparrow,\downarrow} \sum_{x = q,c} \left[ f^{\dagger}_{\ell x\sigma} f^{\pdag}_{\ell+1 x\sigma}  + \mr{h.c.} \right] \, ,
\end{align}
where $c_{\sigma}$ and $q_{\sigma} = c_{\sigma}(2n_{\bar{\sigma}} - 1)$ act on a spinful fermionic impurity orbital, coupled to two semi-infinite tight-binding chains with annihilation operators $f^{\pdag}_{\ell x\sigma}$.
The limits $V_q = 0$, $V_c \neq 0$ and $V_q \neq 0$, $V_c = 0$ are exactly solvable and feature a resonance or PG in the impurity spectral function, respectively. 
The QPT is expected at $V_c = V_q$, where $H_{\mr{imp}}$ is left invariant by $\mc{U}$ [Eq.~\eqref{eq:cq_unitary}].

We solve Eq.~\eqref{eq:Himp} at arbitrary $V_{x}$ with the Numerical Renormalization Group~(NRG)~\cite{Wilson1975,Bulla2008,Weichselbaum2007,Weichselbaum2012,Kugler2022}, using the QSpace-based~\cite{Weichselbaum2012a,Weichselbaum2020,Weichselbaum2024} MuNRG package~\cite{Lee2016,Lee2017}; see the SM~\cite{supplement} for details. 
The resulting $T=0$ impurity spectral functions are shown in Fig.~\ref{fig:QC_transition}(a). As expected, we find a resonance with a height that fulfills the Friedel sum rule ($A(0)V_c^2\tfrac{\pi}{2} = 1$) at $V_c > V_q$ (red)
and a PG at $V_c < V_q$ (blue). At $V_c = V_q$ (green), the spectral function features a resonance with exactly half the height of the Friedel sum rule value, $A(0)V_c^2\tfrac{\pi}{2} = \tfrac{1}{2}$. 
At this point, we find two-channel Kondo type non-FL behavior, see the SM~\cite{supplement} for more detail. 

Second, we consider Eq.~\eqref{eq:H_generic} on
the $d\to\infty$ Bethe lattice 
with $H_2 = 0$, at half-filling and nearest-neighbor hoppings. It
can be solved via generalized dynamical mean-field theory~(GDMFT)~\cite{Stanescu2004,supplement}, which maps the lattice model to an impurity model of the form~\eqref{eq:Himp}, with self-consistently determined bath parameters. 
The resulting local spectral functions for selected parameters $t^c$ and $t^{q}$ are shown in Fig.~\ref{fig:QC_transition}(b).
As expected, we find a PG-FL for $t^q \gg t^c$ (not shown). Using the deep PG-FL solution as initialization for GDMFT, we find a PG-FL solution at both $t^{q} > t^c = 0.9 t^q$ (blue) \textit{and} $t^{q} < t^c = t^q/0.9$ (green). 
For a larger ratio of $t^{q} < t^c = t^q/0.85$ (red) eventually gives a L-FL. The stability of the PG-FL
in the $t^q < t^c$ region indicates that the QPT is first order.

Finally, we study Eq.~\eqref{eq:H_generic} with $H_2 = 0$, at half-filling and nearest-neighbor $t^c$ and $t^q$ on a one-dimensional $L=8$ chain with periodic boundary conditions. Figure~\ref{fig:QC_transition}(c) shows the entropy versus temperature and hopping ratios $t^c/t^q$. We find an accumulation of low-temperature entropy when approaching $t^c/t^q = 1$; at 
$t^c/t^q = 1$, we find an extensive $T\to 0$ entropy of $S(T\to0) = L\ln 2$. 
Interestingly, the transition point at $t^c/t^q = 1$ is exactly solvable and has been studied in detail in Ref.~\cite{Arrachea1994}.
Phase separation seems to be absent in the
thermodynamic limit, indicating a second-order QPT. 
Further Ref.~\cite{Arrachea1994} shows that both doping and a local Hubbard interaction do not immediately destabilize the quantum critical phase,
indicating that particle-hole symmetry is not necessary for the existence of a QPT. 

\textit{Outlook.---}
Interesting future avenues include exploring $q$-particle interactions and symmetry breaking out of PG-FLs, particularly superconductivity. 
It also remains to be seen under what circumstances $q$-particles emerge in microscopic settings lacking explicit CH, and whether they are present in known PG metals. 
As a first step, the ``End Matter'' confirms a $q$-particle resonance within the PG phase of the two-impurity Anderson model~\cite{Zhu2013}. 

Finally, we conjecture that $q$-particles may play a role in Mott PG metals, which phenomenologically arise via the hybridization of electrons with an emergent fermion~\cite{Zhu2013,Sakai2015,Sakai2016,Imada2019,Ido2020,Ledwith2025,Ledwith2025a}. 
Through equations of motion, this emergent fermion can be related to the $q$-particle in Hubbard models~\cite{Zhu2013}, suggesting a significant $q$-particle character along hidden FS segments. 
If present, such deconfined $q$-particles could explain some of the spectral features observed in Mott PG metals.

\begin{acknowledgments}
We are especially grateful to Michele Fabrizio for a very insightful discussion and valuable comments.
We also thank Aaditya Panigrahi, Jan von Delft, Natan Andrei, Patrick Ledwith, Piers Coleman, Pradip Kattel, Sayantan Roy, Seung-Sup Lee, and Yicheng Tang for insightful discussions. 
GK was supported by the National Science Foundation Grant No. DMR-1733071.
AG acknowledges support from the Abrahams Postdoctoral Fellowship of the Center for Materials Theory at Rutgers University.
\end{acknowledgments}
\bibliography{references} 

\begin{thebibliography}{155}%
\makeatletter
\providecommand \@ifxundefined [1]{%
 \@ifx{#1\undefined}
}%
\providecommand \@ifnum [1]{%
 \ifnum #1\expandafter \@firstoftwo
 \else \expandafter \@secondoftwo
 \fi
}%
\providecommand \@ifx [1]{%
 \ifx #1\expandafter \@firstoftwo
 \else \expandafter \@secondoftwo
 \fi
}%
\providecommand \natexlab [1]{#1}%
\providecommand \enquote  [1]{``#1''}%
\providecommand \bibnamefont  [1]{#1}%
\providecommand \bibfnamefont [1]{#1}%
\providecommand \citenamefont [1]{#1}%
\providecommand \href@noop [0]{\@secondoftwo}%
\providecommand \href [0]{\begingroup \@sanitize@url \@href}%
\providecommand \@href[1]{\@@startlink{#1}\@@href}%
\providecommand \@@href[1]{\endgroup#1\@@endlink}%
\providecommand \@sanitize@url [0]{\catcode `\\12\catcode `\$12\catcode
  `\&12\catcode `\#12\catcode `\^12\catcode `\_12\catcode `\%12\relax}%
\providecommand \@@startlink[1]{}%
\providecommand \@@endlink[0]{}%
\providecommand \url  [0]{\begingroup\@sanitize@url \@url }%
\providecommand \@url [1]{\endgroup\@href {#1}{\urlprefix }}%
\providecommand \urlprefix  [0]{URL }%
\providecommand \Eprint [0]{\href }%
\providecommand \doibase [0]{https://doi.org/}%
\providecommand \selectlanguage [0]{\@gobble}%
\providecommand \bibinfo  [0]{\@secondoftwo}%
\providecommand \bibfield  [0]{\@secondoftwo}%
\providecommand \translation [1]{[#1]}%
\providecommand \BibitemOpen [0]{}%
\providecommand \bibitemStop [0]{}%
\providecommand \bibitemNoStop [0]{.\EOS\space}%
\providecommand \EOS [0]{\spacefactor3000\relax}%
\providecommand \BibitemShut  [1]{\csname bibitem#1\endcsname}%
\let\auto@bib@innerbib\@empty
\bibitem [{\citenamefont {Loeser}\ \emph {et~al.}(1996)\citenamefont {Loeser},
  \citenamefont {Shen}, \citenamefont {Dessau}, \citenamefont {Marshall},
  \citenamefont {Park}, \citenamefont {Fournier},\ and\ \citenamefont
  {Kapitulnik}}]{Loeser1996}%
  \BibitemOpen
  \bibfield  {author} {\bibinfo {author} {\bibfnamefont {A.~G.}\ \bibnamefont
  {Loeser}}, \bibinfo {author} {\bibfnamefont {Z.-X.}\ \bibnamefont {Shen}},
  \bibinfo {author} {\bibfnamefont {D.~S.}\ \bibnamefont {Dessau}}, \bibinfo
  {author} {\bibfnamefont {D.~S.}\ \bibnamefont {Marshall}}, \bibinfo {author}
  {\bibfnamefont {C.~H.}\ \bibnamefont {Park}}, \bibinfo {author}
  {\bibfnamefont {P.}~\bibnamefont {Fournier}},\ and\ \bibinfo {author}
  {\bibfnamefont {A.}~\bibnamefont {Kapitulnik}},\ }\bibfield  {title}
  {\bibinfo {title} {Excitation gap in the normal state of underdoped
  {Bi${}_2$Sr${}_2$CaCu${}_2$O${}_{8+\delta}$}},\ }\href
  {https://doi.org/10.1126/science.273.5273.325} {\bibfield  {journal}
  {\bibinfo  {journal} {Science}\ }\textbf {\bibinfo {volume} {273}},\ \bibinfo
  {pages} {325} (\bibinfo {year} {1996})}\BibitemShut {NoStop}%
\bibitem [{\citenamefont {Ding}\ \emph {et~al.}(1996)\citenamefont {Ding},
  \citenamefont {Yokoya}, \citenamefont {Campuzano}, \citenamefont {Takahashi},
  \citenamefont {Randeria}, \citenamefont {Norman}, \citenamefont {Mochiku},
  \citenamefont {Kadowaki},\ and\ \citenamefont {Giapintzakis}}]{Ding1996}%
  \BibitemOpen
  \bibfield  {author} {\bibinfo {author} {\bibfnamefont {H.}~\bibnamefont
  {Ding}}, \bibinfo {author} {\bibfnamefont {T.}~\bibnamefont {Yokoya}},
  \bibinfo {author} {\bibfnamefont {J.~C.}\ \bibnamefont {Campuzano}}, \bibinfo
  {author} {\bibfnamefont {T.}~\bibnamefont {Takahashi}}, \bibinfo {author}
  {\bibfnamefont {M.}~\bibnamefont {Randeria}}, \bibinfo {author}
  {\bibfnamefont {M.~R.}\ \bibnamefont {Norman}}, \bibinfo {author}
  {\bibfnamefont {T.}~\bibnamefont {Mochiku}}, \bibinfo {author} {\bibfnamefont
  {K.}~\bibnamefont {Kadowaki}},\ and\ \bibinfo {author} {\bibfnamefont
  {J.}~\bibnamefont {Giapintzakis}},\ }\bibfield  {title} {\bibinfo {title}
  {Spectroscopic evidence for a pseudogap in the normal state of underdoped
  high-${T}_{c}$ superconductors},\ }\href {https://doi.org/10.1038/382051a0}
  {\bibfield  {journal} {\bibinfo  {journal} {Nature}\ }\textbf {\bibinfo
  {volume} {382}},\ \bibinfo {pages} {51} (\bibinfo {year} {1996})}\BibitemShut
  {NoStop}%
\bibitem [{\citenamefont {Norman}\ \emph {et~al.}(1998)\citenamefont {Norman},
  \citenamefont {Ding}, \citenamefont {Randeria}, \citenamefont {Campuzano},
  \citenamefont {Yokoya}, \citenamefont {Takeuchi}, \citenamefont {Takahashi},
  \citenamefont {Mochiku}, \citenamefont {Kadowaki}, \citenamefont
  {Guptasarma},\ and\ \citenamefont {Hinks}}]{Norman1998}%
  \BibitemOpen
  \bibfield  {author} {\bibinfo {author} {\bibfnamefont {M.~R.}\ \bibnamefont
  {Norman}}, \bibinfo {author} {\bibfnamefont {H.}~\bibnamefont {Ding}},
  \bibinfo {author} {\bibfnamefont {M.}~\bibnamefont {Randeria}}, \bibinfo
  {author} {\bibfnamefont {J.~C.}\ \bibnamefont {Campuzano}}, \bibinfo {author}
  {\bibfnamefont {T.}~\bibnamefont {Yokoya}}, \bibinfo {author} {\bibfnamefont
  {T.}~\bibnamefont {Takeuchi}}, \bibinfo {author} {\bibfnamefont
  {T.}~\bibnamefont {Takahashi}}, \bibinfo {author} {\bibfnamefont
  {T.}~\bibnamefont {Mochiku}}, \bibinfo {author} {\bibfnamefont
  {K.}~\bibnamefont {Kadowaki}}, \bibinfo {author} {\bibfnamefont
  {P.}~\bibnamefont {Guptasarma}},\ and\ \bibinfo {author} {\bibfnamefont
  {D.~G.}\ \bibnamefont {Hinks}},\ }\bibfield  {title} {\bibinfo {title}
  {Destruction of the {Fermi} surface in underdoped high-${T}_{c}$
  superconductors},\ }\href {https://doi.org/10.1038/32366} {\bibfield
  {journal} {\bibinfo  {journal} {Nature}\ }\textbf {\bibinfo {volume} {392}},\
  \bibinfo {pages} {157} (\bibinfo {year} {1998})}\BibitemShut {NoStop}%
\bibitem [{\citenamefont {Timusk}\ and\ \citenamefont
  {Statt}(1999)}]{Timusk1999}%
  \BibitemOpen
  \bibfield  {author} {\bibinfo {author} {\bibfnamefont {T.}~\bibnamefont
  {Timusk}}\ and\ \bibinfo {author} {\bibfnamefont {B.}~\bibnamefont {Statt}},\
  }\bibfield  {title} {\bibinfo {title} {The pseudogap in high-temperature
  superconductors: An experimental survey},\ }\href
  {https://doi.org/10.1088/0034-4885/62/1/002} {\bibfield  {journal} {\bibinfo
  {journal} {Reports on Progress in Physics}\ }\textbf {\bibinfo {volume}
  {62}},\ \bibinfo {pages} {61} (\bibinfo {year} {1999})}\BibitemShut {NoStop}%
\bibitem [{\citenamefont {Damascelli}\ \emph {et~al.}(2003)\citenamefont
  {Damascelli}, \citenamefont {Hussain},\ and\ \citenamefont
  {Shen}}]{Damascelli2003}%
  \BibitemOpen
  \bibfield  {author} {\bibinfo {author} {\bibfnamefont {A.}~\bibnamefont
  {Damascelli}}, \bibinfo {author} {\bibfnamefont {Z.}~\bibnamefont
  {Hussain}},\ and\ \bibinfo {author} {\bibfnamefont {Z.-X.}\ \bibnamefont
  {Shen}},\ }\bibfield  {title} {\bibinfo {title} {Angle-resolved photoemission
  studies of the cuprate superconductors},\ }\href
  {https://doi.org/10.1103/RevModPhys.75.473} {\bibfield  {journal} {\bibinfo
  {journal} {Rev. Mod. Phys.}\ }\textbf {\bibinfo {volume} {75}},\ \bibinfo
  {pages} {473} (\bibinfo {year} {2003})}\BibitemShut {NoStop}%
\bibitem [{\citenamefont {Kanigel}\ \emph {et~al.}(2006)\citenamefont
  {Kanigel}, \citenamefont {Norman}, \citenamefont {Randeria}, \citenamefont
  {Chatterjee}, \citenamefont {Souma}, \citenamefont {Kaminski}, \citenamefont
  {Fretwell}, \citenamefont {Rosenkranz}, \citenamefont {Shi}, \citenamefont
  {Sato}, \citenamefont {Takahashi}, \citenamefont {Li}, \citenamefont {Raffy},
  \citenamefont {Kadowaki}, \citenamefont {Hinks}, \citenamefont {Ozyuzer},\
  and\ \citenamefont {Campuzano}}]{Kanigel2006}%
  \BibitemOpen
  \bibfield  {author} {\bibinfo {author} {\bibfnamefont {A.}~\bibnamefont
  {Kanigel}}, \bibinfo {author} {\bibfnamefont {M.~R.}\ \bibnamefont {Norman}},
  \bibinfo {author} {\bibfnamefont {M.}~\bibnamefont {Randeria}}, \bibinfo
  {author} {\bibfnamefont {U.}~\bibnamefont {Chatterjee}}, \bibinfo {author}
  {\bibfnamefont {S.}~\bibnamefont {Souma}}, \bibinfo {author} {\bibfnamefont
  {A.}~\bibnamefont {Kaminski}}, \bibinfo {author} {\bibfnamefont {H.~M.}\
  \bibnamefont {Fretwell}}, \bibinfo {author} {\bibfnamefont {S.}~\bibnamefont
  {Rosenkranz}}, \bibinfo {author} {\bibfnamefont {M.}~\bibnamefont {Shi}},
  \bibinfo {author} {\bibfnamefont {T.}~\bibnamefont {Sato}}, \bibinfo {author}
  {\bibfnamefont {T.}~\bibnamefont {Takahashi}}, \bibinfo {author}
  {\bibfnamefont {Z.~Z.}\ \bibnamefont {Li}}, \bibinfo {author} {\bibfnamefont
  {H.}~\bibnamefont {Raffy}}, \bibinfo {author} {\bibfnamefont
  {K.}~\bibnamefont {Kadowaki}}, \bibinfo {author} {\bibfnamefont
  {D.}~\bibnamefont {Hinks}}, \bibinfo {author} {\bibfnamefont
  {L.}~\bibnamefont {Ozyuzer}},\ and\ \bibinfo {author} {\bibfnamefont {J.~C.}\
  \bibnamefont {Campuzano}},\ }\bibfield  {title} {\bibinfo {title} {Evolution
  of the pseudogap from {Fermi} arcs to the nodal liquid},\ }\href
  {https://doi.org/10.1038/nphys334} {\bibfield  {journal} {\bibinfo  {journal}
  {Nature Physics}\ }\textbf {\bibinfo {volume} {2}},\ \bibinfo {pages} {447}
  (\bibinfo {year} {2006})}\BibitemShut {NoStop}%
\bibitem [{\citenamefont {Sobota}\ \emph {et~al.}(2021)\citenamefont {Sobota},
  \citenamefont {He},\ and\ \citenamefont {Shen}}]{Sobota2021}%
  \BibitemOpen
  \bibfield  {author} {\bibinfo {author} {\bibfnamefont {J.~A.}\ \bibnamefont
  {Sobota}}, \bibinfo {author} {\bibfnamefont {Y.}~\bibnamefont {He}},\ and\
  \bibinfo {author} {\bibfnamefont {Z.-X.}\ \bibnamefont {Shen}},\ }\bibfield
  {title} {\bibinfo {title} {Angle-resolved photoemission studies of quantum
  materials},\ }\href {https://doi.org/10.1103/RevModPhys.93.025006} {\bibfield
   {journal} {\bibinfo  {journal} {Rev. Mod. Phys.}\ }\textbf {\bibinfo
  {volume} {93}},\ \bibinfo {pages} {025006} (\bibinfo {year}
  {2021})}\BibitemShut {NoStop}%
\bibitem [{\citenamefont {Armanno}\ \emph {et~al.}(2025)\citenamefont
  {Armanno}, \citenamefont {Gingras}, \citenamefont {Goto}, \citenamefont
  {Parent}, \citenamefont {Longa}, \citenamefont {Jabed}, \citenamefont
  {Frimpong}, \citenamefont {Zhong}, \citenamefont {Schneeloch}, \citenamefont
  {Gu}, \citenamefont {Jargot}, \citenamefont {Ibrahim}, \citenamefont
  {Legare}, \citenamefont {Siwick}, \citenamefont {Gauthier}, \citenamefont
  {Georges}, \citenamefont {Millis},\ and\ \citenamefont
  {Boschini}}]{Armanno2025}%
  \BibitemOpen
  \bibfield  {author} {\bibinfo {author} {\bibfnamefont {D.}~\bibnamefont
  {Armanno}}, \bibinfo {author} {\bibfnamefont {O.}~\bibnamefont {Gingras}},
  \bibinfo {author} {\bibfnamefont {F.}~\bibnamefont {Goto}}, \bibinfo {author}
  {\bibfnamefont {J.-M.}\ \bibnamefont {Parent}}, \bibinfo {author}
  {\bibfnamefont {A.}~\bibnamefont {Longa}}, \bibinfo {author} {\bibfnamefont
  {A.}~\bibnamefont {Jabed}}, \bibinfo {author} {\bibfnamefont
  {B.}~\bibnamefont {Frimpong}}, \bibinfo {author} {\bibfnamefont {R.~D.}\
  \bibnamefont {Zhong}}, \bibinfo {author} {\bibfnamefont {J.}~\bibnamefont
  {Schneeloch}}, \bibinfo {author} {\bibfnamefont {G.~D.}\ \bibnamefont {Gu}},
  \bibinfo {author} {\bibfnamefont {G.}~\bibnamefont {Jargot}}, \bibinfo
  {author} {\bibfnamefont {H.}~\bibnamefont {Ibrahim}}, \bibinfo {author}
  {\bibfnamefont {F.}~\bibnamefont {Legare}}, \bibinfo {author} {\bibfnamefont
  {B.~J.}\ \bibnamefont {Siwick}}, \bibinfo {author} {\bibfnamefont
  {N.}~\bibnamefont {Gauthier}}, \bibinfo {author} {\bibfnamefont
  {A.}~\bibnamefont {Georges}}, \bibinfo {author} {\bibfnamefont {A.~J.}\
  \bibnamefont {Millis}},\ and\ \bibinfo {author} {\bibfnamefont
  {F.}~\bibnamefont {Boschini}},\ }\href@noop {} {\bibinfo {title}
  {Light-induced asymmetric pseudogap below ${T}_{c}$ in cuprates}} (\bibinfo
  {year} {2025}),\ \Eprint {https://arxiv.org/abs/2511.20768} {arXiv:2511.20768
  [cond-mat.supr-con]} \BibitemShut {NoStop}%
\bibitem [{\citenamefont {Uchida}\ \emph {et~al.}(2011)\citenamefont {Uchida},
  \citenamefont {Ishizaka}, \citenamefont {Hansmann}, \citenamefont {Kaneko},
  \citenamefont {Ishida}, \citenamefont {Yang}, \citenamefont {Kumai},
  \citenamefont {Toschi}, \citenamefont {Onose}, \citenamefont {Arita},
  \citenamefont {Held}, \citenamefont {Andersen}, \citenamefont {Shin},\ and\
  \citenamefont {Tokura}}]{Uchida2011}%
  \BibitemOpen
  \bibfield  {author} {\bibinfo {author} {\bibfnamefont {M.}~\bibnamefont
  {Uchida}}, \bibinfo {author} {\bibfnamefont {K.}~\bibnamefont {Ishizaka}},
  \bibinfo {author} {\bibfnamefont {P.}~\bibnamefont {Hansmann}}, \bibinfo
  {author} {\bibfnamefont {Y.}~\bibnamefont {Kaneko}}, \bibinfo {author}
  {\bibfnamefont {Y.}~\bibnamefont {Ishida}}, \bibinfo {author} {\bibfnamefont
  {X.}~\bibnamefont {Yang}}, \bibinfo {author} {\bibfnamefont {R.}~\bibnamefont
  {Kumai}}, \bibinfo {author} {\bibfnamefont {A.}~\bibnamefont {Toschi}},
  \bibinfo {author} {\bibfnamefont {Y.}~\bibnamefont {Onose}}, \bibinfo
  {author} {\bibfnamefont {R.}~\bibnamefont {Arita}}, \bibinfo {author}
  {\bibfnamefont {K.}~\bibnamefont {Held}}, \bibinfo {author} {\bibfnamefont
  {O.~K.}\ \bibnamefont {Andersen}}, \bibinfo {author} {\bibfnamefont
  {S.}~\bibnamefont {Shin}},\ and\ \bibinfo {author} {\bibfnamefont
  {Y.}~\bibnamefont {Tokura}},\ }\bibfield  {title} {\bibinfo {title}
  {Pseudogap of metallic layered nickelate
  {${R}_{2-x}{\mathrm{Sr}}_{x}{\mathrm{NiO}}_{4}$
  (${R}=\mathrm{Nd},\mathrm{Eu}$}) crystals measured using angle-resolved
  photoemission spectroscopy},\ }\href
  {https://doi.org/10.1103/PhysRevLett.106.027001} {\bibfield  {journal}
  {\bibinfo  {journal} {Phys. Rev. Lett.}\ }\textbf {\bibinfo {volume} {106}},\
  \bibinfo {pages} {027001} (\bibinfo {year} {2011})}\BibitemShut {NoStop}%
\bibitem [{\citenamefont {Peng}\ \emph {et~al.}(2022)\citenamefont {Peng},
  \citenamefont {Lane}, \citenamefont {Hu}, \citenamefont {Guo}, \citenamefont
  {Chen}, \citenamefont {Sun}, \citenamefont {Hashimoto}, \citenamefont {Lu},
  \citenamefont {Shen}, \citenamefont {Wu}, \citenamefont {Chen}, \citenamefont
  {Markiewicz}, \citenamefont {Wang}, \citenamefont {Bansil}, \citenamefont
  {Wilson},\ and\ \citenamefont {He}}]{Peng2022}%
  \BibitemOpen
  \bibfield  {author} {\bibinfo {author} {\bibfnamefont {S.}~\bibnamefont
  {Peng}}, \bibinfo {author} {\bibfnamefont {C.}~\bibnamefont {Lane}}, \bibinfo
  {author} {\bibfnamefont {Y.}~\bibnamefont {Hu}}, \bibinfo {author}
  {\bibfnamefont {M.}~\bibnamefont {Guo}}, \bibinfo {author} {\bibfnamefont
  {X.}~\bibnamefont {Chen}}, \bibinfo {author} {\bibfnamefont {Z.}~\bibnamefont
  {Sun}}, \bibinfo {author} {\bibfnamefont {M.}~\bibnamefont {Hashimoto}},
  \bibinfo {author} {\bibfnamefont {D.}~\bibnamefont {Lu}}, \bibinfo {author}
  {\bibfnamefont {Z.-X.}\ \bibnamefont {Shen}}, \bibinfo {author}
  {\bibfnamefont {T.}~\bibnamefont {Wu}}, \bibinfo {author} {\bibfnamefont
  {X.}~\bibnamefont {Chen}}, \bibinfo {author} {\bibfnamefont {R.~S.}\
  \bibnamefont {Markiewicz}}, \bibinfo {author} {\bibfnamefont
  {Y.}~\bibnamefont {Wang}}, \bibinfo {author} {\bibfnamefont {A.}~\bibnamefont
  {Bansil}}, \bibinfo {author} {\bibfnamefont {S.~D.}\ \bibnamefont {Wilson}},\
  and\ \bibinfo {author} {\bibfnamefont {J.}~\bibnamefont {He}},\ }\bibfield
  {title} {\bibinfo {title} {Electronic nature of the pseudogap in
  electron-doped {Sr${}_2$IrO${}_4$}},\ }\href
  {https://doi.org/10.1038/s41535-022-00467-1} {\bibfield  {journal} {\bibinfo
  {journal} {npj Quantum Materials}\ }\textbf {\bibinfo {volume} {7}},\
  \bibinfo {pages} {58} (\bibinfo {year} {2022})}\BibitemShut {NoStop}%
\bibitem [{\citenamefont {Alexanian}\ \emph {et~al.}(2025)\citenamefont
  {Alexanian}, \citenamefont {de~la Torre}, \citenamefont {McKeown~Walker},
  \citenamefont {Straub}, \citenamefont {Gatti}, \citenamefont {Hunter},
  \citenamefont {Mandloi}, \citenamefont {Cappelli}, \citenamefont {Ricc\`o},
  \citenamefont {Bruno}, \citenamefont {Radovic}, \citenamefont {Plumb},
  \citenamefont {Shi}, \citenamefont {Osiecki}, \citenamefont {Polley},
  \citenamefont {Kim}, \citenamefont {Dudin}, \citenamefont {Hoesch},
  \citenamefont {Perry}, \citenamefont {Tamai},\ and\ \citenamefont
  {Baumberger}}]{Alexanian2025}%
  \BibitemOpen
  \bibfield  {author} {\bibinfo {author} {\bibfnamefont {Y.}~\bibnamefont
  {Alexanian}}, \bibinfo {author} {\bibfnamefont {A.}~\bibnamefont {de~la
  Torre}}, \bibinfo {author} {\bibfnamefont {S.}~\bibnamefont
  {McKeown~Walker}}, \bibinfo {author} {\bibfnamefont {M.}~\bibnamefont
  {Straub}}, \bibinfo {author} {\bibfnamefont {G.}~\bibnamefont {Gatti}},
  \bibinfo {author} {\bibfnamefont {A.}~\bibnamefont {Hunter}}, \bibinfo
  {author} {\bibfnamefont {S.}~\bibnamefont {Mandloi}}, \bibinfo {author}
  {\bibfnamefont {E.}~\bibnamefont {Cappelli}}, \bibinfo {author}
  {\bibfnamefont {S.}~\bibnamefont {Ricc\`o}}, \bibinfo {author} {\bibfnamefont
  {F.~Y.}\ \bibnamefont {Bruno}}, \bibinfo {author} {\bibfnamefont
  {M.}~\bibnamefont {Radovic}}, \bibinfo {author} {\bibfnamefont {N.~C.}\
  \bibnamefont {Plumb}}, \bibinfo {author} {\bibfnamefont {M.}~\bibnamefont
  {Shi}}, \bibinfo {author} {\bibfnamefont {J.}~\bibnamefont {Osiecki}},
  \bibinfo {author} {\bibfnamefont {C.}~\bibnamefont {Polley}}, \bibinfo
  {author} {\bibfnamefont {T.~K.}\ \bibnamefont {Kim}}, \bibinfo {author}
  {\bibfnamefont {P.}~\bibnamefont {Dudin}}, \bibinfo {author} {\bibfnamefont
  {M.}~\bibnamefont {Hoesch}}, \bibinfo {author} {\bibfnamefont {R.~S.}\
  \bibnamefont {Perry}}, \bibinfo {author} {\bibfnamefont {A.}~\bibnamefont
  {Tamai}},\ and\ \bibinfo {author} {\bibfnamefont {F.}~\bibnamefont
  {Baumberger}},\ }\bibfield  {title} {\bibinfo {title} {{Fermi} surface and
  pseudogap in highly doped {Sr${}_2$IrO${}_4$}},\ }\href
  {https://doi.org/10.1038/s41535-025-00817-9} {\bibfield  {journal} {\bibinfo
  {journal} {npj Quantum Materials}\ }\textbf {\bibinfo {volume} {10}},\
  \bibinfo {pages} {100} (\bibinfo {year} {2025})}\BibitemShut {NoStop}%
\bibitem [{\citenamefont {Oh}\ \emph {et~al.}(2021)\citenamefont {Oh},
  \citenamefont {Nuckolls}, \citenamefont {Wong}, \citenamefont {Lee},
  \citenamefont {Liu}, \citenamefont {Watanabe}, \citenamefont {Taniguchi},\
  and\ \citenamefont {Yazdani}}]{Oh2021}%
  \BibitemOpen
  \bibfield  {author} {\bibinfo {author} {\bibfnamefont {M.}~\bibnamefont
  {Oh}}, \bibinfo {author} {\bibfnamefont {K.~P.}\ \bibnamefont {Nuckolls}},
  \bibinfo {author} {\bibfnamefont {D.}~\bibnamefont {Wong}}, \bibinfo {author}
  {\bibfnamefont {R.~L.}\ \bibnamefont {Lee}}, \bibinfo {author} {\bibfnamefont
  {X.}~\bibnamefont {Liu}}, \bibinfo {author} {\bibfnamefont {K.}~\bibnamefont
  {Watanabe}}, \bibinfo {author} {\bibfnamefont {T.}~\bibnamefont
  {Taniguchi}},\ and\ \bibinfo {author} {\bibfnamefont {A.}~\bibnamefont
  {Yazdani}},\ }\bibfield  {title} {\bibinfo {title} {Evidence for
  unconventional superconductivity in twisted bilayer graphene},\ }\href
  {https://doi.org/10.1038/s41586-021-04121-x} {\bibfield  {journal} {\bibinfo
  {journal} {Nature}\ }\textbf {\bibinfo {volume} {600}},\ \bibinfo {pages}
  {240} (\bibinfo {year} {2021})}\BibitemShut {NoStop}%
\bibitem [{\citenamefont {Chan}\ \emph {et~al.}(2025)\citenamefont {Chan},
  \citenamefont {Schreiber}, \citenamefont {Ayala-Valenzuela}, \citenamefont
  {Bauer}, \citenamefont {Shekhter},\ and\ \citenamefont
  {Harrison}}]{Chan2025}%
  \BibitemOpen
  \bibfield  {author} {\bibinfo {author} {\bibfnamefont {M.~K.}\ \bibnamefont
  {Chan}}, \bibinfo {author} {\bibfnamefont {K.~A.}\ \bibnamefont {Schreiber}},
  \bibinfo {author} {\bibfnamefont {O.~E.}\ \bibnamefont {Ayala-Valenzuela}},
  \bibinfo {author} {\bibfnamefont {E.~D.}\ \bibnamefont {Bauer}}, \bibinfo
  {author} {\bibfnamefont {A.}~\bibnamefont {Shekhter}},\ and\ \bibinfo
  {author} {\bibfnamefont {N.}~\bibnamefont {Harrison}},\ }\bibfield  {title}
  {\bibinfo {title} {Observation of the {Yamaji} effect in a cuprate
  superconductor},\ }\bibfield  {journal} {\bibinfo  {journal} {Nature
  Physics}\ }\href {https://doi.org/10.1038/s41567-025-03032-2}
  {10.1038/s41567-025-03032-2} (\bibinfo {year} {2025})\BibitemShut {NoStop}%
\bibitem [{\citenamefont {Meng}\ \emph {et~al.}(2009)\citenamefont {Meng},
  \citenamefont {Liu}, \citenamefont {Zhang}, \citenamefont {Zhao},
  \citenamefont {Liu}, \citenamefont {Jia}, \citenamefont {Mu}, \citenamefont
  {Liu}, \citenamefont {Dong}, \citenamefont {Zhang}, \citenamefont {Lu},
  \citenamefont {Wang}, \citenamefont {Zhou}, \citenamefont {Zhu},
  \citenamefont {Wang}, \citenamefont {Xu}, \citenamefont {Chen},\ and\
  \citenamefont {Zhou}}]{Meng2009}%
  \BibitemOpen
  \bibfield  {author} {\bibinfo {author} {\bibfnamefont {J.}~\bibnamefont
  {Meng}}, \bibinfo {author} {\bibfnamefont {G.}~\bibnamefont {Liu}}, \bibinfo
  {author} {\bibfnamefont {W.}~\bibnamefont {Zhang}}, \bibinfo {author}
  {\bibfnamefont {L.}~\bibnamefont {Zhao}}, \bibinfo {author} {\bibfnamefont
  {H.}~\bibnamefont {Liu}}, \bibinfo {author} {\bibfnamefont {X.}~\bibnamefont
  {Jia}}, \bibinfo {author} {\bibfnamefont {D.}~\bibnamefont {Mu}}, \bibinfo
  {author} {\bibfnamefont {S.}~\bibnamefont {Liu}}, \bibinfo {author}
  {\bibfnamefont {X.}~\bibnamefont {Dong}}, \bibinfo {author} {\bibfnamefont
  {J.}~\bibnamefont {Zhang}}, \bibinfo {author} {\bibfnamefont
  {W.}~\bibnamefont {Lu}}, \bibinfo {author} {\bibfnamefont {G.}~\bibnamefont
  {Wang}}, \bibinfo {author} {\bibfnamefont {Y.}~\bibnamefont {Zhou}}, \bibinfo
  {author} {\bibfnamefont {Y.}~\bibnamefont {Zhu}}, \bibinfo {author}
  {\bibfnamefont {X.}~\bibnamefont {Wang}}, \bibinfo {author} {\bibfnamefont
  {Z.}~\bibnamefont {Xu}}, \bibinfo {author} {\bibfnamefont {C.}~\bibnamefont
  {Chen}},\ and\ \bibinfo {author} {\bibfnamefont {X.~J.}\ \bibnamefont
  {Zhou}},\ }\bibfield  {title} {\bibinfo {title} {Coexistence of {Fermi} arcs
  and {Fermi} pockets in a high-${T}_{c}$ copper oxide superconductor},\ }\href
  {https://doi.org/10.1038/nature08521} {\bibfield  {journal} {\bibinfo
  {journal} {Nature}\ }\textbf {\bibinfo {volume} {462}},\ \bibinfo {pages}
  {335} (\bibinfo {year} {2009})}\BibitemShut {NoStop}%
\bibitem [{\citenamefont {Chang}\ \emph {et~al.}(2008)\citenamefont {Chang},
  \citenamefont {Sassa}, \citenamefont {Guerrero}, \citenamefont {Månsson},
  \citenamefont {Shi}, \citenamefont {Pailhés}, \citenamefont {Bendounan},
  \citenamefont {Mottl}, \citenamefont {Claesson}, \citenamefont {Tjernberg},
  \citenamefont {Patthey}, \citenamefont {Ido}, \citenamefont {Oda},
  \citenamefont {Momono}, \citenamefont {Mudry},\ and\ \citenamefont
  {Mesot}}]{Chang2008}%
  \BibitemOpen
  \bibfield  {author} {\bibinfo {author} {\bibfnamefont {J.}~\bibnamefont
  {Chang}}, \bibinfo {author} {\bibfnamefont {Y.}~\bibnamefont {Sassa}},
  \bibinfo {author} {\bibfnamefont {S.}~\bibnamefont {Guerrero}}, \bibinfo
  {author} {\bibfnamefont {M.}~\bibnamefont {Månsson}}, \bibinfo {author}
  {\bibfnamefont {M.}~\bibnamefont {Shi}}, \bibinfo {author} {\bibfnamefont
  {S.}~\bibnamefont {Pailhés}}, \bibinfo {author} {\bibfnamefont
  {A.}~\bibnamefont {Bendounan}}, \bibinfo {author} {\bibfnamefont
  {R.}~\bibnamefont {Mottl}}, \bibinfo {author} {\bibfnamefont
  {T.}~\bibnamefont {Claesson}}, \bibinfo {author} {\bibfnamefont
  {O.}~\bibnamefont {Tjernberg}}, \bibinfo {author} {\bibfnamefont
  {L.}~\bibnamefont {Patthey}}, \bibinfo {author} {\bibfnamefont
  {M.}~\bibnamefont {Ido}}, \bibinfo {author} {\bibfnamefont {M.}~\bibnamefont
  {Oda}}, \bibinfo {author} {\bibfnamefont {N.}~\bibnamefont {Momono}},
  \bibinfo {author} {\bibfnamefont {C.}~\bibnamefont {Mudry}},\ and\ \bibinfo
  {author} {\bibfnamefont {J.}~\bibnamefont {Mesot}},\ }\bibfield  {title}
  {\bibinfo {title} {Electronic structure near the 1/8-anomaly in {La}-based
  cuprates},\ }\href {https://doi.org/10.1088/1367-2630/10/10/103016}
  {\bibfield  {journal} {\bibinfo  {journal} {New Journal of Physics}\ }\textbf
  {\bibinfo {volume} {10}},\ \bibinfo {pages} {103016} (\bibinfo {year}
  {2008})}\BibitemShut {NoStop}%
\bibitem [{\citenamefont {King}\ \emph {et~al.}(2011)\citenamefont {King},
  \citenamefont {Rosen}, \citenamefont {Meevasana}, \citenamefont {Tamai},
  \citenamefont {Rozbicki}, \citenamefont {Comin}, \citenamefont {Levy},
  \citenamefont {Fournier}, \citenamefont {Yoshida}, \citenamefont {Eisaki},
  \citenamefont {Shen}, \citenamefont {Ingle}, \citenamefont {Damascelli},\
  and\ \citenamefont {Baumberger}}]{King2011}%
  \BibitemOpen
  \bibfield  {author} {\bibinfo {author} {\bibfnamefont {P.~D.~C.}\
  \bibnamefont {King}}, \bibinfo {author} {\bibfnamefont {J.~A.}\ \bibnamefont
  {Rosen}}, \bibinfo {author} {\bibfnamefont {W.}~\bibnamefont {Meevasana}},
  \bibinfo {author} {\bibfnamefont {A.}~\bibnamefont {Tamai}}, \bibinfo
  {author} {\bibfnamefont {E.}~\bibnamefont {Rozbicki}}, \bibinfo {author}
  {\bibfnamefont {R.}~\bibnamefont {Comin}}, \bibinfo {author} {\bibfnamefont
  {G.}~\bibnamefont {Levy}}, \bibinfo {author} {\bibfnamefont {D.}~\bibnamefont
  {Fournier}}, \bibinfo {author} {\bibfnamefont {Y.}~\bibnamefont {Yoshida}},
  \bibinfo {author} {\bibfnamefont {H.}~\bibnamefont {Eisaki}}, \bibinfo
  {author} {\bibfnamefont {K.~M.}\ \bibnamefont {Shen}}, \bibinfo {author}
  {\bibfnamefont {N.~J.~C.}\ \bibnamefont {Ingle}}, \bibinfo {author}
  {\bibfnamefont {A.}~\bibnamefont {Damascelli}},\ and\ \bibinfo {author}
  {\bibfnamefont {F.}~\bibnamefont {Baumberger}},\ }\bibfield  {title}
  {\bibinfo {title} {Structural origin of apparent {Fermi} surface pockets in
  angle-resolved photoemission of
  {${\mathrm{Bi}}_{2}{\mathrm{Sr}}_{2-x}{\mathrm{La}}_{x}{\mathrm{CuO}}_{6+\delta}$}},\
  }\href {https://doi.org/10.1103/PhysRevLett.106.127005} {\bibfield  {journal}
  {\bibinfo  {journal} {Phys. Rev. Lett.}\ }\textbf {\bibinfo {volume} {106}},\
  \bibinfo {pages} {127005} (\bibinfo {year} {2011})}\BibitemShut {NoStop}%
\bibitem [{\citenamefont {Freutel}\ \emph {et~al.}(2019)\citenamefont
  {Freutel}, \citenamefont {Rameau}, \citenamefont {Rettig}, \citenamefont
  {Avigo}, \citenamefont {Ligges}, \citenamefont {Yoshida}, \citenamefont
  {Eisaki}, \citenamefont {Schneeloch}, \citenamefont {Zhong}, \citenamefont
  {Xu}, \citenamefont {Gu}, \citenamefont {Bovensiepen},\ and\ \citenamefont
  {Johnson}}]{Freutel2019}%
  \BibitemOpen
  \bibfield  {author} {\bibinfo {author} {\bibfnamefont {S.}~\bibnamefont
  {Freutel}}, \bibinfo {author} {\bibfnamefont {J.~D.}\ \bibnamefont {Rameau}},
  \bibinfo {author} {\bibfnamefont {L.}~\bibnamefont {Rettig}}, \bibinfo
  {author} {\bibfnamefont {I.}~\bibnamefont {Avigo}}, \bibinfo {author}
  {\bibfnamefont {M.}~\bibnamefont {Ligges}}, \bibinfo {author} {\bibfnamefont
  {Y.}~\bibnamefont {Yoshida}}, \bibinfo {author} {\bibfnamefont
  {H.}~\bibnamefont {Eisaki}}, \bibinfo {author} {\bibfnamefont
  {J.}~\bibnamefont {Schneeloch}}, \bibinfo {author} {\bibfnamefont {R.~D.}\
  \bibnamefont {Zhong}}, \bibinfo {author} {\bibfnamefont {Z.~J.}\ \bibnamefont
  {Xu}}, \bibinfo {author} {\bibfnamefont {G.~D.}\ \bibnamefont {Gu}}, \bibinfo
  {author} {\bibfnamefont {U.}~\bibnamefont {Bovensiepen}},\ and\ \bibinfo
  {author} {\bibfnamefont {P.~D.}\ \bibnamefont {Johnson}},\ }\bibfield
  {title} {\bibinfo {title} {Optical perturbation of the hole pockets in the
  underdoped high-${T}_{c}$ superconducting cuprates},\ }\href
  {https://doi.org/10.1103/PhysRevB.99.081116} {\bibfield  {journal} {\bibinfo
  {journal} {Phys. Rev. B}\ }\textbf {\bibinfo {volume} {99}},\ \bibinfo
  {pages} {081116} (\bibinfo {year} {2019})}\BibitemShut {NoStop}%
\bibitem [{\citenamefont {Minamitani}\ \emph {et~al.}(2012)\citenamefont
  {Minamitani}, \citenamefont {Tsukahara}, \citenamefont {Matsunaka},
  \citenamefont {Kim}, \citenamefont {Takagi},\ and\ \citenamefont
  {Kawai}}]{Minamitani2012}%
  \BibitemOpen
  \bibfield  {author} {\bibinfo {author} {\bibfnamefont {E.}~\bibnamefont
  {Minamitani}}, \bibinfo {author} {\bibfnamefont {N.}~\bibnamefont
  {Tsukahara}}, \bibinfo {author} {\bibfnamefont {D.}~\bibnamefont
  {Matsunaka}}, \bibinfo {author} {\bibfnamefont {Y.}~\bibnamefont {Kim}},
  \bibinfo {author} {\bibfnamefont {N.}~\bibnamefont {Takagi}},\ and\ \bibinfo
  {author} {\bibfnamefont {M.}~\bibnamefont {Kawai}},\ }\bibfield  {title}
  {\bibinfo {title} {Symmetry-driven novel {Kondo} effect in a molecule},\
  }\href {https://doi.org/10.1103/PhysRevLett.109.086602} {\bibfield  {journal}
  {\bibinfo  {journal} {Phys. Rev. Lett.}\ }\textbf {\bibinfo {volume} {109}},\
  \bibinfo {pages} {086602} (\bibinfo {year} {2012})}\BibitemShut {NoStop}%
\bibitem [{\citenamefont {Hiraoka}\ \emph {et~al.}(2017)\citenamefont
  {Hiraoka}, \citenamefont {Minamitani}, \citenamefont {Arafune}, \citenamefont
  {Tsukahara}, \citenamefont {Watanabe}, \citenamefont {Kawai},\ and\
  \citenamefont {Takagi}}]{Hiraoka2017}%
  \BibitemOpen
  \bibfield  {author} {\bibinfo {author} {\bibfnamefont {R.}~\bibnamefont
  {Hiraoka}}, \bibinfo {author} {\bibfnamefont {E.}~\bibnamefont {Minamitani}},
  \bibinfo {author} {\bibfnamefont {R.}~\bibnamefont {Arafune}}, \bibinfo
  {author} {\bibfnamefont {N.}~\bibnamefont {Tsukahara}}, \bibinfo {author}
  {\bibfnamefont {S.}~\bibnamefont {Watanabe}}, \bibinfo {author}
  {\bibfnamefont {M.}~\bibnamefont {Kawai}},\ and\ \bibinfo {author}
  {\bibfnamefont {N.}~\bibnamefont {Takagi}},\ }\bibfield  {title} {\bibinfo
  {title} {Single-molecule quantum dot as a {Kondo} simulator},\ }\href
  {https://doi.org/10.1038/ncomms16012} {\bibfield  {journal} {\bibinfo
  {journal} {Nature Communications}\ }\textbf {\bibinfo {volume} {8}},\
  \bibinfo {pages} {16012} (\bibinfo {year} {2017})}\BibitemShut {NoStop}%
\bibitem [{\citenamefont {Yang}\ \emph {et~al.}(2019)\citenamefont {Yang},
  \citenamefont {Chen}, \citenamefont {Pope}, \citenamefont {Hu}, \citenamefont
  {Liu}, \citenamefont {Wang}, \citenamefont {Tao}, \citenamefont {Xiao},
  \citenamefont {Fei}, \citenamefont {Zhang}, \citenamefont {Luo},
  \citenamefont {Du}, \citenamefont {Xiang}, \citenamefont {Hofer},\ and\
  \citenamefont {Gao}}]{Yang2019}%
  \BibitemOpen
  \bibfield  {author} {\bibinfo {author} {\bibfnamefont {K.}~\bibnamefont
  {Yang}}, \bibinfo {author} {\bibfnamefont {H.}~\bibnamefont {Chen}}, \bibinfo
  {author} {\bibfnamefont {T.}~\bibnamefont {Pope}}, \bibinfo {author}
  {\bibfnamefont {Y.}~\bibnamefont {Hu}}, \bibinfo {author} {\bibfnamefont
  {L.}~\bibnamefont {Liu}}, \bibinfo {author} {\bibfnamefont {D.}~\bibnamefont
  {Wang}}, \bibinfo {author} {\bibfnamefont {L.}~\bibnamefont {Tao}}, \bibinfo
  {author} {\bibfnamefont {W.}~\bibnamefont {Xiao}}, \bibinfo {author}
  {\bibfnamefont {X.}~\bibnamefont {Fei}}, \bibinfo {author} {\bibfnamefont
  {Y.-Y.}\ \bibnamefont {Zhang}}, \bibinfo {author} {\bibfnamefont {H.-G.}\
  \bibnamefont {Luo}}, \bibinfo {author} {\bibfnamefont {S.}~\bibnamefont
  {Du}}, \bibinfo {author} {\bibfnamefont {T.}~\bibnamefont {Xiang}}, \bibinfo
  {author} {\bibfnamefont {W.~A.}\ \bibnamefont {Hofer}},\ and\ \bibinfo
  {author} {\bibfnamefont {H.-J.}\ \bibnamefont {Gao}},\ }\bibfield  {title}
  {\bibinfo {title} {Tunable giant magnetoresistance in a single-molecule
  junction},\ }\href {https://doi.org/10.1038/s41467-019-11587-x} {\bibfield
  {journal} {\bibinfo  {journal} {Nature Communications}\ }\textbf {\bibinfo
  {volume} {10}},\ \bibinfo {pages} {3599} (\bibinfo {year}
  {2019})}\BibitemShut {NoStop}%
\bibitem [{\citenamefont {Guo}\ \emph {et~al.}(2021)\citenamefont {Guo},
  \citenamefont {Zhu}, \citenamefont {Zhou}, \citenamefont {Yu}, \citenamefont
  {Lu},\ and\ \citenamefont {Liang}}]{Guo2021}%
  \BibitemOpen
  \bibfield  {author} {\bibinfo {author} {\bibfnamefont {X.}~\bibnamefont
  {Guo}}, \bibinfo {author} {\bibfnamefont {Q.}~\bibnamefont {Zhu}}, \bibinfo
  {author} {\bibfnamefont {L.}~\bibnamefont {Zhou}}, \bibinfo {author}
  {\bibfnamefont {W.}~\bibnamefont {Yu}}, \bibinfo {author} {\bibfnamefont
  {W.}~\bibnamefont {Lu}},\ and\ \bibinfo {author} {\bibfnamefont
  {W.}~\bibnamefont {Liang}},\ }\bibfield  {title} {\bibinfo {title} {Evolution
  and universality of two-stage {Kondo} effect in single manganese
  phthalocyanine molecule transistors},\ }\href
  {https://doi.org/10.1038/s41467-021-21492-x} {\bibfield  {journal} {\bibinfo
  {journal} {Nature Communications}\ }\textbf {\bibinfo {volume} {12}},\
  \bibinfo {pages} {1566} (\bibinfo {year} {2021})}\BibitemShut {NoStop}%
\bibitem [{\citenamefont {Jones}\ and\ \citenamefont
  {Varma}(1987)}]{Jones1987}%
  \BibitemOpen
  \bibfield  {author} {\bibinfo {author} {\bibfnamefont {B.~A.}\ \bibnamefont
  {Jones}}\ and\ \bibinfo {author} {\bibfnamefont {C.~M.}\ \bibnamefont
  {Varma}},\ }\bibfield  {title} {\bibinfo {title} {Study of two magnetic
  impurities in a {Fermi} gas},\ }\href
  {https://doi.org/10.1103/PhysRevLett.58.843} {\bibfield  {journal} {\bibinfo
  {journal} {Phys. Rev. Lett.}\ }\textbf {\bibinfo {volume} {58}},\ \bibinfo
  {pages} {843} (\bibinfo {year} {1987})}\BibitemShut {NoStop}%
\bibitem [{\citenamefont {De~Leo}\ and\ \citenamefont
  {Fabrizio}(2004)}]{DeLeo2004}%
  \BibitemOpen
  \bibfield  {author} {\bibinfo {author} {\bibfnamefont {L.}~\bibnamefont
  {De~Leo}}\ and\ \bibinfo {author} {\bibfnamefont {M.}~\bibnamefont
  {Fabrizio}},\ }\bibfield  {title} {\bibinfo {title} {Spectral properties of a
  two-orbital {Anderson} impurity model across a non-{Fermi}-liquid fixed
  point},\ }\href {https://doi.org/10.1103/PhysRevB.69.245114} {\bibfield
  {journal} {\bibinfo  {journal} {Phys. Rev. B}\ }\textbf {\bibinfo {volume}
  {69}},\ \bibinfo {pages} {245114} (\bibinfo {year} {2004})}\BibitemShut
  {NoStop}%
\bibitem [{\citenamefont {Sakai}\ and\ \citenamefont
  {Shimizu}(1992)}]{Sakai1992}%
  \BibitemOpen
  \bibfield  {author} {\bibinfo {author} {\bibfnamefont {O.}~\bibnamefont
  {Sakai}}\ and\ \bibinfo {author} {\bibfnamefont {Y.}~\bibnamefont
  {Shimizu}},\ }\bibfield  {title} {\bibinfo {title} {Excitation spectra of the
  two impurity {Anderson} model. i. critical transition in the two magnetic
  impurity problem and the roles of the parity splitting},\ }\href
  {https://doi.org/10.1143/JPSJ.61.2333} {\bibfield  {journal} {\bibinfo
  {journal} {Journal of the Physical Society of Japan}\ }\textbf {\bibinfo
  {volume} {61}},\ \bibinfo {pages} {2333} (\bibinfo {year}
  {1992})}\BibitemShut {NoStop}%
\bibitem [{\citenamefont {Zar\'and}\ \emph {et~al.}(2006)\citenamefont
  {Zar\'and}, \citenamefont {Chung}, \citenamefont {Simon},\ and\ \citenamefont
  {Vojta}}]{Zarand2006}%
  \BibitemOpen
  \bibfield  {author} {\bibinfo {author} {\bibfnamefont {G.}~\bibnamefont
  {Zar\'and}}, \bibinfo {author} {\bibfnamefont {C.-H.}\ \bibnamefont {Chung}},
  \bibinfo {author} {\bibfnamefont {P.}~\bibnamefont {Simon}},\ and\ \bibinfo
  {author} {\bibfnamefont {M.}~\bibnamefont {Vojta}},\ }\bibfield  {title}
  {\bibinfo {title} {Quantum criticality in a double-quantum-dot system},\
  }\href {https://doi.org/10.1103/PhysRevLett.97.166802} {\bibfield  {journal}
  {\bibinfo  {journal} {Phys. Rev. Lett.}\ }\textbf {\bibinfo {volume} {97}},\
  \bibinfo {pages} {166802} (\bibinfo {year} {2006})}\BibitemShut {NoStop}%
\bibitem [{\citenamefont {Chung}\ and\ \citenamefont
  {Hofstetter}(2007)}]{Chung2007}%
  \BibitemOpen
  \bibfield  {author} {\bibinfo {author} {\bibfnamefont {C.-H.}\ \bibnamefont
  {Chung}}\ and\ \bibinfo {author} {\bibfnamefont {W.}~\bibnamefont
  {Hofstetter}},\ }\bibfield  {title} {\bibinfo {title} {{Kondo} effect in
  coupled quantum dots with {RKKY} interaction: Effects of finite temperature
  and magnetic field},\ }\href {https://doi.org/10.1103/PhysRevB.76.045329}
  {\bibfield  {journal} {\bibinfo  {journal} {Phys. Rev. B}\ }\textbf {\bibinfo
  {volume} {76}},\ \bibinfo {pages} {045329} (\bibinfo {year}
  {2007})}\BibitemShut {NoStop}%
\bibitem [{\citenamefont {Nishikawa}\ \emph
  {et~al.}(2012{\natexlab{a}})\citenamefont {Nishikawa}, \citenamefont {Crow},\
  and\ \citenamefont {Hewson}}]{Nishikawa2012}%
  \BibitemOpen
  \bibfield  {author} {\bibinfo {author} {\bibfnamefont {Y.}~\bibnamefont
  {Nishikawa}}, \bibinfo {author} {\bibfnamefont {D.~J.~G.}\ \bibnamefont
  {Crow}},\ and\ \bibinfo {author} {\bibfnamefont {A.~C.}\ \bibnamefont
  {Hewson}},\ }\bibfield  {title} {\bibinfo {title} {Convergence of energy
  scales on the approach to a local quantum critical point},\ }\href
  {https://doi.org/10.1103/PhysRevLett.108.056402} {\bibfield  {journal}
  {\bibinfo  {journal} {Phys. Rev. Lett.}\ }\textbf {\bibinfo {volume} {108}},\
  \bibinfo {pages} {056402} (\bibinfo {year} {2012}{\natexlab{a}})}\BibitemShut
  {NoStop}%
\bibitem [{\citenamefont {Nishikawa}\ \emph
  {et~al.}(2012{\natexlab{b}})\citenamefont {Nishikawa}, \citenamefont {Crow},\
  and\ \citenamefont {Hewson}}]{Nishikawa2012a}%
  \BibitemOpen
  \bibfield  {author} {\bibinfo {author} {\bibfnamefont {Y.}~\bibnamefont
  {Nishikawa}}, \bibinfo {author} {\bibfnamefont {D.~J.~G.}\ \bibnamefont
  {Crow}},\ and\ \bibinfo {author} {\bibfnamefont {A.~C.}\ \bibnamefont
  {Hewson}},\ }\bibfield  {title} {\bibinfo {title} {Phase diagram and critical
  points of a double quantum dot},\ }\href
  {https://doi.org/10.1103/PhysRevB.86.125134} {\bibfield  {journal} {\bibinfo
  {journal} {Phys. Rev. B}\ }\textbf {\bibinfo {volume} {86}},\ \bibinfo
  {pages} {125134} (\bibinfo {year} {2012}{\natexlab{b}})}\BibitemShut
  {NoStop}%
\bibitem [{\citenamefont {Nishikawa}\ \emph {et~al.}(2018)\citenamefont
  {Nishikawa}, \citenamefont {Curtin}, \citenamefont {Hewson},\ and\
  \citenamefont {Crow}}]{Nishikawa2018}%
  \BibitemOpen
  \bibfield  {author} {\bibinfo {author} {\bibfnamefont {Y.}~\bibnamefont
  {Nishikawa}}, \bibinfo {author} {\bibfnamefont {O.~J.}\ \bibnamefont
  {Curtin}}, \bibinfo {author} {\bibfnamefont {A.~C.}\ \bibnamefont {Hewson}},\
  and\ \bibinfo {author} {\bibfnamefont {D.~J.~G.}\ \bibnamefont {Crow}},\
  }\bibfield  {title} {\bibinfo {title} {Magnetic field induced quantum
  criticality and the {Luttinger} sum rule},\ }\href
  {https://doi.org/10.1103/PhysRevB.98.104419} {\bibfield  {journal} {\bibinfo
  {journal} {Phys. Rev. B}\ }\textbf {\bibinfo {volume} {98}},\ \bibinfo
  {pages} {104419} (\bibinfo {year} {2018})}\BibitemShut {NoStop}%
\bibitem [{\citenamefont {Blesio}\ \emph {et~al.}(2018)\citenamefont {Blesio},
  \citenamefont {Manuel}, \citenamefont {Roura-Bas},\ and\ \citenamefont
  {Aligia}}]{Blesio2018}%
  \BibitemOpen
  \bibfield  {author} {\bibinfo {author} {\bibfnamefont {G.~G.}\ \bibnamefont
  {Blesio}}, \bibinfo {author} {\bibfnamefont {L.~O.}\ \bibnamefont {Manuel}},
  \bibinfo {author} {\bibfnamefont {P.}~\bibnamefont {Roura-Bas}},\ and\
  \bibinfo {author} {\bibfnamefont {A.~A.}\ \bibnamefont {Aligia}},\ }\bibfield
   {title} {\bibinfo {title} {Topological quantum phase transition between
  {Fermi} liquid phases in an {Anderson} impurity model},\ }\href
  {https://doi.org/10.1103/PhysRevB.98.195435} {\bibfield  {journal} {\bibinfo
  {journal} {Phys. Rev. B}\ }\textbf {\bibinfo {volume} {98}},\ \bibinfo
  {pages} {195435} (\bibinfo {year} {2018})}\BibitemShut {NoStop}%
\bibitem [{\citenamefont {Žitko}\ \emph {et~al.}(2021)\citenamefont {Žitko},
  \citenamefont {Blesio}, \citenamefont {Manuel},\ and\ \citenamefont
  {Aligia}}]{Zitko2021}%
  \BibitemOpen
  \bibfield  {author} {\bibinfo {author} {\bibfnamefont {R.}~\bibnamefont
  {Žitko}}, \bibinfo {author} {\bibfnamefont {G.~G.}\ \bibnamefont {Blesio}},
  \bibinfo {author} {\bibfnamefont {L.~O.}\ \bibnamefont {Manuel}},\ and\
  \bibinfo {author} {\bibfnamefont {A.~A.}\ \bibnamefont {Aligia}},\ }\bibfield
   {title} {\bibinfo {title} {Iron phthalocyanine on {Au(111)} is a
  ``non-{Landau}'' {Fermi} liquid},\ }\href
  {https://doi.org/10.1038/s41467-021-26339-z} {\bibfield  {journal} {\bibinfo
  {journal} {Nature Communications}\ }\textbf {\bibinfo {volume} {12}},\
  \bibinfo {pages} {6027} (\bibinfo {year} {2021})}\BibitemShut {NoStop}%
\bibitem [{\citenamefont {Emery}\ and\ \citenamefont
  {Kivelson}(1995)}]{Emery1995}%
  \BibitemOpen
  \bibfield  {author} {\bibinfo {author} {\bibfnamefont {V.~J.}\ \bibnamefont
  {Emery}}\ and\ \bibinfo {author} {\bibfnamefont {S.~A.}\ \bibnamefont
  {Kivelson}},\ }\bibfield  {title} {\bibinfo {title} {Importance of phase
  fluctuations in superconductors with small superfluid density},\ }\href
  {https://doi.org/10.1038/374434a0} {\bibfield  {journal} {\bibinfo  {journal}
  {Nature}\ }\textbf {\bibinfo {volume} {374}},\ \bibinfo {pages} {434}
  (\bibinfo {year} {1995})}\BibitemShut {NoStop}%
\bibitem [{\citenamefont {Eberlein}\ \emph {et~al.}(2016)\citenamefont
  {Eberlein}, \citenamefont {Metzner}, \citenamefont {Sachdev},\ and\
  \citenamefont {Yamase}}]{Eberlein2016}%
  \BibitemOpen
  \bibfield  {author} {\bibinfo {author} {\bibfnamefont {A.}~\bibnamefont
  {Eberlein}}, \bibinfo {author} {\bibfnamefont {W.}~\bibnamefont {Metzner}},
  \bibinfo {author} {\bibfnamefont {S.}~\bibnamefont {Sachdev}},\ and\ \bibinfo
  {author} {\bibfnamefont {H.}~\bibnamefont {Yamase}},\ }\bibfield  {title}
  {\bibinfo {title} {{Fermi} surface reconstruction and drop in the {Hall}
  number due to spiral antiferromagnetism in high-${T}_{c}$ cuprates},\ }\href
  {https://doi.org/10.1103/PhysRevLett.117.187001} {\bibfield  {journal}
  {\bibinfo  {journal} {Phys. Rev. Lett.}\ }\textbf {\bibinfo {volume} {117}},\
  \bibinfo {pages} {187001} (\bibinfo {year} {2016})}\BibitemShut {NoStop}%
\bibitem [{\citenamefont {Verret}\ \emph {et~al.}(2017)\citenamefont {Verret},
  \citenamefont {Simard}, \citenamefont {Charlebois}, \citenamefont
  {S\'en\'echal},\ and\ \citenamefont {Tremblay}}]{Verret2017}%
  \BibitemOpen
  \bibfield  {author} {\bibinfo {author} {\bibfnamefont {S.}~\bibnamefont
  {Verret}}, \bibinfo {author} {\bibfnamefont {O.}~\bibnamefont {Simard}},
  \bibinfo {author} {\bibfnamefont {M.}~\bibnamefont {Charlebois}}, \bibinfo
  {author} {\bibfnamefont {D.}~\bibnamefont {S\'en\'echal}},\ and\ \bibinfo
  {author} {\bibfnamefont {A.-M.~S.}\ \bibnamefont {Tremblay}},\ }\bibfield
  {title} {\bibinfo {title} {Phenomenological theories of the low-temperature
  pseudogap: {Hall} number, specific heat, and {Seebeck} coefficient},\ }\href
  {https://doi.org/10.1103/PhysRevB.96.125139} {\bibfield  {journal} {\bibinfo
  {journal} {Phys. Rev. B}\ }\textbf {\bibinfo {volume} {96}},\ \bibinfo
  {pages} {125139} (\bibinfo {year} {2017})}\BibitemShut {NoStop}%
\bibitem [{\citenamefont {Bonetti}\ \emph {et~al.}(2020)\citenamefont
  {Bonetti}, \citenamefont {Mitscherling}, \citenamefont {Vilardi},\ and\
  \citenamefont {Metzner}}]{Bonetti2020}%
  \BibitemOpen
  \bibfield  {author} {\bibinfo {author} {\bibfnamefont {P.~M.}\ \bibnamefont
  {Bonetti}}, \bibinfo {author} {\bibfnamefont {J.}~\bibnamefont
  {Mitscherling}}, \bibinfo {author} {\bibfnamefont {D.}~\bibnamefont
  {Vilardi}},\ and\ \bibinfo {author} {\bibfnamefont {W.}~\bibnamefont
  {Metzner}},\ }\bibfield  {title} {\bibinfo {title} {Charge carrier drop at
  the onset of pseudogap behavior in the two-dimensional {Hubbard} model},\
  }\href {https://doi.org/10.1103/PhysRevB.101.165142} {\bibfield  {journal}
  {\bibinfo  {journal} {Phys. Rev. B}\ }\textbf {\bibinfo {volume} {101}},\
  \bibinfo {pages} {165142} (\bibinfo {year} {2020})}\BibitemShut {NoStop}%
\bibitem [{\citenamefont {Bonetti}\ and\ \citenamefont
  {Metzner}(2022)}]{Bonetti2022}%
  \BibitemOpen
  \bibfield  {author} {\bibinfo {author} {\bibfnamefont {P.~M.}\ \bibnamefont
  {Bonetti}}\ and\ \bibinfo {author} {\bibfnamefont {W.}~\bibnamefont
  {Metzner}},\ }\bibfield  {title} {\bibinfo {title} {{SU(2)} gauge theory of
  the pseudogap phase in the two-dimensional {Hubbard} model},\ }\href
  {https://doi.org/10.1103/PhysRevB.106.205152} {\bibfield  {journal} {\bibinfo
   {journal} {Phys. Rev. B}\ }\textbf {\bibinfo {volume} {106}},\ \bibinfo
  {pages} {205152} (\bibinfo {year} {2022})}\BibitemShut {NoStop}%
\bibitem [{\citenamefont {Klett}\ \emph {et~al.}(2022)\citenamefont {Klett},
  \citenamefont {Hansmann},\ and\ \citenamefont {Schäfer}}]{Klett2022}%
  \BibitemOpen
  \bibfield  {author} {\bibinfo {author} {\bibfnamefont {M.}~\bibnamefont
  {Klett}}, \bibinfo {author} {\bibfnamefont {P.}~\bibnamefont {Hansmann}},\
  and\ \bibinfo {author} {\bibfnamefont {T.}~\bibnamefont {Schäfer}},\
  }\bibfield  {title} {\bibinfo {title} {Magnetic properties and pseudogap
  formation in infinite-layer nickelates: Insights from the single-band
  {Hubbard} model},\ }\bibfield  {journal} {\bibinfo  {journal} {Frontiers in
  Physics}\ }\textbf {\bibinfo {volume} {10}},\ \href
  {https://doi.org/10.3389/fphy.2022.834682} {10.3389/fphy.2022.834682}
  (\bibinfo {year} {2022})\BibitemShut {NoStop}%
\bibitem [{\citenamefont {Lihm}\ \emph {et~al.}(2026)\citenamefont {Lihm},
  \citenamefont {Kiese}, \citenamefont {Lee},\ and\ \citenamefont
  {Kugler}}]{Lihm2026}%
  \BibitemOpen
  \bibfield  {author} {\bibinfo {author} {\bibfnamefont {J.-M.}\ \bibnamefont
  {Lihm}}, \bibinfo {author} {\bibfnamefont {D.}~\bibnamefont {Kiese}},
  \bibinfo {author} {\bibfnamefont {S.-S.~B.}\ \bibnamefont {Lee}},\ and\
  \bibinfo {author} {\bibfnamefont {F.~B.}\ \bibnamefont {Kugler}},\ }\bibfield
   {title} {\bibinfo {title} {The finite-difference parquet method: Enhanced
  electron-paramagnon scattering opens a pseudogap},\ }\href
  {https://doi.org/10.1073/pnas.2525308123} {\bibfield  {journal} {\bibinfo
  {journal} {Proceedings of the National Academy of Sciences}\ }\textbf
  {\bibinfo {volume} {123}},\ \bibinfo {pages} {e2525308123} (\bibinfo {year}
  {2026})}\BibitemShut {NoStop}%
\bibitem [{\citenamefont {Forni}\ \emph {et~al.}(2026)\citenamefont {Forni},
  \citenamefont {Bonetti}, \citenamefont {M\"uller-Groeling}, \citenamefont
  {Vilardi},\ and\ \citenamefont {Metzner}}]{Forni2026}%
  \BibitemOpen
  \bibfield  {author} {\bibinfo {author} {\bibfnamefont {P.}~\bibnamefont
  {Forni}}, \bibinfo {author} {\bibfnamefont {P.~M.}\ \bibnamefont {Bonetti}},
  \bibinfo {author} {\bibfnamefont {H.}~\bibnamefont {M\"uller-Groeling}},
  \bibinfo {author} {\bibfnamefont {D.}~\bibnamefont {Vilardi}},\ and\ \bibinfo
  {author} {\bibfnamefont {W.}~\bibnamefont {Metzner}},\ }\bibfield  {title}
  {\bibinfo {title} {Spin susceptibility in a pseudogap state with fluctuating
  spiral magnetic order},\ }\href {https://doi.org/10.1103/zm7b-jdzf}
  {\bibfield  {journal} {\bibinfo  {journal} {Phys. Rev. B}\ }\textbf {\bibinfo
  {volume} {113}},\ \bibinfo {pages} {045144} (\bibinfo {year}
  {2026})}\BibitemShut {NoStop}%
\bibitem [{\citenamefont {Hubbard}(1963)}]{Hubbard1963}%
  \BibitemOpen
  \bibfield  {author} {\bibinfo {author} {\bibfnamefont {J.}~\bibnamefont
  {Hubbard}},\ }\bibfield  {title} {\bibinfo {title} {Electron correlations in
  narrow energy bands},\ }\href {https://doi.org/10.1098/rspa.1963.0204}
  {\bibfield  {journal} {\bibinfo  {journal} {Proceedings of the Royal Society
  of London. Series A. Mathematical and Physical Sciences}\ }\textbf {\bibinfo
  {volume} {276}},\ \bibinfo {pages} {238} (\bibinfo {year}
  {1963})}\BibitemShut {NoStop}%
\bibitem [{\citenamefont {Yang}\ \emph {et~al.}(2006)\citenamefont {Yang},
  \citenamefont {Rice},\ and\ \citenamefont {Zhang}}]{Yang2006}%
  \BibitemOpen
  \bibfield  {author} {\bibinfo {author} {\bibfnamefont {K.-Y.}\ \bibnamefont
  {Yang}}, \bibinfo {author} {\bibfnamefont {T.~M.}\ \bibnamefont {Rice}},\
  and\ \bibinfo {author} {\bibfnamefont {F.-C.}\ \bibnamefont {Zhang}},\
  }\bibfield  {title} {\bibinfo {title} {Phenomenological theory of the
  pseudogap state},\ }\href {https://doi.org/10.1103/PhysRevB.73.174501}
  {\bibfield  {journal} {\bibinfo  {journal} {Phys. Rev. B}\ }\textbf {\bibinfo
  {volume} {73}},\ \bibinfo {pages} {174501} (\bibinfo {year}
  {2006})}\BibitemShut {NoStop}%
\bibitem [{\citenamefont {Sakai}\ \emph {et~al.}(2010)\citenamefont {Sakai},
  \citenamefont {Motome},\ and\ \citenamefont {Imada}}]{Sakai2010}%
  \BibitemOpen
  \bibfield  {author} {\bibinfo {author} {\bibfnamefont {S.}~\bibnamefont
  {Sakai}}, \bibinfo {author} {\bibfnamefont {Y.}~\bibnamefont {Motome}},\ and\
  \bibinfo {author} {\bibfnamefont {M.}~\bibnamefont {Imada}},\ }\bibfield
  {title} {\bibinfo {title} {Doped high-${T}_{c}$ cuprate superconductors
  elucidated in the light of zeros and poles of the electronic {Green}'s
  function},\ }\href {https://doi.org/10.1103/PhysRevB.82.134505} {\bibfield
  {journal} {\bibinfo  {journal} {Phys. Rev. B}\ }\textbf {\bibinfo {volume}
  {82}},\ \bibinfo {pages} {134505} (\bibinfo {year} {2010})}\BibitemShut
  {NoStop}%
\bibitem [{\citenamefont {Yamaji}\ and\ \citenamefont
  {Imada}(2011)}]{Yamaji2011}%
  \BibitemOpen
  \bibfield  {author} {\bibinfo {author} {\bibfnamefont {Y.}~\bibnamefont
  {Yamaji}}\ and\ \bibinfo {author} {\bibfnamefont {M.}~\bibnamefont {Imada}},\
  }\bibfield  {title} {\bibinfo {title} {Composite-fermion theory for
  pseudogap, {Fermi} arc, hole pocket, and non-{Fermi} liquid of underdoped
  cuprate superconductors},\ }\href
  {https://doi.org/10.1103/PhysRevLett.106.016404} {\bibfield  {journal}
  {\bibinfo  {journal} {Phys. Rev. Lett.}\ }\textbf {\bibinfo {volume} {106}},\
  \bibinfo {pages} {016404} (\bibinfo {year} {2011})}\BibitemShut {NoStop}%
\bibitem [{\citenamefont {Robinson}\ \emph {et~al.}(2019)\citenamefont
  {Robinson}, \citenamefont {Johnson}, \citenamefont {Rice},\ and\
  \citenamefont {Tsvelik}}]{Robinson2019}%
  \BibitemOpen
  \bibfield  {author} {\bibinfo {author} {\bibfnamefont {N.~J.}\ \bibnamefont
  {Robinson}}, \bibinfo {author} {\bibfnamefont {P.~D.}\ \bibnamefont
  {Johnson}}, \bibinfo {author} {\bibfnamefont {T.~M.}\ \bibnamefont {Rice}},\
  and\ \bibinfo {author} {\bibfnamefont {A.~M.}\ \bibnamefont {Tsvelik}},\
  }\bibfield  {title} {\bibinfo {title} {Anomalies in the pseudogap phase of
  the cuprates: Competing ground states and the role of umklapp scattering},\
  }\href {https://doi.org/10.1088/1361-6633/ab31ed} {\bibfield  {journal}
  {\bibinfo  {journal} {Reports on Progress in Physics}\ }\textbf {\bibinfo
  {volume} {82}},\ \bibinfo {pages} {126501} (\bibinfo {year}
  {2019})}\BibitemShut {NoStop}%
\bibitem [{\citenamefont {Scheurer}\ \emph {et~al.}(2018)\citenamefont
  {Scheurer}, \citenamefont {Chatterjee}, \citenamefont {Wu}, \citenamefont
  {Ferrero}, \citenamefont {Georges},\ and\ \citenamefont
  {Sachdev}}]{Scheurer2018}%
  \BibitemOpen
  \bibfield  {author} {\bibinfo {author} {\bibfnamefont {M.~S.}\ \bibnamefont
  {Scheurer}}, \bibinfo {author} {\bibfnamefont {S.}~\bibnamefont
  {Chatterjee}}, \bibinfo {author} {\bibfnamefont {W.}~\bibnamefont {Wu}},
  \bibinfo {author} {\bibfnamefont {M.}~\bibnamefont {Ferrero}}, \bibinfo
  {author} {\bibfnamefont {A.}~\bibnamefont {Georges}},\ and\ \bibinfo {author}
  {\bibfnamefont {S.}~\bibnamefont {Sachdev}},\ }\bibfield  {title} {\bibinfo
  {title} {Topological order in the pseudogap metal},\ }\href
  {https://doi.org/10.1073/pnas.1720580115} {\bibfield  {journal} {\bibinfo
  {journal} {Proceedings of the National Academy of Sciences}\ }\textbf
  {\bibinfo {volume} {115}},\ \bibinfo {pages} {E3665} (\bibinfo {year}
  {2018})}\BibitemShut {NoStop}%
\bibitem [{\citenamefont {Wu}\ \emph {et~al.}(2018)\citenamefont {Wu},
  \citenamefont {Scheurer}, \citenamefont {Chatterjee}, \citenamefont
  {Sachdev}, \citenamefont {Georges},\ and\ \citenamefont {Ferrero}}]{Wu2018}%
  \BibitemOpen
  \bibfield  {author} {\bibinfo {author} {\bibfnamefont {W.}~\bibnamefont
  {Wu}}, \bibinfo {author} {\bibfnamefont {M.~S.}\ \bibnamefont {Scheurer}},
  \bibinfo {author} {\bibfnamefont {S.}~\bibnamefont {Chatterjee}}, \bibinfo
  {author} {\bibfnamefont {S.}~\bibnamefont {Sachdev}}, \bibinfo {author}
  {\bibfnamefont {A.}~\bibnamefont {Georges}},\ and\ \bibinfo {author}
  {\bibfnamefont {M.}~\bibnamefont {Ferrero}},\ }\bibfield  {title} {\bibinfo
  {title} {Pseudogap and {Fermi}-surface topology in the two-dimensional
  {Hubbard} model},\ }\href {https://doi.org/10.1103/PhysRevX.8.021048}
  {\bibfield  {journal} {\bibinfo  {journal} {Phys. Rev. X}\ }\textbf {\bibinfo
  {volume} {8}},\ \bibinfo {pages} {021048} (\bibinfo {year}
  {2018})}\BibitemShut {NoStop}%
\bibitem [{\citenamefont {Zhang}\ and\ \citenamefont
  {Sachdev}(2020)}]{Zhang2020}%
  \BibitemOpen
  \bibfield  {author} {\bibinfo {author} {\bibfnamefont {Y.-H.}\ \bibnamefont
  {Zhang}}\ and\ \bibinfo {author} {\bibfnamefont {S.}~\bibnamefont
  {Sachdev}},\ }\bibfield  {title} {\bibinfo {title} {From the pseudogap metal
  to the {Fermi} liquid using ancilla qubits},\ }\href
  {https://doi.org/10.1103/PhysRevResearch.2.023172} {\bibfield  {journal}
  {\bibinfo  {journal} {Phys. Rev. Res.}\ }\textbf {\bibinfo {volume} {2}},\
  \bibinfo {pages} {023172} (\bibinfo {year} {2020})}\BibitemShut {NoStop}%
\bibitem [{\citenamefont {Mascot}\ \emph {et~al.}(2022)\citenamefont {Mascot},
  \citenamefont {Nikolaenko}, \citenamefont {Tikhanovskaya}, \citenamefont
  {Zhang}, \citenamefont {Morr},\ and\ \citenamefont {Sachdev}}]{Mascot2022}%
  \BibitemOpen
  \bibfield  {author} {\bibinfo {author} {\bibfnamefont {E.}~\bibnamefont
  {Mascot}}, \bibinfo {author} {\bibfnamefont {A.}~\bibnamefont {Nikolaenko}},
  \bibinfo {author} {\bibfnamefont {M.}~\bibnamefont {Tikhanovskaya}}, \bibinfo
  {author} {\bibfnamefont {Y.-H.}\ \bibnamefont {Zhang}}, \bibinfo {author}
  {\bibfnamefont {D.~K.}\ \bibnamefont {Morr}},\ and\ \bibinfo {author}
  {\bibfnamefont {S.}~\bibnamefont {Sachdev}},\ }\bibfield  {title} {\bibinfo
  {title} {Electronic spectra with paramagnon fractionalization in the
  single-band {Hubbard} model},\ }\href
  {https://doi.org/10.1103/PhysRevB.105.075146} {\bibfield  {journal} {\bibinfo
   {journal} {Phys. Rev. B}\ }\textbf {\bibinfo {volume} {105}},\ \bibinfo
  {pages} {075146} (\bibinfo {year} {2022})}\BibitemShut {NoStop}%
\bibitem [{\citenamefont {Zhou}\ \emph {et~al.}(2024)\citenamefont {Zhou},
  \citenamefont {Jin},\ and\ \citenamefont {Zhang}}]{Zhou2024}%
  \BibitemOpen
  \bibfield  {author} {\bibinfo {author} {\bibfnamefont {B.}~\bibnamefont
  {Zhou}}, \bibinfo {author} {\bibfnamefont {H.-K.}\ \bibnamefont {Jin}},\ and\
  \bibinfo {author} {\bibfnamefont {Y.-H.}\ \bibnamefont {Zhang}},\ }\href@noop
  {} {\bibinfo {title} {Variational wavefunction for {Mott} insulator at finite
  ${U}$ using ancilla qubits}} (\bibinfo {year} {2024}),\ \Eprint
  {https://arxiv.org/abs/2409.07512} {arXiv:2409.07512 [cond-mat.str-el]}
  \BibitemShut {NoStop}%
\bibitem [{\citenamefont {Ledwith}\ \emph
  {et~al.}(2025{\natexlab{a}})\citenamefont {Ledwith}, \citenamefont {Dong},
  \citenamefont {Vishwanath},\ and\ \citenamefont {Khalaf}}]{Ledwith2025}%
  \BibitemOpen
  \bibfield  {author} {\bibinfo {author} {\bibfnamefont {P.~J.}\ \bibnamefont
  {Ledwith}}, \bibinfo {author} {\bibfnamefont {J.}~\bibnamefont {Dong}},
  \bibinfo {author} {\bibfnamefont {A.}~\bibnamefont {Vishwanath}},\ and\
  \bibinfo {author} {\bibfnamefont {E.}~\bibnamefont {Khalaf}},\ }\bibfield
  {title} {\bibinfo {title} {Nonlocal moments and {Mott} semimetal in the
  {Chern} bands of twisted bilayer graphene},\ }\href
  {https://doi.org/10.1103/PhysRevX.15.021087} {\bibfield  {journal} {\bibinfo
  {journal} {Phys. Rev. X}\ }\textbf {\bibinfo {volume} {15}},\ \bibinfo
  {pages} {021087} (\bibinfo {year} {2025}{\natexlab{a}})}\BibitemShut
  {NoStop}%
\bibitem [{\citenamefont {Ledwith}\ \emph
  {et~al.}(2025{\natexlab{b}})\citenamefont {Ledwith}, \citenamefont
  {Vishwanath},\ and\ \citenamefont {Khalaf}}]{Ledwith2025a}%
  \BibitemOpen
  \bibfield  {author} {\bibinfo {author} {\bibfnamefont {P.~J.}\ \bibnamefont
  {Ledwith}}, \bibinfo {author} {\bibfnamefont {A.}~\bibnamefont
  {Vishwanath}},\ and\ \bibinfo {author} {\bibfnamefont {E.}~\bibnamefont
  {Khalaf}},\ }\href@noop {} {\bibinfo {title} {Exotic carriers from
  concentrated topology: {Dirac} trions as the origin of the missing spectral
  weight in twisted bilayer graphene}} (\bibinfo {year} {2025}{\natexlab{b}}),\
  \Eprint {https://arxiv.org/abs/2505.08779} {arXiv:2505.08779
  [cond-mat.str-el]} \BibitemShut {NoStop}%
\bibitem [{\citenamefont {Zhu}\ and\ \citenamefont {Zhu}(2013)}]{Zhu2013}%
  \BibitemOpen
  \bibfield  {author} {\bibinfo {author} {\bibfnamefont {L.}~\bibnamefont
  {Zhu}}\ and\ \bibinfo {author} {\bibfnamefont {J.-X.}\ \bibnamefont {Zhu}},\
  }\bibfield  {title} {\bibinfo {title} {Singularity in self-energy and
  composite fermion excitations of interacting electrons},\ }\href
  {https://doi.org/10.1103/PhysRevB.87.085120} {\bibfield  {journal} {\bibinfo
  {journal} {Phys. Rev. B}\ }\textbf {\bibinfo {volume} {87}},\ \bibinfo
  {pages} {085120} (\bibinfo {year} {2013})}\BibitemShut {NoStop}%
\bibitem [{\citenamefont {Sakai}\ \emph {et~al.}(2015)\citenamefont {Sakai},
  \citenamefont {Civelli}, \citenamefont {Nomura},\ and\ \citenamefont
  {Imada}}]{Sakai2015}%
  \BibitemOpen
  \bibfield  {author} {\bibinfo {author} {\bibfnamefont {S.}~\bibnamefont
  {Sakai}}, \bibinfo {author} {\bibfnamefont {M.}~\bibnamefont {Civelli}},
  \bibinfo {author} {\bibfnamefont {Y.}~\bibnamefont {Nomura}},\ and\ \bibinfo
  {author} {\bibfnamefont {M.}~\bibnamefont {Imada}},\ }\bibfield  {title}
  {\bibinfo {title} {Hidden fermionic excitation in the superconductivity of
  the strongly attractive {Hubbard} model},\ }\href
  {https://doi.org/10.1103/PhysRevB.92.180503} {\bibfield  {journal} {\bibinfo
  {journal} {Phys. Rev. B}\ }\textbf {\bibinfo {volume} {92}},\ \bibinfo
  {pages} {180503} (\bibinfo {year} {2015})}\BibitemShut {NoStop}%
\bibitem [{\citenamefont {Sakai}\ \emph {et~al.}(2016)\citenamefont {Sakai},
  \citenamefont {Civelli},\ and\ \citenamefont {Imada}}]{Sakai2016}%
  \BibitemOpen
  \bibfield  {author} {\bibinfo {author} {\bibfnamefont {S.}~\bibnamefont
  {Sakai}}, \bibinfo {author} {\bibfnamefont {M.}~\bibnamefont {Civelli}},\
  and\ \bibinfo {author} {\bibfnamefont {M.}~\bibnamefont {Imada}},\ }\bibfield
   {title} {\bibinfo {title} {Hidden fermionic excitation boosting
  high-temperature superconductivity in cuprates},\ }\href
  {https://doi.org/10.1103/PhysRevLett.116.057003} {\bibfield  {journal}
  {\bibinfo  {journal} {Phys. Rev. Lett.}\ }\textbf {\bibinfo {volume} {116}},\
  \bibinfo {pages} {057003} (\bibinfo {year} {2016})}\BibitemShut {NoStop}%
\bibitem [{\citenamefont {Imada}\ and\ \citenamefont
  {Suzuki}(2019)}]{Imada2019}%
  \BibitemOpen
  \bibfield  {author} {\bibinfo {author} {\bibfnamefont {M.}~\bibnamefont
  {Imada}}\ and\ \bibinfo {author} {\bibfnamefont {T.~J.}\ \bibnamefont
  {Suzuki}},\ }\bibfield  {title} {\bibinfo {title} {Excitons and dark fermions
  as origins of {Mott} gap, pseudogap and superconductivity in cuprate
  superconductors---general concept and basic formalism based on gap physics},\
  }\href {https://doi.org/10.7566/JPSJ.88.024701} {\bibfield  {journal}
  {\bibinfo  {journal} {Journal of the Physical Society of Japan}\ }\textbf
  {\bibinfo {volume} {88}},\ \bibinfo {pages} {024701} (\bibinfo {year}
  {2019})}\BibitemShut {NoStop}%
\bibitem [{\citenamefont {Ido}\ \emph {et~al.}(2020)\citenamefont {Ido},
  \citenamefont {Imada},\ and\ \citenamefont {Misawa}}]{Ido2020}%
  \BibitemOpen
  \bibfield  {author} {\bibinfo {author} {\bibfnamefont {K.}~\bibnamefont
  {Ido}}, \bibinfo {author} {\bibfnamefont {M.}~\bibnamefont {Imada}},\ and\
  \bibinfo {author} {\bibfnamefont {T.}~\bibnamefont {Misawa}},\ }\bibfield
  {title} {\bibinfo {title} {Charge dynamics of correlated electrons:
  Variational description with inclusion of composite fermions},\ }\href
  {https://doi.org/10.1103/PhysRevB.101.075124} {\bibfield  {journal} {\bibinfo
   {journal} {Phys. Rev. B}\ }\textbf {\bibinfo {volume} {101}},\ \bibinfo
  {pages} {075124} (\bibinfo {year} {2020})}\BibitemShut {NoStop}%
\bibitem [{\citenamefont {Dzyaloshinskii}(2003)}]{Dzyaloshinskii2003}%
  \BibitemOpen
  \bibfield  {author} {\bibinfo {author} {\bibfnamefont {I.}~\bibnamefont
  {Dzyaloshinskii}},\ }\bibfield  {title} {\bibinfo {title} {Some consequences
  of the {Luttinger} theorem: The {Luttinger} surfaces in non-{Fermi} liquids
  and {Mott} insulators},\ }\href {https://doi.org/10.1103/physrevb.68.085113}
  {\bibfield  {journal} {\bibinfo  {journal} {Phys. Rev. B}\ }\textbf {\bibinfo
  {volume} {68}},\ \bibinfo {pages} {085113} (\bibinfo {year}
  {2003})}\BibitemShut {NoStop}%
\bibitem [{\citenamefont {Fabrizio}(2020)}]{Fabrizio2020}%
  \BibitemOpen
  \bibfield  {author} {\bibinfo {author} {\bibfnamefont {M.}~\bibnamefont
  {Fabrizio}},\ }\bibfield  {title} {\bibinfo {title} {{Landau-Fermi} liquids
  without quasiparticles},\ }\href
  {https://doi.org/10.1103/physrevb.102.155122} {\bibfield  {journal} {\bibinfo
   {journal} {Phys. Rev. B}\ }\textbf {\bibinfo {volume} {102}},\ \bibinfo
  {pages} {155122} (\bibinfo {year} {2020})}\BibitemShut {NoStop}%
\bibitem [{\citenamefont {Fabrizio}(2022)}]{Fabrizio2022}%
  \BibitemOpen
  \bibfield  {author} {\bibinfo {author} {\bibfnamefont {M.}~\bibnamefont
  {Fabrizio}},\ }\bibfield  {title} {\bibinfo {title} {Emergent quasiparticles
  at {Luttinger} surfaces},\ }\href
  {https://doi.org/10.1038/s41467-022-29190-y} {\bibfield  {journal} {\bibinfo
  {journal} {Nat. Commun.}\ }\textbf {\bibinfo {volume} {13}},\ \bibinfo
  {pages} {1561} (\bibinfo {year} {2022})}\BibitemShut {NoStop}%
\bibitem [{\citenamefont {Fabrizio}(2023)}]{Fabrizio2023}%
  \BibitemOpen
  \bibfield  {author} {\bibinfo {author} {\bibfnamefont {M.}~\bibnamefont
  {Fabrizio}},\ }\bibfield  {title} {\bibinfo {title} {Spin-liquid insulators
  can be {Landau}'s {Fermi} liquids},\ }\href
  {https://doi.org/10.1103/physrevlett.130.156702} {\bibfield  {journal}
  {\bibinfo  {journal} {Phys. Rev. Lett.}\ }\textbf {\bibinfo {volume} {130}},\
  \bibinfo {pages} {156702} (\bibinfo {year} {2023})}\BibitemShut {NoStop}%
\bibitem [{\citenamefont {Abrikosov}\ \emph {et~al.}(1963)\citenamefont
  {Abrikosov}, \citenamefont {Gorkov}, \citenamefont {Dzyaloshinski},\ and\
  \citenamefont {Silverman}}]{Abrikosov1963}%
  \BibitemOpen
  \bibfield  {author} {\bibinfo {author} {\bibfnamefont {A.}~\bibnamefont
  {Abrikosov}}, \bibinfo {author} {\bibfnamefont {L.}~\bibnamefont {Gorkov}},
  \bibinfo {author} {\bibfnamefont {I.}~\bibnamefont {Dzyaloshinski}},\ and\
  \bibinfo {author} {\bibfnamefont {R.}~\bibnamefont {Silverman}},\ }\href@noop
  {} {\emph {\bibinfo {title} {Methods of quantum field theory in statistical
  physics}}}\ (\bibinfo {year} {1963})\BibitemShut {NoStop}%
\bibitem [{\citenamefont {Altland}\ and\ \citenamefont
  {Simons}(2023)}]{Altland2023}%
  \BibitemOpen
  \bibfield  {author} {\bibinfo {author} {\bibfnamefont {A.}~\bibnamefont
  {Altland}}\ and\ \bibinfo {author} {\bibfnamefont {B.}~\bibnamefont
  {Simons}},\ }\href {https://doi.org/10.1017/9781108781244} {\emph {\bibinfo
  {title} {Condensed matter field theory}}}\ (\bibinfo {year}
  {2023})\BibitemShut {NoStop}%
\bibitem [{\citenamefont {Hirsch}(1989)}]{Hirsch1989}%
  \BibitemOpen
  \bibfield  {author} {\bibinfo {author} {\bibfnamefont {J.}~\bibnamefont
  {Hirsch}},\ }\bibfield  {title} {\bibinfo {title} {Bond-charge repulsion and
  hole superconductivity},\ }\href
  {https://doi.org/10.1016/0921-4534(89)90225-6} {\bibfield  {journal}
  {\bibinfo  {journal} {Physica C: Superconductivity and its Applications}\
  }\textbf {\bibinfo {volume} {158}},\ \bibinfo {pages} {326} (\bibinfo {year}
  {1989})}\BibitemShut {NoStop}%
\bibitem [{\citenamefont {Hirsch}\ and\ \citenamefont
  {Marsiglio}(1989)}]{Hirsch1989a}%
  \BibitemOpen
  \bibfield  {author} {\bibinfo {author} {\bibfnamefont {J.}~\bibnamefont
  {Hirsch}}\ and\ \bibinfo {author} {\bibfnamefont {F.}~\bibnamefont
  {Marsiglio}},\ }\bibfield  {title} {\bibinfo {title} {Hole superconductivity:
  Review and some new results},\ }\href
  {https://doi.org/10.1016/0921-4534(89)91165-9} {\bibfield  {journal}
  {\bibinfo  {journal} {Physica C: Superconductivity and its Applications}\
  }\textbf {\bibinfo {volume} {162--164}},\ \bibinfo {pages} {591} (\bibinfo
  {year} {1989})}\BibitemShut {NoStop}%
\bibitem [{\citenamefont {Essler}\ \emph {et~al.}(1992)\citenamefont {Essler},
  \citenamefont {Korepin},\ and\ \citenamefont {Schoutens}}]{Essler1992}%
  \BibitemOpen
  \bibfield  {author} {\bibinfo {author} {\bibfnamefont {F.~H.~L.}\
  \bibnamefont {Essler}}, \bibinfo {author} {\bibfnamefont {V.~E.}\
  \bibnamefont {Korepin}},\ and\ \bibinfo {author} {\bibfnamefont
  {K.}~\bibnamefont {Schoutens}},\ }\bibfield  {title} {\bibinfo {title} {New
  exactly solvable model of strongly correlated electrons motivated by
  high-${T}_{c}$ superconductivity},\ }\href
  {https://doi.org/10.1103/physrevlett.68.2960} {\bibfield  {journal} {\bibinfo
   {journal} {Physical Review Letters}\ }\textbf {\bibinfo {volume} {68}},\
  \bibinfo {pages} {2960} (\bibinfo {year} {1992})}\BibitemShut {NoStop}%
\bibitem [{\citenamefont {Bariev}\ \emph {et~al.}(1993)\citenamefont {Bariev},
  \citenamefont {Klumper}, \citenamefont {Schadschneider},\ and\ \citenamefont
  {Zittartz}}]{Bariev1993}%
  \BibitemOpen
  \bibfield  {author} {\bibinfo {author} {\bibfnamefont {R.~Z.}\ \bibnamefont
  {Bariev}}, \bibinfo {author} {\bibfnamefont {A.}~\bibnamefont {Klumper}},
  \bibinfo {author} {\bibfnamefont {A.}~\bibnamefont {Schadschneider}},\ and\
  \bibinfo {author} {\bibfnamefont {J.}~\bibnamefont {Zittartz}},\ }\bibfield
  {title} {\bibinfo {title} {Excitation spectrum and critical exponents of a
  one-dimensional integrable model of fermions with correlated hopping},\
  }\href {https://doi.org/10.1088/0305-4470/26/19/019} {\bibfield  {journal}
  {\bibinfo  {journal} {Journal of Physics A: Mathematical and General}\
  }\textbf {\bibinfo {volume} {26}},\ \bibinfo {pages} {4863} (\bibinfo {year}
  {1993})}\BibitemShut {NoStop}%
\bibitem [{\citenamefont {Simón}\ and\ \citenamefont
  {Aligia}(1993)}]{Simon1993}%
  \BibitemOpen
  \bibfield  {author} {\bibinfo {author} {\bibfnamefont {M.~E.}\ \bibnamefont
  {Simón}}\ and\ \bibinfo {author} {\bibfnamefont {A.~A.}\ \bibnamefont
  {Aligia}},\ }\bibfield  {title} {\bibinfo {title} {{Brinkman-Rice} transition
  in layered perovskites},\ }\href {https://doi.org/10.1103/physrevb.48.7471}
  {\bibfield  {journal} {\bibinfo  {journal} {Physical Review B}\ }\textbf
  {\bibinfo {volume} {48}},\ \bibinfo {pages} {7471} (\bibinfo {year}
  {1993})}\BibitemShut {NoStop}%
\bibitem [{\citenamefont {Arrachea}\ and\ \citenamefont
  {Aligia}(1994)}]{Arrachea1994}%
  \BibitemOpen
  \bibfield  {author} {\bibinfo {author} {\bibfnamefont {L.}~\bibnamefont
  {Arrachea}}\ and\ \bibinfo {author} {\bibfnamefont {A.~A.}\ \bibnamefont
  {Aligia}},\ }\bibfield  {title} {\bibinfo {title} {Exact solution of a
  {Hubbard} chain with bond-charge interaction},\ }\href
  {https://doi.org/10.1103/physrevlett.73.2240} {\bibfield  {journal} {\bibinfo
   {journal} {Physical Review Letters}\ }\textbf {\bibinfo {volume} {73}},\
  \bibinfo {pages} {2240} (\bibinfo {year} {1994})}\BibitemShut {NoStop}%
\bibitem [{\citenamefont {Gagliano}\ \emph {et~al.}(1995)\citenamefont
  {Gagliano}, \citenamefont {Aligia}, \citenamefont {Arrachea},\ and\
  \citenamefont {Avignon}}]{Gagliano1995}%
  \BibitemOpen
  \bibfield  {author} {\bibinfo {author} {\bibfnamefont {E.~R.}\ \bibnamefont
  {Gagliano}}, \bibinfo {author} {\bibfnamefont {A.~A.}\ \bibnamefont
  {Aligia}}, \bibinfo {author} {\bibfnamefont {L.}~\bibnamefont {Arrachea}},\
  and\ \bibinfo {author} {\bibfnamefont {M.}~\bibnamefont {Avignon}},\
  }\bibfield  {title} {\bibinfo {title} {Single-particle spectral function of a
  generalized {Hubbard} model: Metal-insulator transition},\ }\href
  {https://doi.org/10.1103/PhysRevB.51.14012} {\bibfield  {journal} {\bibinfo
  {journal} {Phys. Rev. B}\ }\textbf {\bibinfo {volume} {51}},\ \bibinfo
  {pages} {14012} (\bibinfo {year} {1995})}\BibitemShut {NoStop}%
\bibitem [{\citenamefont {Schulz}\ and\ \citenamefont
  {Shastry}(1998)}]{Schulz1998}%
  \BibitemOpen
  \bibfield  {author} {\bibinfo {author} {\bibfnamefont {H.~J.}\ \bibnamefont
  {Schulz}}\ and\ \bibinfo {author} {\bibfnamefont {B.~S.}\ \bibnamefont
  {Shastry}},\ }\bibfield  {title} {\bibinfo {title} {A new class of exactly
  solvable interacting fermion models in one dimension},\ }\href
  {https://doi.org/10.1103/PhysRevLett.80.1924} {\bibfield  {journal} {\bibinfo
   {journal} {Phys. Rev. Lett.}\ }\textbf {\bibinfo {volume} {80}},\ \bibinfo
  {pages} {1924} (\bibinfo {year} {1998})}\BibitemShut {NoStop}%
\bibitem [{\citenamefont {Japaridze}\ and\ \citenamefont
  {Kampf}(1999)}]{Japaridze1999}%
  \BibitemOpen
  \bibfield  {author} {\bibinfo {author} {\bibfnamefont {G.~I.}\ \bibnamefont
  {Japaridze}}\ and\ \bibinfo {author} {\bibfnamefont {A.~P.}\ \bibnamefont
  {Kampf}},\ }\bibfield  {title} {\bibinfo {title} {Weak-coupling phase diagram
  of the extended {Hubbard} model with correlated-hopping interaction},\ }\href
  {https://doi.org/10.1103/physrevb.59.12822} {\bibfield  {journal} {\bibinfo
  {journal} {Physical Review B}\ }\textbf {\bibinfo {volume} {59}},\ \bibinfo
  {pages} {12822} (\bibinfo {year} {1999})}\BibitemShut {NoStop}%
\bibitem [{\citenamefont {Arrachea}\ and\ \citenamefont
  {Aligia}(2000)}]{Arrachea2000}%
  \BibitemOpen
  \bibfield  {author} {\bibinfo {author} {\bibfnamefont {L.}~\bibnamefont
  {Arrachea}}\ and\ \bibinfo {author} {\bibfnamefont {A.~A.}\ \bibnamefont
  {Aligia}},\ }\bibfield  {title} {\bibinfo {title} {Pairing correlations in a
  generalized {Hubbard} model for the cuprates},\ }\href
  {https://doi.org/10.1103/physrevb.61.9686} {\bibfield  {journal} {\bibinfo
  {journal} {Physical Review B}\ }\textbf {\bibinfo {volume} {61}},\ \bibinfo
  {pages} {9686} (\bibinfo {year} {2000})}\BibitemShut {NoStop}%
\bibitem [{\citenamefont {Aligia}\ \emph {et~al.}(2000)\citenamefont {Aligia},
  \citenamefont {Hallberg}, \citenamefont {Batista},\ and\ \citenamefont
  {Ortiz}}]{Aligia2000}%
  \BibitemOpen
  \bibfield  {author} {\bibinfo {author} {\bibfnamefont {A.~A.}\ \bibnamefont
  {Aligia}}, \bibinfo {author} {\bibfnamefont {K.}~\bibnamefont {Hallberg}},
  \bibinfo {author} {\bibfnamefont {C.~D.}\ \bibnamefont {Batista}},\ and\
  \bibinfo {author} {\bibfnamefont {G.}~\bibnamefont {Ortiz}},\ }\bibfield
  {title} {\bibinfo {title} {Phase diagrams from topological transitions: The
  {Hubbard} chain with correlated hopping},\ }\href
  {https://doi.org/10.1103/physrevb.61.7883} {\bibfield  {journal} {\bibinfo
  {journal} {Physical Review B}\ }\textbf {\bibinfo {volume} {61}},\ \bibinfo
  {pages} {7883} (\bibinfo {year} {2000})}\BibitemShut {NoStop}%
\bibitem [{\citenamefont {Aligia}\ \emph {et~al.}(2007)\citenamefont {Aligia},
  \citenamefont {Anfossi}, \citenamefont {Arrachea}, \citenamefont {Degli
  Esposti~Boschi}, \citenamefont {Dobry}, \citenamefont {Gazza}, \citenamefont
  {Montorsi}, \citenamefont {Ortolani},\ and\ \citenamefont
  {Torio}}]{Aligia2007}%
  \BibitemOpen
  \bibfield  {author} {\bibinfo {author} {\bibfnamefont {A.~A.}\ \bibnamefont
  {Aligia}}, \bibinfo {author} {\bibfnamefont {A.}~\bibnamefont {Anfossi}},
  \bibinfo {author} {\bibfnamefont {L.}~\bibnamefont {Arrachea}}, \bibinfo
  {author} {\bibfnamefont {C.}~\bibnamefont {Degli Esposti~Boschi}}, \bibinfo
  {author} {\bibfnamefont {A.~O.}\ \bibnamefont {Dobry}}, \bibinfo {author}
  {\bibfnamefont {C.}~\bibnamefont {Gazza}}, \bibinfo {author} {\bibfnamefont
  {A.}~\bibnamefont {Montorsi}}, \bibinfo {author} {\bibfnamefont
  {F.}~\bibnamefont {Ortolani}},\ and\ \bibinfo {author} {\bibfnamefont
  {M.~E.}\ \bibnamefont {Torio}},\ }\bibfield  {title} {\bibinfo {title}
  {Incommensurability and unconventional superconductor to insulator transition
  in the {Hubbard} model with bond-charge interaction},\ }\href
  {https://doi.org/10.1103/physrevlett.99.206401} {\bibfield  {journal}
  {\bibinfo  {journal} {Physical Review Letters}\ }\textbf {\bibinfo {volume}
  {99}},\ \bibinfo {pages} {206401} (\bibinfo {year} {2007})}\BibitemShut
  {NoStop}%
\bibitem [{\citenamefont {Jiang}\ \emph {et~al.}(2023)\citenamefont {Jiang},
  \citenamefont {Scalapino},\ and\ \citenamefont {White}}]{Jiang2023}%
  \BibitemOpen
  \bibfield  {author} {\bibinfo {author} {\bibfnamefont {S.}~\bibnamefont
  {Jiang}}, \bibinfo {author} {\bibfnamefont {D.~J.}\ \bibnamefont
  {Scalapino}},\ and\ \bibinfo {author} {\bibfnamefont {S.~R.}\ \bibnamefont
  {White}},\ }\bibfield  {title} {\bibinfo {title} {Density matrix
  renormalization group based downfolding of the three-band {Hubbard} model:
  Importance of density-assisted hopping},\ }\href
  {https://doi.org/10.1103/physrevb.108.l161111} {\bibfield  {journal}
  {\bibinfo  {journal} {Physical Review B}\ }\textbf {\bibinfo {volume}
  {108}},\ \bibinfo {pages} {l161111} (\bibinfo {year} {2023})}\BibitemShut
  {NoStop}%
\bibitem [{\citenamefont {Kovalska}\ \emph {et~al.}(2025)\citenamefont
  {Kovalska}, \citenamefont {von Delft},\ and\ \citenamefont
  {Gleis}}]{Kovalska2025}%
  \BibitemOpen
  \bibfield  {author} {\bibinfo {author} {\bibfnamefont {O.}~\bibnamefont
  {Kovalska}}, \bibinfo {author} {\bibfnamefont {J.}~\bibnamefont {von
  Delft}},\ and\ \bibinfo {author} {\bibfnamefont {A.}~\bibnamefont {Gleis}},\
  }\href@noop {} {\bibinfo {title} {Tangent space {Krylov} computation of
  real-frequency spectral functions: Influence of density-assisted hopping on
  2d {Mott} physics}} (\bibinfo {year} {2025}),\ \Eprint
  {https://arxiv.org/abs/2510.07279} {arXiv:2510.07279 [cond-mat.str-el]}
  \BibitemShut {NoStop}%
\bibitem [{\citenamefont {Greschner}\ \emph {et~al.}(2014)\citenamefont
  {Greschner}, \citenamefont {Sun}, \citenamefont {Poletti},\ and\
  \citenamefont {Santos}}]{Greschner2014}%
  \BibitemOpen
  \bibfield  {author} {\bibinfo {author} {\bibfnamefont {S.}~\bibnamefont
  {Greschner}}, \bibinfo {author} {\bibfnamefont {G.}~\bibnamefont {Sun}},
  \bibinfo {author} {\bibfnamefont {D.}~\bibnamefont {Poletti}},\ and\ \bibinfo
  {author} {\bibfnamefont {L.}~\bibnamefont {Santos}},\ }\bibfield  {title}
  {\bibinfo {title} {Density-dependent synthetic gauge fields using
  periodically modulated interactions},\ }\href
  {https://doi.org/10.1103/physrevlett.113.215303} {\bibfield  {journal}
  {\bibinfo  {journal} {Physical Review Letters}\ }\textbf {\bibinfo {volume}
  {113}},\ \bibinfo {pages} {215303} (\bibinfo {year} {2014})}\BibitemShut
  {NoStop}%
\bibitem [{\citenamefont {Bermudez}\ and\ \citenamefont
  {Porras}(2015)}]{Bermudez2015}%
  \BibitemOpen
  \bibfield  {author} {\bibinfo {author} {\bibfnamefont {A.}~\bibnamefont
  {Bermudez}}\ and\ \bibinfo {author} {\bibfnamefont {D.}~\bibnamefont
  {Porras}},\ }\bibfield  {title} {\bibinfo {title} {Interaction-dependent
  photon-assisted tunneling in optical lattices: a quantum simulator of
  strongly-correlated electrons and dynamical gauge fields},\ }\href
  {https://doi.org/10.1088/1367-2630/17/10/103021} {\bibfield  {journal}
  {\bibinfo  {journal} {New Journal of Physics}\ }\textbf {\bibinfo {volume}
  {17}},\ \bibinfo {pages} {103021} (\bibinfo {year} {2015})}\BibitemShut
  {NoStop}%
\bibitem [{\citenamefont {Barbiero}\ \emph {et~al.}(2019)\citenamefont
  {Barbiero}, \citenamefont {Schweizer}, \citenamefont {Aidelsburger},
  \citenamefont {Demler}, \citenamefont {Goldman},\ and\ \citenamefont
  {Grusdt}}]{Barbiero2019}%
  \BibitemOpen
  \bibfield  {author} {\bibinfo {author} {\bibfnamefont {L.}~\bibnamefont
  {Barbiero}}, \bibinfo {author} {\bibfnamefont {C.}~\bibnamefont {Schweizer}},
  \bibinfo {author} {\bibfnamefont {M.}~\bibnamefont {Aidelsburger}}, \bibinfo
  {author} {\bibfnamefont {E.}~\bibnamefont {Demler}}, \bibinfo {author}
  {\bibfnamefont {N.}~\bibnamefont {Goldman}},\ and\ \bibinfo {author}
  {\bibfnamefont {F.}~\bibnamefont {Grusdt}},\ }\bibfield  {title} {\bibinfo
  {title} {Coupling ultracold matter to dynamical gauge fields in optical
  lattices: From flux attachment to $\mathbb{Z}_2$ lattice gauge theories},\
  }\bibfield  {journal} {\bibinfo  {journal} {Science Advances}\ }\textbf
  {\bibinfo {volume} {5}},\ \href {https://doi.org/10.1126/sciadv.aav7444}
  {10.1126/sciadv.aav7444} (\bibinfo {year} {2019})\BibitemShut {NoStop}%
\bibitem [{\citenamefont {Görg}\ \emph {et~al.}(2019)\citenamefont {Görg},
  \citenamefont {Sandholzer}, \citenamefont {Minguzzi}, \citenamefont
  {Desbuquois}, \citenamefont {Messer},\ and\ \citenamefont
  {Esslinger}}]{Goerg2019}%
  \BibitemOpen
  \bibfield  {author} {\bibinfo {author} {\bibfnamefont {F.}~\bibnamefont
  {Görg}}, \bibinfo {author} {\bibfnamefont {K.}~\bibnamefont {Sandholzer}},
  \bibinfo {author} {\bibfnamefont {J.}~\bibnamefont {Minguzzi}}, \bibinfo
  {author} {\bibfnamefont {R.}~\bibnamefont {Desbuquois}}, \bibinfo {author}
  {\bibfnamefont {M.}~\bibnamefont {Messer}},\ and\ \bibinfo {author}
  {\bibfnamefont {T.}~\bibnamefont {Esslinger}},\ }\bibfield  {title} {\bibinfo
  {title} {Realization of density-dependent {Peierls} phases to engineer
  quantized gauge fields coupled to ultracold matter},\ }\href
  {https://doi.org/10.1038/s41567-019-0615-4} {\bibfield  {journal} {\bibinfo
  {journal} {Nature Physics}\ }\textbf {\bibinfo {volume} {15}},\ \bibinfo
  {pages} {1161} (\bibinfo {year} {2019})}\BibitemShut {NoStop}%
\bibitem [{\citenamefont {Schweizer}\ \emph {et~al.}(2019)\citenamefont
  {Schweizer}, \citenamefont {Grusdt}, \citenamefont {Berngruber},
  \citenamefont {Barbiero}, \citenamefont {Demler}, \citenamefont {Goldman},
  \citenamefont {Bloch},\ and\ \citenamefont {Aidelsburger}}]{Schweizer2019}%
  \BibitemOpen
  \bibfield  {author} {\bibinfo {author} {\bibfnamefont {C.}~\bibnamefont
  {Schweizer}}, \bibinfo {author} {\bibfnamefont {F.}~\bibnamefont {Grusdt}},
  \bibinfo {author} {\bibfnamefont {M.}~\bibnamefont {Berngruber}}, \bibinfo
  {author} {\bibfnamefont {L.}~\bibnamefont {Barbiero}}, \bibinfo {author}
  {\bibfnamefont {E.}~\bibnamefont {Demler}}, \bibinfo {author} {\bibfnamefont
  {N.}~\bibnamefont {Goldman}}, \bibinfo {author} {\bibfnamefont
  {I.}~\bibnamefont {Bloch}},\ and\ \bibinfo {author} {\bibfnamefont
  {M.}~\bibnamefont {Aidelsburger}},\ }\bibfield  {title} {\bibinfo {title}
  {{Floquet} approach to $\mathbb{Z}_2$ lattice gauge theories with ultracold
  atoms in optical lattices},\ }\href
  {https://doi.org/10.1038/s41567-019-0649-7} {\bibfield  {journal} {\bibinfo
  {journal} {Nature Physics}\ }\textbf {\bibinfo {volume} {15}},\ \bibinfo
  {pages} {1168} (\bibinfo {year} {2019})}\BibitemShut {NoStop}%
\bibitem [{\citenamefont {Lienhard}\ \emph {et~al.}(2020)\citenamefont
  {Lienhard}, \citenamefont {Scholl}, \citenamefont {Weber}, \citenamefont
  {Barredo}, \citenamefont {de~Léséleuc}, \citenamefont {Bai}, \citenamefont
  {Lang}, \citenamefont {Fleischhauer}, \citenamefont {Büchler}, \citenamefont
  {Lahaye},\ and\ \citenamefont {Browaeys}}]{Lienhard2020}%
  \BibitemOpen
  \bibfield  {author} {\bibinfo {author} {\bibfnamefont {V.}~\bibnamefont
  {Lienhard}}, \bibinfo {author} {\bibfnamefont {P.}~\bibnamefont {Scholl}},
  \bibinfo {author} {\bibfnamefont {S.}~\bibnamefont {Weber}}, \bibinfo
  {author} {\bibfnamefont {D.}~\bibnamefont {Barredo}}, \bibinfo {author}
  {\bibfnamefont {S.}~\bibnamefont {de~Léséleuc}}, \bibinfo {author}
  {\bibfnamefont {R.}~\bibnamefont {Bai}}, \bibinfo {author} {\bibfnamefont
  {N.}~\bibnamefont {Lang}}, \bibinfo {author} {\bibfnamefont {M.}~\bibnamefont
  {Fleischhauer}}, \bibinfo {author} {\bibfnamefont {H.~P.}\ \bibnamefont
  {Büchler}}, \bibinfo {author} {\bibfnamefont {T.}~\bibnamefont {Lahaye}},\
  and\ \bibinfo {author} {\bibfnamefont {A.}~\bibnamefont {Browaeys}},\
  }\bibfield  {title} {\bibinfo {title} {Realization of a density-dependent
  {Peierls} phase in a synthetic, spin-orbit coupled {Rydberg} system},\ }\href
  {https://doi.org/10.1103/physrevx.10.021031} {\bibfield  {journal} {\bibinfo
  {journal} {Physical Review X}\ }\textbf {\bibinfo {volume} {10}},\ \bibinfo
  {pages} {021031} (\bibinfo {year} {2020})}\BibitemShut {NoStop}%
\bibitem [{\citenamefont {Montorsi}\ \emph {et~al.}(2022)\citenamefont
  {Montorsi}, \citenamefont {Bhattacharya}, \citenamefont {González-Cuadra},
  \citenamefont {Lewenstein}, \citenamefont {Palumbo},\ and\ \citenamefont
  {Barbiero}}]{Montorsi2022}%
  \BibitemOpen
  \bibfield  {author} {\bibinfo {author} {\bibfnamefont {A.}~\bibnamefont
  {Montorsi}}, \bibinfo {author} {\bibfnamefont {U.}~\bibnamefont
  {Bhattacharya}}, \bibinfo {author} {\bibfnamefont {D.}~\bibnamefont
  {González-Cuadra}}, \bibinfo {author} {\bibfnamefont {M.}~\bibnamefont
  {Lewenstein}}, \bibinfo {author} {\bibfnamefont {G.}~\bibnamefont
  {Palumbo}},\ and\ \bibinfo {author} {\bibfnamefont {L.}~\bibnamefont
  {Barbiero}},\ }\bibfield  {title} {\bibinfo {title} {Interacting second-order
  topological insulators in one-dimensional fermions with correlated hopping},\
  }\href {https://doi.org/10.1103/physrevb.106.l241115} {\bibfield  {journal}
  {\bibinfo  {journal} {Physical Review B}\ }\textbf {\bibinfo {volume}
  {106}},\ \bibinfo {pages} {l241115} (\bibinfo {year} {2022})}\BibitemShut
  {NoStop}%
\bibitem [{\citenamefont {Jamotte}\ \emph {et~al.}(2022)\citenamefont
  {Jamotte}, \citenamefont {Goldman},\ and\ \citenamefont
  {Di~Liberto}}]{Jamotte2022}%
  \BibitemOpen
  \bibfield  {author} {\bibinfo {author} {\bibfnamefont {M.}~\bibnamefont
  {Jamotte}}, \bibinfo {author} {\bibfnamefont {N.}~\bibnamefont {Goldman}},\
  and\ \bibinfo {author} {\bibfnamefont {M.}~\bibnamefont {Di~Liberto}},\
  }\bibfield  {title} {\bibinfo {title} {Strain and pseudo-magnetic fields in
  optical lattices from density-assisted tunneling},\ }\bibfield  {journal}
  {\bibinfo  {journal} {Communications Physics}\ }\textbf {\bibinfo {volume}
  {5}},\ \href {https://doi.org/10.1038/s42005-022-00802-9}
  {10.1038/s42005-022-00802-9} (\bibinfo {year} {2022})\BibitemShut {NoStop}%
\bibitem [{\citenamefont {Segura}\ \emph {et~al.}(2023)\citenamefont {Segura},
  \citenamefont {Hallberg},\ and\ \citenamefont {Aligia}}]{Segura2023}%
  \BibitemOpen
  \bibfield  {author} {\bibinfo {author} {\bibfnamefont {O.~A.~M.}\
  \bibnamefont {Segura}}, \bibinfo {author} {\bibfnamefont {K.}~\bibnamefont
  {Hallberg}},\ and\ \bibinfo {author} {\bibfnamefont {A.~A.}\ \bibnamefont
  {Aligia}},\ }\bibfield  {title} {\bibinfo {title} {Charge and spin gaps in
  the ionic {Hubbard} model with density-dependent hopping},\ }\href
  {https://doi.org/10.1103/physrevb.108.195135} {\bibfield  {journal} {\bibinfo
   {journal} {Physical Review B}\ }\textbf {\bibinfo {volume} {108}},\ \bibinfo
  {pages} {195135} (\bibinfo {year} {2023})}\BibitemShut {NoStop}%
\bibitem [{\citenamefont {Roura-Bas}\ and\ \citenamefont
  {Aligia}(2023)}]{RouraBas2023}%
  \BibitemOpen
  \bibfield  {author} {\bibinfo {author} {\bibfnamefont {P.}~\bibnamefont
  {Roura-Bas}}\ and\ \bibinfo {author} {\bibfnamefont {A.~A.}\ \bibnamefont
  {Aligia}},\ }\bibfield  {title} {\bibinfo {title} {Phase diagram of the ionic
  {Hubbard} model with density-dependent hopping},\ }\href
  {https://doi.org/10.1103/physrevb.108.115132} {\bibfield  {journal} {\bibinfo
   {journal} {Physical Review B}\ }\textbf {\bibinfo {volume} {108}},\ \bibinfo
  {pages} {115132} (\bibinfo {year} {2023})}\BibitemShut {NoStop}%
\bibitem [{\citenamefont {Hirsch}(1991)}]{Hirsch1991}%
  \BibitemOpen
  \bibfield  {author} {\bibinfo {author} {\bibfnamefont {J.~E.}\ \bibnamefont
  {Hirsch}},\ }\bibfield  {title} {\bibinfo {title} {Pairing of holes in a
  tight-binding model with repulsive {Coulomb} interactions},\ }\href
  {https://doi.org/10.1103/PhysRevB.43.11400} {\bibfield  {journal} {\bibinfo
  {journal} {Phys. Rev. B}\ }\textbf {\bibinfo {volume} {43}},\ \bibinfo
  {pages} {11400} (\bibinfo {year} {1991})}\BibitemShut {NoStop}%
\bibitem [{\citenamefont {Schüttler}\ and\ \citenamefont
  {Fedro}(1992)}]{Schuettler1992}%
  \BibitemOpen
  \bibfield  {author} {\bibinfo {author} {\bibfnamefont {H.-B.}\ \bibnamefont
  {Schüttler}}\ and\ \bibinfo {author} {\bibfnamefont {A.~J.}\ \bibnamefont
  {Fedro}},\ }\bibfield  {title} {\bibinfo {title} {Copper-oxygen charge
  excitations and the effective-single-band theory of cuprate
  superconductors},\ }\href {https://doi.org/10.1103/physrevb.45.7588}
  {\bibfield  {journal} {\bibinfo  {journal} {Physical Review B}\ }\textbf
  {\bibinfo {volume} {45}},\ \bibinfo {pages} {7588} (\bibinfo {year}
  {1992})}\BibitemShut {NoStop}%
\bibitem [{\citenamefont {Simón}\ \emph {et~al.}(1993)\citenamefont {Simón},
  \citenamefont {Baliña},\ and\ \citenamefont {Aligia}}]{Simon1993a}%
  \BibitemOpen
  \bibfield  {author} {\bibinfo {author} {\bibfnamefont {M.}~\bibnamefont
  {Simón}}, \bibinfo {author} {\bibfnamefont {M.}~\bibnamefont {Baliña}},\
  and\ \bibinfo {author} {\bibfnamefont {A.}~\bibnamefont {Aligia}},\
  }\bibfield  {title} {\bibinfo {title} {Effective one-band hamiltonian for
  cuprate superconductor metal-insulator transition},\ }\href
  {https://doi.org/10.1016/0921-4534(93)90529-y} {\bibfield  {journal}
  {\bibinfo  {journal} {Physica C: Superconductivity}\ }\textbf {\bibinfo
  {volume} {206}},\ \bibinfo {pages} {297} (\bibinfo {year}
  {1993})}\BibitemShut {NoStop}%
\bibitem [{\citenamefont {Feiner}\ \emph {et~al.}(1996)\citenamefont {Feiner},
  \citenamefont {Jefferson},\ and\ \citenamefont {Raimondi}}]{Feiner1996}%
  \BibitemOpen
  \bibfield  {author} {\bibinfo {author} {\bibfnamefont {L.~F.}\ \bibnamefont
  {Feiner}}, \bibinfo {author} {\bibfnamefont {J.~H.}\ \bibnamefont
  {Jefferson}},\ and\ \bibinfo {author} {\bibfnamefont {R.}~\bibnamefont
  {Raimondi}},\ }\bibfield  {title} {\bibinfo {title} {Effective single-band
  models for the high-${T}_{c}$ cuprates. i. {Coulomb} interactions},\ }\href
  {https://doi.org/10.1103/PhysRevB.53.8751} {\bibfield  {journal} {\bibinfo
  {journal} {Phys. Rev. B}\ }\textbf {\bibinfo {volume} {53}},\ \bibinfo
  {pages} {8751} (\bibinfo {year} {1996})}\BibitemShut {NoStop}%
\bibitem [{\citenamefont {Raimondi}\ \emph {et~al.}(1996)\citenamefont
  {Raimondi}, \citenamefont {Jefferson},\ and\ \citenamefont
  {Feiner}}]{Raimondi1996}%
  \BibitemOpen
  \bibfield  {author} {\bibinfo {author} {\bibfnamefont {R.}~\bibnamefont
  {Raimondi}}, \bibinfo {author} {\bibfnamefont {J.~H.}\ \bibnamefont
  {Jefferson}},\ and\ \bibinfo {author} {\bibfnamefont {L.~F.}\ \bibnamefont
  {Feiner}},\ }\bibfield  {title} {\bibinfo {title} {Effective single-band
  models for the high-${T}_{c}$ cuprates. ii. role of apical oxygen},\ }\href
  {https://doi.org/10.1103/PhysRevB.53.8774} {\bibfield  {journal} {\bibinfo
  {journal} {Phys. Rev. B}\ }\textbf {\bibinfo {volume} {53}},\ \bibinfo
  {pages} {8774} (\bibinfo {year} {1996})}\BibitemShut {NoStop}%
\bibitem [{\citenamefont {Sim\'on}\ \emph {et~al.}(1997)\citenamefont
  {Sim\'on}, \citenamefont {Aligia},\ and\ \citenamefont
  {Gagliano}}]{Simon1997}%
  \BibitemOpen
  \bibfield  {author} {\bibinfo {author} {\bibfnamefont {M.~E.}\ \bibnamefont
  {Sim\'on}}, \bibinfo {author} {\bibfnamefont {A.~A.}\ \bibnamefont
  {Aligia}},\ and\ \bibinfo {author} {\bibfnamefont {E.~R.}\ \bibnamefont
  {Gagliano}},\ }\bibfield  {title} {\bibinfo {title} {Optical properties of an
  effective one-band {Hubbard} model for the cuprates},\ }\href
  {https://doi.org/10.1103/PhysRevB.56.5637} {\bibfield  {journal} {\bibinfo
  {journal} {Phys. Rev. B}\ }\textbf {\bibinfo {volume} {56}},\ \bibinfo
  {pages} {5637} (\bibinfo {year} {1997})}\BibitemShut {NoStop}%
\bibitem [{\citenamefont {Wilson}(1983)}]{Wilson1983}%
  \BibitemOpen
  \bibfield  {author} {\bibinfo {author} {\bibfnamefont {K.~G.}\ \bibnamefont
  {Wilson}},\ }\bibfield  {title} {\bibinfo {title} {The renormalization group
  and critical phenomena},\ }\href {https://doi.org/10.1103/RevModPhys.55.583}
  {\bibfield  {journal} {\bibinfo  {journal} {Rev. Mod. Phys.}\ }\textbf
  {\bibinfo {volume} {55}},\ \bibinfo {pages} {583} (\bibinfo {year}
  {1983})}\BibitemShut {NoStop}%
\bibitem [{\citenamefont {Profe}\ \emph {et~al.}(2025)\citenamefont {Profe},
  \citenamefont {Vučičević}, \citenamefont {Stavropoulos}, \citenamefont
  {Rösner}, \citenamefont {Valentí},\ and\ \citenamefont
  {Klebl}}]{Profe2025}%
  \BibitemOpen
  \bibfield  {author} {\bibinfo {author} {\bibfnamefont {J.~B.}\ \bibnamefont
  {Profe}}, \bibinfo {author} {\bibfnamefont {J.}~\bibnamefont {Vučičević}},
  \bibinfo {author} {\bibfnamefont {P.~P.}\ \bibnamefont {Stavropoulos}},
  \bibinfo {author} {\bibfnamefont {M.}~\bibnamefont {Rösner}}, \bibinfo
  {author} {\bibfnamefont {R.}~\bibnamefont {Valentí}},\ and\ \bibinfo
  {author} {\bibfnamefont {L.}~\bibnamefont {Klebl}},\ }\href@noop {} {\bibinfo
  {title} {Exact downfolding and its perturbative approximation}} (\bibinfo
  {year} {2025}),\ \Eprint {https://arxiv.org/abs/2507.16916} {arXiv:2507.16916
  [cond-mat.str-el]} \BibitemShut {NoStop}%
\bibitem [{Hub()}]{Hubbard_commutator}%
  \BibitemOpen
  \href@noop {} {\bibinfo {title} {As a side note, these operators show up when
  commuting an electron annihilation operator $c_{i\sigma}$ with the {Hubbard}
  interaction $h_u \propto (n_{i\uparrow} - 1/2)(n_{i\downarrow} - 1/2)$,
  $[h_u,c_{i\sigma}] \propto q_{i\sigma}$. $h_q$ therefore describes itinerant,
  {Hubbard}-dressed electrons}}\BibitemShut {NoStop}%
\bibitem [{Note1()}]{Note1}%
  \BibitemOpen
  \bibinfo {note} {The canonical anticommutation relations $\{q_{i\sigma
  },q_{j\sigma '}\} = 0$ and $\{q^{\dagger }_{i\sigma }, q_{j\sigma '}\} =
  \delta _{ij} \delta _{\sigma \sigma '}$ follow from $(2 n_{i\protect \bar
  {\sigma }} - 1)^2 = 1$. The $q$-operators and electronic operators are
  \protect \textit {not} mutually canonical.}\BibitemShut {Stop}%
\bibitem [{Note2()}]{Note2}%
  \BibitemOpen
  \bibinfo {note} {This follows from $n_{i\protect \bar \sigma }^2 =
  n_{i\protect \bar \sigma }$ and $(2 n_{i\protect \bar {\sigma }} - 1)(2
  n_{j\protect \bar {\sigma }} - 1) = 4 n_{i\protect \bar {\sigma }}
  n_{j\protect \bar {\sigma }} - 2 n_{i\protect \bar {\sigma }} - 2
  n_{j\protect \bar {\sigma }} + 1$.}\BibitemShut {Stop}%
\bibitem [{\citenamefont {Shankar}(1994)}]{Shankar1994}%
  \BibitemOpen
  \bibfield  {author} {\bibinfo {author} {\bibfnamefont {R.}~\bibnamefont
  {Shankar}},\ }\bibfield  {title} {\bibinfo {title} {Renormalization-group
  approach to interacting fermions},\ }\href
  {https://doi.org/10.1103/revmodphys.66.129} {\bibfield  {journal} {\bibinfo
  {journal} {Reviews of Modern Physics}\ }\textbf {\bibinfo {volume} {66}},\
  \bibinfo {pages} {129} (\bibinfo {year} {1994})}\BibitemShut {NoStop}%
\bibitem [{sup()}]{supplement}%
  \BibitemOpen
  \bibfield  {title} {\bibinfo {title} {{See the Supplemental Material at [url]
  for additional information on $q$-particle coherent states and the path
  integral; details on the formal calculation of the single-electron Green's
  function and details on its numerical evaluation; more information on the
  dynamical quasiparticle weight and quasiparticle propagators; a discussion of
  the non-freeness of the PG-FL; a proof of the existence of at least one
  quantum phase transition separating PG-FL from Landau FL for a real,
  particle-hole symmetric Hamiltonian; details on GDMFT; and details on
  impurity model calculations. The Supplemental Material includes
  Refs.~\cite{Rojas1995,Rieger1999,Zlatic2000,Mori1965,Haydock1980a,Viswanath1994,Tiegel2014,julien2008_EROS,Auerbach2018,Auerbach2019,Foley2024_Liouvillian,Pelz2026,Bulla1998,Nozieres1980,Andrei1984,Tsvelick1985,Affleck1991,Ludwig1991,Anders2005}}},\
  }\href@noop {} {\ }\BibitemShut {NoStop}%
\bibitem [{\citenamefont {Rojas}\ \emph {et~al.}(1995)\citenamefont {Rojas},
  \citenamefont {Godby},\ and\ \citenamefont {Needs}}]{Rojas1995}%
  \BibitemOpen
  \bibfield  {author} {\bibinfo {author} {\bibfnamefont {H.~N.}\ \bibnamefont
  {Rojas}}, \bibinfo {author} {\bibfnamefont {R.~W.}\ \bibnamefont {Godby}},\
  and\ \bibinfo {author} {\bibfnamefont {R.~J.}\ \bibnamefont {Needs}},\
  }\bibfield  {title} {\bibinfo {title} {Space-time method for ab initio
  calculations of self-energies and dielectric response functions of solids},\
  }\href {https://doi.org/10.1103/PhysRevLett.74.1827} {\bibfield  {journal}
  {\bibinfo  {journal} {Phys. Rev. Lett.}\ }\textbf {\bibinfo {volume} {74}},\
  \bibinfo {pages} {1827} (\bibinfo {year} {1995})}\BibitemShut {NoStop}%
\bibitem [{\citenamefont {Rieger}\ \emph {et~al.}(1999)\citenamefont {Rieger},
  \citenamefont {Steinbeck}, \citenamefont {White}, \citenamefont {Rojas},\
  and\ \citenamefont {Godby}}]{Rieger1999}%
  \BibitemOpen
  \bibfield  {author} {\bibinfo {author} {\bibfnamefont {M.~M.}\ \bibnamefont
  {Rieger}}, \bibinfo {author} {\bibfnamefont {L.}~\bibnamefont {Steinbeck}},
  \bibinfo {author} {\bibfnamefont {I.~D.}\ \bibnamefont {White}}, \bibinfo
  {author} {\bibfnamefont {H.~N.}\ \bibnamefont {Rojas}},\ and\ \bibinfo
  {author} {\bibfnamefont {R.~W.}\ \bibnamefont {Godby}},\ }\bibfield  {title}
  {\bibinfo {title} {The {GW} space-time method for the self-energy of large
  systems},\ }\href {https://doi.org/10.1016/S0010-4655(98)00174-X} {\bibfield
  {journal} {\bibinfo  {journal} {Comput. Phys. Commun.}\ }\textbf {\bibinfo
  {volume} {117}},\ \bibinfo {pages} {211} (\bibinfo {year}
  {1999})}\BibitemShut {NoStop}%
\bibitem [{\citenamefont {Zlati\'c}\ \emph {et~al.}(2000)\citenamefont
  {Zlati\'c}, \citenamefont {Horvati\'c}, \citenamefont {Doli\v{c}ki},
  \citenamefont {Grabowski}, \citenamefont {Entel},\ and\ \citenamefont
  {Schotte}}]{Zlatic2000}%
  \BibitemOpen
  \bibfield  {author} {\bibinfo {author} {\bibfnamefont {V.}~\bibnamefont
  {Zlati\'c}}, \bibinfo {author} {\bibfnamefont {B.}~\bibnamefont
  {Horvati\'c}}, \bibinfo {author} {\bibfnamefont {B.}~\bibnamefont
  {Doli\v{c}ki}}, \bibinfo {author} {\bibfnamefont {S.}~\bibnamefont
  {Grabowski}}, \bibinfo {author} {\bibfnamefont {P.}~\bibnamefont {Entel}},\
  and\ \bibinfo {author} {\bibfnamefont {K.-D.}\ \bibnamefont {Schotte}},\
  }\bibfield  {title} {\bibinfo {title} {Perturbation expansion for the
  two-dimensional {Hubbard} model},\ }\href
  {https://doi.org/10.1103/PhysRevB.63.035104} {\bibfield  {journal} {\bibinfo
  {journal} {Phys. Rev. B}\ }\textbf {\bibinfo {volume} {63}},\ \bibinfo
  {pages} {035104} (\bibinfo {year} {2000})}\BibitemShut {NoStop}%
\bibitem [{\citenamefont {Mori}(1965)}]{Mori1965}%
  \BibitemOpen
  \bibfield  {author} {\bibinfo {author} {\bibfnamefont {H.}~\bibnamefont
  {Mori}},\ }\bibfield  {title} {\bibinfo {title} {A continued-fraction
  representation of the time-correlation functions},\ }\href
  {https://doi.org/10.1143/PTP.34.399} {\bibfield  {journal} {\bibinfo
  {journal} {Progress of Theoretical Physics}\ }\textbf {\bibinfo {volume}
  {34}},\ \bibinfo {pages} {399} (\bibinfo {year} {1965})}\BibitemShut
  {NoStop}%
\bibitem [{\citenamefont {Haydock}(1980)}]{Haydock1980a}%
  \BibitemOpen
  \bibfield  {author} {\bibinfo {author} {\bibfnamefont {R.}~\bibnamefont
  {Haydock}},\ }\bibfield  {title} {\bibinfo {title} {The recursive solution to
  the {Schrodinger} equation},\ }\href@noop {} {\bibfield  {journal} {\bibinfo
  {journal} {Solid State Physics}\ }\textbf {\bibinfo {volume} {35}},\ \bibinfo
  {pages} {215} (\bibinfo {year} {1980})}\BibitemShut {NoStop}%
\bibitem [{\citenamefont {Viswanath}\ and\ \citenamefont
  {Müller}(1994)}]{Viswanath1994}%
  \BibitemOpen
  \bibfield  {author} {\bibinfo {author} {\bibfnamefont {V.~S.}\ \bibnamefont
  {Viswanath}}\ and\ \bibinfo {author} {\bibfnamefont {G.}~\bibnamefont
  {Müller}},\ }\href {https://doi.org/10.1007/978-3-540-48651-0} {\emph
  {\bibinfo {title} {The recursion method}}}\ (\bibinfo  {publisher} {Springer
  Berlin, Heidelberg},\ \bibinfo {year} {1994})\BibitemShut {NoStop}%
\bibitem [{\citenamefont {Tiegel}\ \emph {et~al.}(2014)\citenamefont {Tiegel},
  \citenamefont {Manmana}, \citenamefont {Pruschke},\ and\ \citenamefont
  {Honecker}}]{Tiegel2014}%
  \BibitemOpen
  \bibfield  {author} {\bibinfo {author} {\bibfnamefont {A.~C.}\ \bibnamefont
  {Tiegel}}, \bibinfo {author} {\bibfnamefont {S.~R.}\ \bibnamefont {Manmana}},
  \bibinfo {author} {\bibfnamefont {T.}~\bibnamefont {Pruschke}},\ and\
  \bibinfo {author} {\bibfnamefont {A.}~\bibnamefont {Honecker}},\ }\bibfield
  {title} {\bibinfo {title} {Matrix product state formulation of
  frequency-space dynamics at finite temperatures},\ }\href
  {https://doi.org/10.1103/PhysRevB.90.060406} {\bibfield  {journal} {\bibinfo
  {journal} {Phys. Rev. B}\ }\textbf {\bibinfo {volume} {90}},\ \bibinfo
  {pages} {060406} (\bibinfo {year} {2014})}\BibitemShut {NoStop}%
\bibitem [{\citenamefont {Julien}\ and\ \citenamefont
  {Albers}(2008)}]{julien2008_EROS}%
  \BibitemOpen
  \bibfield  {author} {\bibinfo {author} {\bibfnamefont {J.-P.}\ \bibnamefont
  {Julien}}\ and\ \bibinfo {author} {\bibfnamefont {R.~C.}\ \bibnamefont
  {Albers}},\ }\href@noop {} {\bibinfo {title} {Extended recursion in operator
  space {({EROS})}, a new impurity solver for the single impurity {Anderson}
  model}} (\bibinfo {year} {2008}),\ \Eprint {https://arxiv.org/abs/0810.3302}
  {arXiv:0810.3302 [cond-mat.str-el]} \BibitemShut {NoStop}%
\bibitem [{\citenamefont {Auerbach}(2018)}]{Auerbach2018}%
  \BibitemOpen
  \bibfield  {author} {\bibinfo {author} {\bibfnamefont {A.}~\bibnamefont
  {Auerbach}},\ }\bibfield  {title} {\bibinfo {title} {Hall number of strongly
  correlated metals},\ }\href {https://doi.org/10.1103/PhysRevLett.121.066601}
  {\bibfield  {journal} {\bibinfo  {journal} {Phys. Rev. Lett.}\ }\textbf
  {\bibinfo {volume} {121}},\ \bibinfo {pages} {066601} (\bibinfo {year}
  {2018})}\BibitemShut {NoStop}%
\bibitem [{\citenamefont {Auerbach}(2019)}]{Auerbach2019}%
  \BibitemOpen
  \bibfield  {author} {\bibinfo {author} {\bibfnamefont {A.}~\bibnamefont
  {Auerbach}},\ }\bibfield  {title} {\bibinfo {title} {Equilibrium formulae for
  transverse magnetotransport of strongly correlated metals},\ }\href
  {https://doi.org/10.1103/PhysRevB.99.115115} {\bibfield  {journal} {\bibinfo
  {journal} {Phys. Rev. B}\ }\textbf {\bibinfo {volume} {99}},\ \bibinfo
  {pages} {115115} (\bibinfo {year} {2019})}\BibitemShut {NoStop}%
\bibitem [{\citenamefont {Foley}(2024)}]{Foley2024_Liouvillian}%
  \BibitemOpen
  \bibfield  {author} {\bibinfo {author} {\bibfnamefont {A.}~\bibnamefont
  {Foley}},\ }\href@noop {} {\bibinfo {title} {{Liouvillian} recursion method
  for the electronic {Green}'s function}} (\bibinfo {year} {2024}),\ \Eprint
  {https://arxiv.org/abs/2401.02527} {arXiv:2401.02527 [cond-mat.str-el]}
  \BibitemShut {NoStop}%
\bibitem [{\citenamefont {Pelz}\ \emph {et~al.}(2026)\citenamefont {Pelz},
  \citenamefont {von Delft},\ and\ \citenamefont {Gleis}}]{Pelz2026}%
  \BibitemOpen
  \bibfield  {author} {\bibinfo {author} {\bibfnamefont {M.}~\bibnamefont
  {Pelz}}, \bibinfo {author} {\bibfnamefont {J.}~\bibnamefont {von Delft}},\
  and\ \bibinfo {author} {\bibfnamefont {A.}~\bibnamefont {Gleis}},\
  }\href@noop {} {\bibinfo {title} {{Liouvillian} interpolation of the
  self-energy of cluster dynamical mean-field theories}} (\bibinfo {year}
  {2026}),\ \Eprint {https://arxiv.org/abs/2602.16351} {arXiv:2602.16351
  [cond-mat.str-el]} \BibitemShut {NoStop}%
\bibitem [{\citenamefont {Bulla}\ \emph {et~al.}(1998)\citenamefont {Bulla},
  \citenamefont {Hewson},\ and\ \citenamefont {Pruschke}}]{Bulla1998}%
  \BibitemOpen
  \bibfield  {author} {\bibinfo {author} {\bibfnamefont {R.}~\bibnamefont
  {Bulla}}, \bibinfo {author} {\bibfnamefont {A.~C.}\ \bibnamefont {Hewson}},\
  and\ \bibinfo {author} {\bibfnamefont {T.}~\bibnamefont {Pruschke}},\
  }\bibfield  {title} {\bibinfo {title} {Numerical renormalization group
  calculations for the self-energy of the impurity {Anderson} model},\ }\href
  {https://doi.org/10.1088/0953-8984/10/37/021} {\bibfield  {journal} {\bibinfo
   {journal} {Journal of Physics: Condensed Matter}\ }\textbf {\bibinfo
  {volume} {10}},\ \bibinfo {pages} {8365} (\bibinfo {year}
  {1998})}\BibitemShut {NoStop}%
\bibitem [{\citenamefont {Nozi\`eres}\ and\ \citenamefont
  {Blandin}(1980)}]{Nozieres1980}%
  \BibitemOpen
  \bibfield  {author} {\bibinfo {author} {\bibfnamefont {P.}~\bibnamefont
  {Nozi\`eres}}\ and\ \bibinfo {author} {\bibfnamefont {A.}~\bibnamefont
  {Blandin}},\ }\bibfield  {title} {\bibinfo {title} {{Kondo} effect in real
  metals},\ }\href {https://doi.org/10.1051/jphys:01980004103019300} {\bibfield
   {journal} {\bibinfo  {journal} {J. Phys. France}\ }\textbf {\bibinfo
  {volume} {41}},\ \bibinfo {pages} {193} (\bibinfo {year} {1980})}\BibitemShut
  {NoStop}%
\bibitem [{\citenamefont {Andrei}\ and\ \citenamefont
  {Destri}(1984)}]{Andrei1984}%
  \BibitemOpen
  \bibfield  {author} {\bibinfo {author} {\bibfnamefont {N.}~\bibnamefont
  {Andrei}}\ and\ \bibinfo {author} {\bibfnamefont {C.}~\bibnamefont
  {Destri}},\ }\bibfield  {title} {\bibinfo {title} {Solution of the
  multichannel {Kondo} problem},\ }\href
  {https://doi.org/10.1103/PhysRevLett.52.364} {\bibfield  {journal} {\bibinfo
  {journal} {Phys. Rev. Lett.}\ }\textbf {\bibinfo {volume} {52}},\ \bibinfo
  {pages} {364} (\bibinfo {year} {1984})}\BibitemShut {NoStop}%
\bibitem [{\citenamefont {Tsvelick}\ and\ \citenamefont
  {Wiegmann}(1985)}]{Tsvelick1985}%
  \BibitemOpen
  \bibfield  {author} {\bibinfo {author} {\bibfnamefont {A.~M.}\ \bibnamefont
  {Tsvelick}}\ and\ \bibinfo {author} {\bibfnamefont {P.~B.}\ \bibnamefont
  {Wiegmann}},\ }\bibfield  {title} {\bibinfo {title} {Exact solution of the
  multichannel {Kondo} problem, scaling, and integrability},\ }\href
  {https://doi.org/10.1007/BF01017853} {\bibfield  {journal} {\bibinfo
  {journal} {Journal of Statistical Physics}\ }\textbf {\bibinfo {volume}
  {38}},\ \bibinfo {pages} {125} (\bibinfo {year} {1985})}\BibitemShut
  {NoStop}%
\bibitem [{\citenamefont {Affleck}\ and\ \citenamefont
  {Ludwig}(1991)}]{Affleck1991}%
  \BibitemOpen
  \bibfield  {author} {\bibinfo {author} {\bibfnamefont {I.}~\bibnamefont
  {Affleck}}\ and\ \bibinfo {author} {\bibfnamefont {A.~W.}\ \bibnamefont
  {Ludwig}},\ }\bibfield  {title} {\bibinfo {title} {Critical theory of
  overscreened {Kondo} fixed points},\ }\href
  {https://doi.org/10.1016/0550-3213(91)90419-X} {\bibfield  {journal}
  {\bibinfo  {journal} {Nuclear Physics B}\ }\textbf {\bibinfo {volume}
  {360}},\ \bibinfo {pages} {641} (\bibinfo {year} {1991})}\BibitemShut
  {NoStop}%
\bibitem [{\citenamefont {Ludwig}\ and\ \citenamefont
  {Affleck}(1991)}]{Ludwig1991}%
  \BibitemOpen
  \bibfield  {author} {\bibinfo {author} {\bibfnamefont {A.~W.~W.}\
  \bibnamefont {Ludwig}}\ and\ \bibinfo {author} {\bibfnamefont
  {I.}~\bibnamefont {Affleck}},\ }\bibfield  {title} {\bibinfo {title} {Exact
  conformal-field-theory results for the multichannel {Kondo} effect:
  Single-fermion {Green's} function, self-energy, and resistivity},\ }\href
  {https://doi.org/10.1103/PhysRevLett.67.3160} {\bibfield  {journal} {\bibinfo
   {journal} {Phys. Rev. Lett.}\ }\textbf {\bibinfo {volume} {67}},\ \bibinfo
  {pages} {3160} (\bibinfo {year} {1991})}\BibitemShut {NoStop}%
\bibitem [{\citenamefont {Anders}(2005)}]{Anders2005}%
  \BibitemOpen
  \bibfield  {author} {\bibinfo {author} {\bibfnamefont {F.~B.}\ \bibnamefont
  {Anders}},\ }\bibfield  {title} {\bibinfo {title} {Renormalization-group
  approach to spectral properties of the two-channel {Anderson} impurity
  model},\ }\href {https://doi.org/10.1103/PhysRevB.71.121101} {\bibfield
  {journal} {\bibinfo  {journal} {Phys. Rev. B}\ }\textbf {\bibinfo {volume}
  {71}},\ \bibinfo {pages} {121101} (\bibinfo {year} {2005})}\BibitemShut
  {NoStop}%
\bibitem [{\citenamefont {Nandkishore}\ \emph {et~al.}(2012)\citenamefont
  {Nandkishore}, \citenamefont {Metlitski},\ and\ \citenamefont
  {Senthil}}]{Nandkishore2012}%
  \BibitemOpen
  \bibfield  {author} {\bibinfo {author} {\bibfnamefont {R.}~\bibnamefont
  {Nandkishore}}, \bibinfo {author} {\bibfnamefont {M.~A.}\ \bibnamefont
  {Metlitski}},\ and\ \bibinfo {author} {\bibfnamefont {T.}~\bibnamefont
  {Senthil}},\ }\bibfield  {title} {\bibinfo {title} {Orthogonal metals: The
  simplest non-{Fermi} liquids},\ }\href
  {https://doi.org/10.1103/physrevb.86.045128} {\bibfield  {journal} {\bibinfo
  {journal} {Physical Review B}\ }\textbf {\bibinfo {volume} {86}},\ \bibinfo
  {pages} {045128} (\bibinfo {year} {2012})}\BibitemShut {NoStop}%
\bibitem [{Note3()}]{Note3}%
  \BibitemOpen
  \bibinfo {note} {The electron is composed of three $q$-particles which are
  deconfined.}\BibitemShut {Stop}%
\bibitem [{\citenamefont {Hodges}\ \emph {et~al.}(1971)\citenamefont {Hodges},
  \citenamefont {Smith},\ and\ \citenamefont {Wilkins}}]{Hodges1971}%
  \BibitemOpen
  \bibfield  {author} {\bibinfo {author} {\bibfnamefont {C.}~\bibnamefont
  {Hodges}}, \bibinfo {author} {\bibfnamefont {H.}~\bibnamefont {Smith}},\ and\
  \bibinfo {author} {\bibfnamefont {J.~W.}\ \bibnamefont {Wilkins}},\
  }\bibfield  {title} {\bibinfo {title} {Effect of {Fermi} surface geometry on
  electron-electron scattering},\ }\href
  {https://doi.org/10.1103/physrevb.4.302} {\bibfield  {journal} {\bibinfo
  {journal} {Physical Review B}\ }\textbf {\bibinfo {volume} {4}},\ \bibinfo
  {pages} {302} (\bibinfo {year} {1971})}\BibitemShut {NoStop}%
\bibitem [{\citenamefont {Galán}\ \emph {et~al.}(1993)\citenamefont {Galán},
  \citenamefont {Vergés},\ and\ \citenamefont {Martin-Rodero}}]{Galan1993}%
  \BibitemOpen
  \bibfield  {author} {\bibinfo {author} {\bibfnamefont {J.}~\bibnamefont
  {Galán}}, \bibinfo {author} {\bibfnamefont {J.~A.}\ \bibnamefont
  {Vergés}},\ and\ \bibinfo {author} {\bibfnamefont {A.}~\bibnamefont
  {Martin-Rodero}},\ }\bibfield  {title} {\bibinfo {title} {Second-order
  self-energy of the {Hubbard} {Hamiltonian}: Absence of quasiparticle
  excitations near half-filling},\ }\href
  {https://doi.org/10.1103/physrevb.48.13654} {\bibfield  {journal} {\bibinfo
  {journal} {Physical Review B}\ }\textbf {\bibinfo {volume} {48}},\ \bibinfo
  {pages} {13654} (\bibinfo {year} {1993})}\BibitemShut {NoStop}%
\bibitem [{\citenamefont {Daul}\ and\ \citenamefont
  {Dzierzawa}(1997)}]{Daul1997}%
  \BibitemOpen
  \bibfield  {author} {\bibinfo {author} {\bibfnamefont {S.}~\bibnamefont
  {Daul}}\ and\ \bibinfo {author} {\bibfnamefont {M.}~\bibnamefont
  {Dzierzawa}},\ }\bibfield  {title} {\bibinfo {title} {Second order
  self-energy of the two-dimensional {Hubbard} model},\ }\href
  {https://doi.org/10.1007/s002570050332} {\bibfield  {journal} {\bibinfo
  {journal} {Zeitschrift für Physik B Condensed Matter}\ }\textbf {\bibinfo
  {volume} {103}},\ \bibinfo {pages} {41} (\bibinfo {year} {1997})}\BibitemShut
  {NoStop}%
\bibitem [{\citenamefont {Das~Sarma}\ and\ \citenamefont
  {Liao}(2021)}]{DasSarma2021}%
  \BibitemOpen
  \bibfield  {author} {\bibinfo {author} {\bibfnamefont {S.}~\bibnamefont
  {Das~Sarma}}\ and\ \bibinfo {author} {\bibfnamefont {Y.}~\bibnamefont
  {Liao}},\ }\bibfield  {title} {\bibinfo {title} {Know the enemy: 2d {Fermi}
  liquids},\ }\href {https://doi.org/10.1016/j.aop.2021.168495} {\bibfield
  {journal} {\bibinfo  {journal} {Annals of Physics}\ }\textbf {\bibinfo
  {volume} {435}},\ \bibinfo {pages} {168495} (\bibinfo {year}
  {2021})}\BibitemShut {NoStop}%
\bibitem [{Note4()}]{Note4}%
  \BibitemOpen
  \bibinfo {note} {Exceptions occur in $d=1$ and at van Hove singularities, but
  $A^{(3)}_{{\protect \mathbf {k}}\sigma }(\omega =0,T=0) = 0$ still holds in
  these cases.}\BibitemShut {Stop}%
\bibitem [{\citenamefont {Gottlieb}\ and\ \citenamefont
  {Mauser}(2005)}]{Gottlieb2005}%
  \BibitemOpen
  \bibfield  {author} {\bibinfo {author} {\bibfnamefont {A.~D.}\ \bibnamefont
  {Gottlieb}}\ and\ \bibinfo {author} {\bibfnamefont {N.~J.}\ \bibnamefont
  {Mauser}},\ }\bibfield  {title} {\bibinfo {title} {New measure of electron
  correlation},\ }\href {https://doi.org/10.1103/physrevlett.95.123003}
  {\bibfield  {journal} {\bibinfo  {journal} {Physical Review Letters}\
  }\textbf {\bibinfo {volume} {95}},\ \bibinfo {pages} {123003} (\bibinfo
  {year} {2005})}\BibitemShut {NoStop}%
\bibitem [{\citenamefont {Gottlieb}\ and\ \citenamefont
  {Mauser}(2014)}]{Gottlieb2014}%
  \BibitemOpen
  \bibfield  {author} {\bibinfo {author} {\bibfnamefont {A.~D.}\ \bibnamefont
  {Gottlieb}}\ and\ \bibinfo {author} {\bibfnamefont {N.~J.}\ \bibnamefont
  {Mauser}},\ }\bibfield  {title} {\bibinfo {title} {Correlation in fermion or
  boson systems as the minimum of entropy relative to all free states}\ }\href
  {https://doi.org/10.48550/ARXIV.1403.7640} {10.48550/ARXIV.1403.7640}
  (\bibinfo {year} {2014})\BibitemShut {NoStop}%
\bibitem [{\citenamefont {Aliverti-Piuri}\ \emph {et~al.}(2024)\citenamefont
  {Aliverti-Piuri}, \citenamefont {Chatterjee}, \citenamefont {Ding},
  \citenamefont {Liao}, \citenamefont {Liebert},\ and\ \citenamefont
  {Schilling}}]{AlivertiPiuri2024}%
  \BibitemOpen
  \bibfield  {author} {\bibinfo {author} {\bibfnamefont {D.}~\bibnamefont
  {Aliverti-Piuri}}, \bibinfo {author} {\bibfnamefont {K.}~\bibnamefont
  {Chatterjee}}, \bibinfo {author} {\bibfnamefont {L.}~\bibnamefont {Ding}},
  \bibinfo {author} {\bibfnamefont {K.}~\bibnamefont {Liao}}, \bibinfo {author}
  {\bibfnamefont {J.}~\bibnamefont {Liebert}},\ and\ \bibinfo {author}
  {\bibfnamefont {C.}~\bibnamefont {Schilling}},\ }\bibfield  {title} {\bibinfo
  {title} {What can quantum information theory offer to quantum chemistry?},\
  }\href {https://doi.org/10.1039/d4fd00059e} {\bibfield  {journal} {\bibinfo
  {journal} {Faraday Discussions}\ }\textbf {\bibinfo {volume} {254}},\
  \bibinfo {pages} {76} (\bibinfo {year} {2024})}\BibitemShut {NoStop}%
\bibitem [{\citenamefont {Coleman}\ \emph {et~al.}(2005)\citenamefont
  {Coleman}, \citenamefont {Paul},\ and\ \citenamefont {Rech}}]{Coleman2005}%
  \BibitemOpen
  \bibfield  {author} {\bibinfo {author} {\bibfnamefont {P.}~\bibnamefont
  {Coleman}}, \bibinfo {author} {\bibfnamefont {I.}~\bibnamefont {Paul}},\ and\
  \bibinfo {author} {\bibfnamefont {J.}~\bibnamefont {Rech}},\ }\bibfield
  {title} {\bibinfo {title} {Sum rules and {Ward} identities in the {Kondo}
  lattice},\ }\href {https://doi.org/10.1103/physrevb.72.094430} {\bibfield
  {journal} {\bibinfo  {journal} {Phys. Rev. B}\ }\textbf {\bibinfo {volume}
  {72}},\ \bibinfo {pages} {094430} (\bibinfo {year} {2005})}\BibitemShut
  {NoStop}%
\bibitem [{\citenamefont {Rosch}(2007)}]{Rosch2007}%
  \BibitemOpen
  \bibfield  {author} {\bibinfo {author} {\bibfnamefont {A.}~\bibnamefont
  {Rosch}},\ }\bibfield  {title} {\bibinfo {title} {Breakdown of {Luttinger}'s
  theorem in two-orbital {Mott} insulators},\ }\href
  {https://doi.org/10.1140/epjb/e2007-00312-3} {\bibfield  {journal} {\bibinfo
  {journal} {Eur. Phys. J. B}\ }\textbf {\bibinfo {volume} {59}},\ \bibinfo
  {pages} {495} (\bibinfo {year} {2007})}\BibitemShut {NoStop}%
\bibitem [{\citenamefont {Farid}(2007)}]{Farid2007}%
  \BibitemOpen
  \bibfield  {author} {\bibinfo {author} {\bibfnamefont {B.}~\bibnamefont
  {Farid}},\ }\href {https://doi.org/10.48550/ARXIV.0711.0952} {\bibinfo
  {title} {On the {Luttinger} theorem concerning number of particles in the
  ground states of systems of interacting fermions}} (\bibinfo {year} {2007}),\
  \Eprint {https://arxiv.org/abs/0711.0952} {arXiv:0711.0952 [cond-mat.str-el]}
  \BibitemShut {NoStop}%
\bibitem [{\citenamefont {Stanescu}\ \emph {et~al.}(2007)\citenamefont
  {Stanescu}, \citenamefont {Phillips},\ and\ \citenamefont
  {Choy}}]{Stanescu2007}%
  \BibitemOpen
  \bibfield  {author} {\bibinfo {author} {\bibfnamefont {T.~D.}\ \bibnamefont
  {Stanescu}}, \bibinfo {author} {\bibfnamefont {P.}~\bibnamefont {Phillips}},\
  and\ \bibinfo {author} {\bibfnamefont {T.-P.}\ \bibnamefont {Choy}},\
  }\bibfield  {title} {\bibinfo {title} {Theory of the {Luttinger} surface in
  doped {Mott} insulators},\ }\href
  {https://doi.org/10.1103/physrevb.75.104503} {\bibfield  {journal} {\bibinfo
  {journal} {Physical Review B}\ }\textbf {\bibinfo {volume} {75}},\ \bibinfo
  {pages} {104503} (\bibinfo {year} {2007})}\BibitemShut {NoStop}%
\bibitem [{\citenamefont {Dave}\ \emph {et~al.}(2013)\citenamefont {Dave},
  \citenamefont {Phillips},\ and\ \citenamefont {Kane}}]{Dave2013}%
  \BibitemOpen
  \bibfield  {author} {\bibinfo {author} {\bibfnamefont {K.~B.}\ \bibnamefont
  {Dave}}, \bibinfo {author} {\bibfnamefont {P.~W.}\ \bibnamefont {Phillips}},\
  and\ \bibinfo {author} {\bibfnamefont {C.~L.}\ \bibnamefont {Kane}},\
  }\bibfield  {title} {\bibinfo {title} {Absence of {Luttinger}'s theorem due
  to zeros in the single-particle {Green} function},\ }\href
  {https://doi.org/10.1103/physrevlett.110.090403} {\bibfield  {journal}
  {\bibinfo  {journal} {Physical Review Letters}\ }\textbf {\bibinfo {volume}
  {110}},\ \bibinfo {pages} {090403} (\bibinfo {year} {2013})}\BibitemShut
  {NoStop}%
\bibitem [{\citenamefont {Seki}\ and\ \citenamefont {Yunoki}(2017)}]{Seki2017}%
  \BibitemOpen
  \bibfield  {author} {\bibinfo {author} {\bibfnamefont {K.}~\bibnamefont
  {Seki}}\ and\ \bibinfo {author} {\bibfnamefont {S.}~\bibnamefont {Yunoki}},\
  }\bibfield  {title} {\bibinfo {title} {Topological interpretation of the
  {Luttinger} theorem},\ }\href {https://doi.org/10.1103/physrevb.96.085124}
  {\bibfield  {journal} {\bibinfo  {journal} {Physical Review B}\ }\textbf
  {\bibinfo {volume} {96}},\ \bibinfo {pages} {085124} (\bibinfo {year}
  {2017})}\BibitemShut {NoStop}%
\bibitem [{\citenamefont {Heath}\ and\ \citenamefont
  {Bedell}(2020)}]{Heath2020}%
  \BibitemOpen
  \bibfield  {author} {\bibinfo {author} {\bibfnamefont {J.~T.}\ \bibnamefont
  {Heath}}\ and\ \bibinfo {author} {\bibfnamefont {K.~S.}\ \bibnamefont
  {Bedell}},\ }\bibfield  {title} {\bibinfo {title} {Necessary and sufficient
  conditions for the validity of {Luttinger}'s theorem},\ }\bibfield  {journal}
  {\bibinfo  {journal} {New J. Phys.}\ }\href
  {https://doi.org/10.1088/1367-2630/ab890e} {10.1088/1367-2630/ab890e}
  (\bibinfo {year} {2020})\BibitemShut {NoStop}%
\bibitem [{\citenamefont {Hazra}\ and\ \citenamefont
  {Coleman}(2021)}]{Hazra2021}%
  \BibitemOpen
  \bibfield  {author} {\bibinfo {author} {\bibfnamefont {T.}~\bibnamefont
  {Hazra}}\ and\ \bibinfo {author} {\bibfnamefont {P.}~\bibnamefont
  {Coleman}},\ }\bibfield  {title} {\bibinfo {title} {{Luttinger} sum rules and
  spin fractionalization in the {SU(N)} {Kondo} lattice},\ }\href
  {https://doi.org/10.1103/physrevresearch.3.033284} {\bibfield  {journal}
  {\bibinfo  {journal} {Phys. Rev. Res.}\ }\textbf {\bibinfo {volume} {3}},\
  \bibinfo {pages} {033284} (\bibinfo {year} {2021})}\BibitemShut {NoStop}%
\bibitem [{\citenamefont {Skolimowski}\ and\ \citenamefont
  {Fabrizio}(2022)}]{Skolimowski2022}%
  \BibitemOpen
  \bibfield  {author} {\bibinfo {author} {\bibfnamefont {J.}~\bibnamefont
  {Skolimowski}}\ and\ \bibinfo {author} {\bibfnamefont {M.}~\bibnamefont
  {Fabrizio}},\ }\bibfield  {title} {\bibinfo {title} {{Luttinger}'s theorem in
  the presence of {Luttinger} surfaces},\ }\href
  {https://doi.org/10.1103/physrevb.106.045109} {\bibfield  {journal} {\bibinfo
   {journal} {Physical Review B}\ }\textbf {\bibinfo {volume} {106}},\ \bibinfo
  {pages} {045109} (\bibinfo {year} {2022})}\BibitemShut {NoStop}%
\bibitem [{\citenamefont {La~Nave}\ \emph {et~al.}(2025)\citenamefont
  {La~Nave}, \citenamefont {Zhao},\ and\ \citenamefont
  {Phillips}}]{LaNave2025}%
  \BibitemOpen
  \bibfield  {author} {\bibinfo {author} {\bibfnamefont {G.}~\bibnamefont
  {La~Nave}}, \bibinfo {author} {\bibfnamefont {J.}~\bibnamefont {Zhao}},\ and\
  \bibinfo {author} {\bibfnamefont {P.~W.}\ \bibnamefont {Phillips}},\ }\href
  {https://doi.org/10.48550/ARXIV.2506.04342} {\bibinfo {title} {The
  {Luttinger} count is the homotopy not the physical charge: Generalized
  anomalies characterize non-{Fermi} liquids}} (\bibinfo {year} {2025}),\
  \Eprint {https://arxiv.org/abs/2506.04342} {arXiv:2506.04342
  [cond-mat.str-el]} \BibitemShut {NoStop}%
\bibitem [{\citenamefont {Luttinger}(1960)}]{Luttinger1960}%
  \BibitemOpen
  \bibfield  {author} {\bibinfo {author} {\bibfnamefont {J.~M.}\ \bibnamefont
  {Luttinger}},\ }\bibfield  {title} {\bibinfo {title} {{Fermi} surface and
  some simple equilibrium properties of a system of interacting fermions},\
  }\href {https://doi.org/10.1103/physrev.119.1153} {\bibfield  {journal}
  {\bibinfo  {journal} {Phys. Rev.}\ }\textbf {\bibinfo {volume} {119}},\
  \bibinfo {pages} {1153} (\bibinfo {year} {1960})}\BibitemShut {NoStop}%
\bibitem [{\citenamefont {Kokalj}\ and\ \citenamefont
  {Prelovšek}(2007)}]{Kokalj2007}%
  \BibitemOpen
  \bibfield  {author} {\bibinfo {author} {\bibfnamefont {J.}~\bibnamefont
  {Kokalj}}\ and\ \bibinfo {author} {\bibfnamefont {P.}~\bibnamefont
  {Prelovšek}},\ }\bibfield  {title} {\bibinfo {title} {{Luttinger} sum rule
  for finite systems of correlated electrons},\ }\href
  {https://doi.org/10.1103/physrevb.75.045111} {\bibfield  {journal} {\bibinfo
  {journal} {Physical Review B}\ }\textbf {\bibinfo {volume} {75}},\ \bibinfo
  {pages} {045111} (\bibinfo {year} {2007})}\BibitemShut {NoStop}%
\bibitem [{\citenamefont {Stanescu}\ and\ \citenamefont
  {Kotliar}(2006)}]{Stanescu2006}%
  \BibitemOpen
  \bibfield  {author} {\bibinfo {author} {\bibfnamefont {T.~D.}\ \bibnamefont
  {Stanescu}}\ and\ \bibinfo {author} {\bibfnamefont {G.}~\bibnamefont
  {Kotliar}},\ }\bibfield  {title} {\bibinfo {title} {{Fermi} arcs and hidden
  zeros of the {Green} function in the pseudogap state},\ }\href
  {https://doi.org/10.1103/PhysRevB.74.125110} {\bibfield  {journal} {\bibinfo
  {journal} {Phys. Rev. B}\ }\textbf {\bibinfo {volume} {74}},\ \bibinfo
  {pages} {125110} (\bibinfo {year} {2006})}\BibitemShut {NoStop}%
\bibitem [{Note5()}]{Note5}%
  \BibitemOpen
  \bibinfo {note} {Same-sign $q$-particle and electronic hopping are frustrated
  close to half-filling. For a free-electron Hamiltonian with hopping
  $t^{c}_{ij}$, we have $t^c_{ij\sigma } \langle c^{\protect \dag }_{i\sigma }
  c^{{\protect \vphantom {dagger}}}_{j\sigma }\rangle < 0$. The corresponding
  $q$-particle bond expectation value is $\langle q^{\protect \dag }_{i\sigma }
  q^{{\protect \vphantom {dagger}}}_{j\sigma } \rangle = \langle c^{\protect
  \dag }_{i\sigma } c^{{\protect \vphantom {dagger}}}_{j\sigma } \rangle [(2
  n_{\protect \bar {\sigma }} - 1)^2 - |\langle c^{\protect \dag }_{i\protect
  \bar {\sigma }} c^{{\protect \vphantom {dagger}}}_{j\protect \bar \sigma }
  \rangle |^2]$. Close to half-filling, its sign is opposite to $\langle
  c^{\protect \dag }_{i\sigma } c^{{\protect \vphantom {dagger}}}_{j\sigma }
  \rangle $, and a small $q$-particle hopping amplitude $t_{ij}$ with the same
  sign as $t^{c}_{ij}$ therefore raises the energy, i.e.\ they are
  frustrated.}\BibitemShut {Stop}%
\bibitem [{\citenamefont {Wilson}(1975)}]{Wilson1975}%
  \BibitemOpen
  \bibfield  {author} {\bibinfo {author} {\bibfnamefont {K.~G.}\ \bibnamefont
  {Wilson}},\ }\bibfield  {title} {\bibinfo {title} {The renormalization group:
  Critical phenomena and the {Kondo} problem},\ }\href
  {https://doi.org/10.1103/revmodphys.47.773} {\bibfield  {journal} {\bibinfo
  {journal} {Rev. Mod. Phys.}\ }\textbf {\bibinfo {volume} {47}},\ \bibinfo
  {pages} {773} (\bibinfo {year} {1975})}\BibitemShut {NoStop}%
\bibitem [{\citenamefont {Bulla}\ \emph {et~al.}(2008)\citenamefont {Bulla},
  \citenamefont {Costi},\ and\ \citenamefont {Pruschke}}]{Bulla2008}%
  \BibitemOpen
  \bibfield  {author} {\bibinfo {author} {\bibfnamefont {R.}~\bibnamefont
  {Bulla}}, \bibinfo {author} {\bibfnamefont {T.~A.}\ \bibnamefont {Costi}},\
  and\ \bibinfo {author} {\bibfnamefont {T.}~\bibnamefont {Pruschke}},\
  }\bibfield  {title} {\bibinfo {title} {Numerical renormalization group method
  for quantum impurity systems},\ }\href
  {https://doi.org/10.1103/revmodphys.80.395} {\bibfield  {journal} {\bibinfo
  {journal} {Rev. Mod. Phys.}\ }\textbf {\bibinfo {volume} {80}},\ \bibinfo
  {pages} {395} (\bibinfo {year} {2008})}\BibitemShut {NoStop}%
\bibitem [{\citenamefont {Weichselbaum}\ and\ \citenamefont
  {Delft}(2007)}]{Weichselbaum2007}%
  \BibitemOpen
  \bibfield  {author} {\bibinfo {author} {\bibfnamefont {A.}~\bibnamefont
  {Weichselbaum}}\ and\ \bibinfo {author} {\bibfnamefont {J.~v.}\ \bibnamefont
  {Delft}},\ }\bibfield  {title} {\bibinfo {title} {Sum-rule conserving
  spectral functions from the numerical renormalization group},\ }\href
  {https://doi.org/10.1103/physrevlett.99.076402} {\bibfield  {journal}
  {\bibinfo  {journal} {Phys. Rev. Lett.}\ }\textbf {\bibinfo {volume} {99}},\
  \bibinfo {pages} {076402} (\bibinfo {year} {2007})}\BibitemShut {NoStop}%
\bibitem [{\citenamefont
  {Weichselbaum}(2012{\natexlab{a}})}]{Weichselbaum2012}%
  \BibitemOpen
  \bibfield  {author} {\bibinfo {author} {\bibfnamefont {A.}~\bibnamefont
  {Weichselbaum}},\ }\bibfield  {title} {\bibinfo {title} {Tensor networks and
  the numerical renormalization group},\ }\href
  {https://doi.org/10.1103/physrevb.86.245124} {\bibfield  {journal} {\bibinfo
  {journal} {Phys. Rev. B}\ }\textbf {\bibinfo {volume} {86}},\ \bibinfo
  {pages} {245124} (\bibinfo {year} {2012}{\natexlab{a}})}\BibitemShut
  {NoStop}%
\bibitem [{\citenamefont {Kugler}(2022)}]{Kugler2022}%
  \BibitemOpen
  \bibfield  {author} {\bibinfo {author} {\bibfnamefont {F.~B.}\ \bibnamefont
  {Kugler}},\ }\bibfield  {title} {\bibinfo {title} {Improved estimator for
  numerical renormalization group calculations of the self-energy},\ }\href
  {https://doi.org/10.1103/physrevb.105.245132} {\bibfield  {journal} {\bibinfo
   {journal} {Phys. Rev. B}\ }\textbf {\bibinfo {volume} {105}},\ \bibinfo
  {pages} {245132} (\bibinfo {year} {2022})}\BibitemShut {NoStop}%
\bibitem [{\citenamefont
  {Weichselbaum}(2012{\natexlab{b}})}]{Weichselbaum2012a}%
  \BibitemOpen
  \bibfield  {author} {\bibinfo {author} {\bibfnamefont {A.}~\bibnamefont
  {Weichselbaum}},\ }\bibfield  {title} {\bibinfo {title} {Non-abelian
  symmetries in tensor networks: A quantum symmetry space approach},\ }\href
  {https://doi.org/10.1016/j.aop.2012.07.009} {\bibfield  {journal} {\bibinfo
  {journal} {Ann. of Phys.}\ }\textbf {\bibinfo {volume} {327}},\ \bibinfo
  {pages} {2972} (\bibinfo {year} {2012}{\natexlab{b}})}\BibitemShut {NoStop}%
\bibitem [{\citenamefont {Weichselbaum}(2020)}]{Weichselbaum2020}%
  \BibitemOpen
  \bibfield  {author} {\bibinfo {author} {\bibfnamefont {A.}~\bibnamefont
  {Weichselbaum}},\ }\bibfield  {title} {\bibinfo {title} {X-symbols for
  non-abelian symmetries in tensor networks},\ }\href
  {https://doi.org/10.1103/physrevresearch.2.023385} {\bibfield  {journal}
  {\bibinfo  {journal} {Phys. Rev. Research}\ }\textbf {\bibinfo {volume}
  {2}},\ \bibinfo {pages} {023385} (\bibinfo {year} {2020})}\BibitemShut
  {NoStop}%
\bibitem [{\citenamefont {Weichselbaum}(2024)}]{Weichselbaum2024}%
  \BibitemOpen
  \bibfield  {author} {\bibinfo {author} {\bibfnamefont {A.}~\bibnamefont
  {Weichselbaum}},\ }\bibfield  {title} {\bibinfo {title} {{QSpace} - an
  open-source tensor library for abelian and non-abelian symmetries},\ }\href
  {https://doi.org/10.21468/scipostphyscodeb.40} {\bibfield  {journal}
  {\bibinfo  {journal} {SciPost Phys. Codebases}\ ,\ \bibinfo {pages} {40}}
  (\bibinfo {year} {2024})}\BibitemShut {NoStop}%
\bibitem [{\citenamefont {Lee}\ and\ \citenamefont
  {Weichselbaum}(2016)}]{Lee2016}%
  \BibitemOpen
  \bibfield  {author} {\bibinfo {author} {\bibfnamefont {S.-S.~B.}\
  \bibnamefont {Lee}}\ and\ \bibinfo {author} {\bibfnamefont {A.}~\bibnamefont
  {Weichselbaum}},\ }\bibfield  {title} {\bibinfo {title} {Adaptive broadening
  to improve spectral resolution in the numerical renormalization group},\
  }\href {https://doi.org/10.1103/physrevb.94.235127} {\bibfield  {journal}
  {\bibinfo  {journal} {Phys. Rev. B}\ }\textbf {\bibinfo {volume} {94}},\
  \bibinfo {pages} {235127} (\bibinfo {year} {2016})}\BibitemShut {NoStop}%
\bibitem [{\citenamefont {Lee}\ \emph {et~al.}(2017)\citenamefont {Lee},
  \citenamefont {Delft},\ and\ \citenamefont {Weichselbaum}}]{Lee2017}%
  \BibitemOpen
  \bibfield  {author} {\bibinfo {author} {\bibfnamefont {S.-S.~B.}\
  \bibnamefont {Lee}}, \bibinfo {author} {\bibfnamefont {J.~v.}\ \bibnamefont
  {Delft}},\ and\ \bibinfo {author} {\bibfnamefont {A.}~\bibnamefont
  {Weichselbaum}},\ }\bibfield  {title} {\bibinfo {title} {Doublon-holon origin
  of the subpeaks at the {Hubbard} band edges},\ }\href
  {https://doi.org/10.1103/physrevlett.119.236402} {\bibfield  {journal}
  {\bibinfo  {journal} {Phys. Rev. Lett.}\ }\textbf {\bibinfo {volume} {119}},\
  \bibinfo {pages} {236402} (\bibinfo {year} {2017})}\BibitemShut {NoStop}%
\bibitem [{\citenamefont {Stanescu}\ and\ \citenamefont
  {Kotliar}(2004)}]{Stanescu2004}%
  \BibitemOpen
  \bibfield  {author} {\bibinfo {author} {\bibfnamefont {T.~D.}\ \bibnamefont
  {Stanescu}}\ and\ \bibinfo {author} {\bibfnamefont {G.}~\bibnamefont
  {Kotliar}},\ }\bibfield  {title} {\bibinfo {title} {Strong coupling theory
  for interacting lattice models},\ }\href
  {https://doi.org/10.1103/physrevb.70.205112} {\bibfield  {journal} {\bibinfo
  {journal} {Physical Review B}\ }\textbf {\bibinfo {volume} {70}},\ \bibinfo
  {pages} {205112} (\bibinfo {year} {2004})}\BibitemShut {NoStop}%
\bibitem [{\citenamefont {Gleis}\ \emph {et~al.}(2024)\citenamefont {Gleis},
  \citenamefont {Lee}, \citenamefont {Kotliar},\ and\ \citenamefont
  {Delft}}]{Gleis2024}%
  \BibitemOpen
  \bibfield  {author} {\bibinfo {author} {\bibfnamefont {A.}~\bibnamefont
  {Gleis}}, \bibinfo {author} {\bibfnamefont {S.-S.~B.}\ \bibnamefont {Lee}},
  \bibinfo {author} {\bibfnamefont {G.}~\bibnamefont {Kotliar}},\ and\ \bibinfo
  {author} {\bibfnamefont {J.~v.}\ \bibnamefont {Delft}},\ }\bibfield  {title}
  {\bibinfo {title} {Emergent properties of the periodic {Anderson} model: A
  high-resolution, real-frequency study of heavy-fermion quantum criticality},\
  }\href {https://doi.org/10.1103/physrevx.14.041036} {\bibfield  {journal}
  {\bibinfo  {journal} {Phys. Rev. X}\ }\textbf {\bibinfo {volume} {14}},\
  \bibinfo {pages} {041036} (\bibinfo {year} {2024})}\BibitemShut {NoStop}%
\bibitem [{\citenamefont {Gleis}\ \emph {et~al.}(2025)\citenamefont {Gleis},
  \citenamefont {Lee}, \citenamefont {Kotliar},\ and\ \citenamefont {von
  Delft}}]{Gleis2025}%
  \BibitemOpen
  \bibfield  {author} {\bibinfo {author} {\bibfnamefont {A.}~\bibnamefont
  {Gleis}}, \bibinfo {author} {\bibfnamefont {S.-S.~B.}\ \bibnamefont {Lee}},
  \bibinfo {author} {\bibfnamefont {G.}~\bibnamefont {Kotliar}},\ and\ \bibinfo
  {author} {\bibfnamefont {J.}~\bibnamefont {von Delft}},\ }\bibfield  {title}
  {\bibinfo {title} {Dynamical scaling and {Planckian} dissipation due to
  heavy-fermion quantum criticality},\ }\href
  {https://doi.org/10.1103/physrevlett.134.106501} {\bibfield  {journal}
  {\bibinfo  {journal} {Physical Review Letters}\ }\textbf {\bibinfo {volume}
  {134}},\ \bibinfo {pages} {106501} (\bibinfo {year} {2025})}\BibitemShut
  {NoStop}%
\end{thebibliography}%

\clearpage

\setcounter{secnumdepth}{2} 

\bigskip
\bigskip
\onecolumngrid
\begin{center}
\textbf{\large End Matter}
\end{center}
\bigskip
\vspace{-0.2cm}
\twocolumngrid

Section~\ref{sec:Luttinger_Cv} of this ``End Matter'' explicitly shows that there is no Luttinger surface contribution to the specific heat in the $q$-particle model,
while Sec.~\ref{sec:2IAM} demonstrates the presence of a $q$-particle resonance in the pseudogapped phase of the two-impurity Anderson model.

\vspace{-0.5cm}
\section{Absence of Luttinger surface contribution to the specific heat}
\label{sec:Luttinger_Cv}
\vspace{-0.2cm}

In this section, we compute the QP density of states~(DOS) as proposed in Refs.~\cite{Fabrizio2020,Fabrizio2022,Fabrizio2023}.
By construction, this QP-DOS contains contributions from the LSs. 
However, due to these LS contributions, this QP-DOS leads to incorrect predictions for the specific heat.

The QP propagator defined in Refs.~\cite{Fabrizio2020,Fabrizio2022} is given by
\vspace{-4pt}
\begin{align}
\nonumber
G^{\ast}_{\vec{k}\sigma}(\omega) = Z^{-1}_{\vec{k}\sigma}(\omega) G_{\vec{k}\sigma}(\omega) \, , \;
Z^{-1}_{\vec{k}\sigma}(\omega) = 1 - \frac{\partial \mr{Re} \, \Sigma_{\vec{k}\sigma}(\omega)}{\partial \omega} \, ,
\end{align}
where the QP weight $Z_{\vec{k}\sigma}(\omega)$ is frequency dependent. 
Close to a FS, $Z_{\vec{k}\sigma}(\omega)$ is constant and identical to the usual QP weight, 
while at the LSs encountered in Refs.~\cite{Fabrizio2020,Fabrizio2022} and in this work, one finds $Z_{\vec{k}\sigma}(\omega) \sim \omega^2$.

In the vicinity of LSs, our self-energy has the low-frequency form
\vspace{-3pt}
\begin{align}
\label{eq:EM_Sigma_lowE}
\Sigma_{\vec{k}\sigma}(\omega) &= \Sigma^{\mr{HF}}_{\vec{k}\sigma} + \frac{\Delta^2_{\vec{k}\sigma}}{\omega - \epsilon^{\ast}_{\vec{k}\sigma} + \mr{i} \gamma^{\ast}_{\vec{k}\sigma} \omega^2} 
\\ \nonumber
&\simeq \Sigma^{\mr{HF}}_{\vec{k}\sigma} + \frac{\Delta^2_{\vec{k}\sigma}}{\omega - \epsilon^{\ast}_{\vec{k}\sigma}} - \mr{i} \gamma^{\ast}_{\vec{k}\sigma} \Delta^2_{\vec{k}\sigma} \frac{\omega^2}{(\omega - \epsilon^{\ast}_{\vec{k}\sigma})^2} \, ,
\end{align}
where $\Sigma^{\mr{HF}}_{\vec{k}\sigma}$ is the Hartree-Fock contribution and the parameters $\Delta_{\vec{k}\sigma}$, $\epsilon^{\ast}_{\vec{k}\sigma}$, and $\gamma^{\ast}_{\vec{k}\sigma}$
can be extracted from the low-energy behavior of $G^{(3)}_{\vec{k}\sigma}(\omega)$ in the vicinity of the LS, see the SM~\cite{supplement} for details.
A self-energy of the form~\eqref{eq:EM_Sigma_lowE} has been considered on phenomenological grounds to study LS quasiparticles in Ref.~\cite{Fabrizio2022}.
Our analysis closely follows Ref.~\cite{Fabrizio2022}, and we refer the reader to this reference for details and justification.

The dynamical quasiparticle weight is
\vspace{-2pt}
\begin{align}
Z^{-1}_{\vec{k}\sigma}(\omega) = \frac{(\omega - \epsilon^{\ast}_{\vec{k}\sigma})^2 + \Delta^2_{\vec{k}\sigma}}{(\omega - \epsilon^{\ast}_{\vec{k}\sigma})^2} \, ,
\end{align}
so that the QP propagator in the vicinity of a LS is
\vspace{-3pt}
\begin{align}
\label{eq:EM_Gast}
&G^{\ast}_{\vec{k}\sigma}(\omega) =
\\[-5pt] \nonumber
&\quad \frac{(\omega - \epsilon^{\ast}_{\vec{k}\sigma})^2 + \Delta^2_{\vec{k}\sigma}}{(\omega - E_{\vec{k}\sigma}) (\omega - \epsilon^{\ast}_{\vec{k}\sigma})^2  -  (\omega -\epsilon^{\ast}_{\vec{k}\sigma})\Delta^2_{\vec{k}\sigma} + \mr{i}\gamma_{\vec{k}\sigma} \Delta^2_{\vec{k}\sigma} \omega^2} \, ,
\end{align}
with $E_{\vec{k}\sigma} = \epsilon_{\vec{k}\sigma} + \Sigma^{\mr{HF}}_{\vec{k}\sigma}$.
Note that Eq.~\eqref{eq:EM_Gast} is only valid in the vicinity of LSs, since Eq.~\eqref{eq:EM_Sigma_lowE} does not hold everywhere.
At a LS, $G^{\ast}_{\vec{k}\sigma}(\omega)$ exhibits a pole of weight 1 and with dispersion $\epsilon^{\ast}_{\vec{k}\sigma}$.
The recently proposed Matsubara axis representation~\cite{Fabrizio2023} of the dynamical QP weight 
$Z_{\vec{k}\sigma}(\mr{i}\omega)$ and QP propagator $G^{\ast}_{\vec{k}\sigma}(\mr{i}\omega)$ leads to results consistent with the real-frequency representation, see the SM~\cite{supplement} for details.

At the FS and if $Z_{\sigma} = (2 n_{\bar{\sigma}} - 1)^2 \neq 0$, we get, as usual, $Z_{\vec{k}\sigma}(0) = Z_{\sigma} \neq 0$, and the QP propagator is
\vspace{-3pt}
\begin{align}
G^{\ast}_{\vec{k}\sigma}(\omega) \simeq \frac{1}{\omega - Z_{\sigma}(\epsilon_{\vec{k}\sigma} + \Sigma_{\vec{k}\sigma}(0))} = \frac{1}{\omega - \epsilon_{\vec{k}\sigma}} \, ,
\end{align}
see the SM~\cite{supplement} for a derivation. At the FS, $G^{\ast}_{\vec{k}\sigma}(\omega)$ is therefore identical to the $q$-QP propagator, and the FS contribution to the QP-DOS already accounts for the correct specific heat.
Note that at $Z_{\sigma} = 0$, the FS contribution in $G^{\ast}_{\vec{k}\sigma}(\omega)$ is missing.

\begin{figure}[t]
\includegraphics[width=\linewidth]{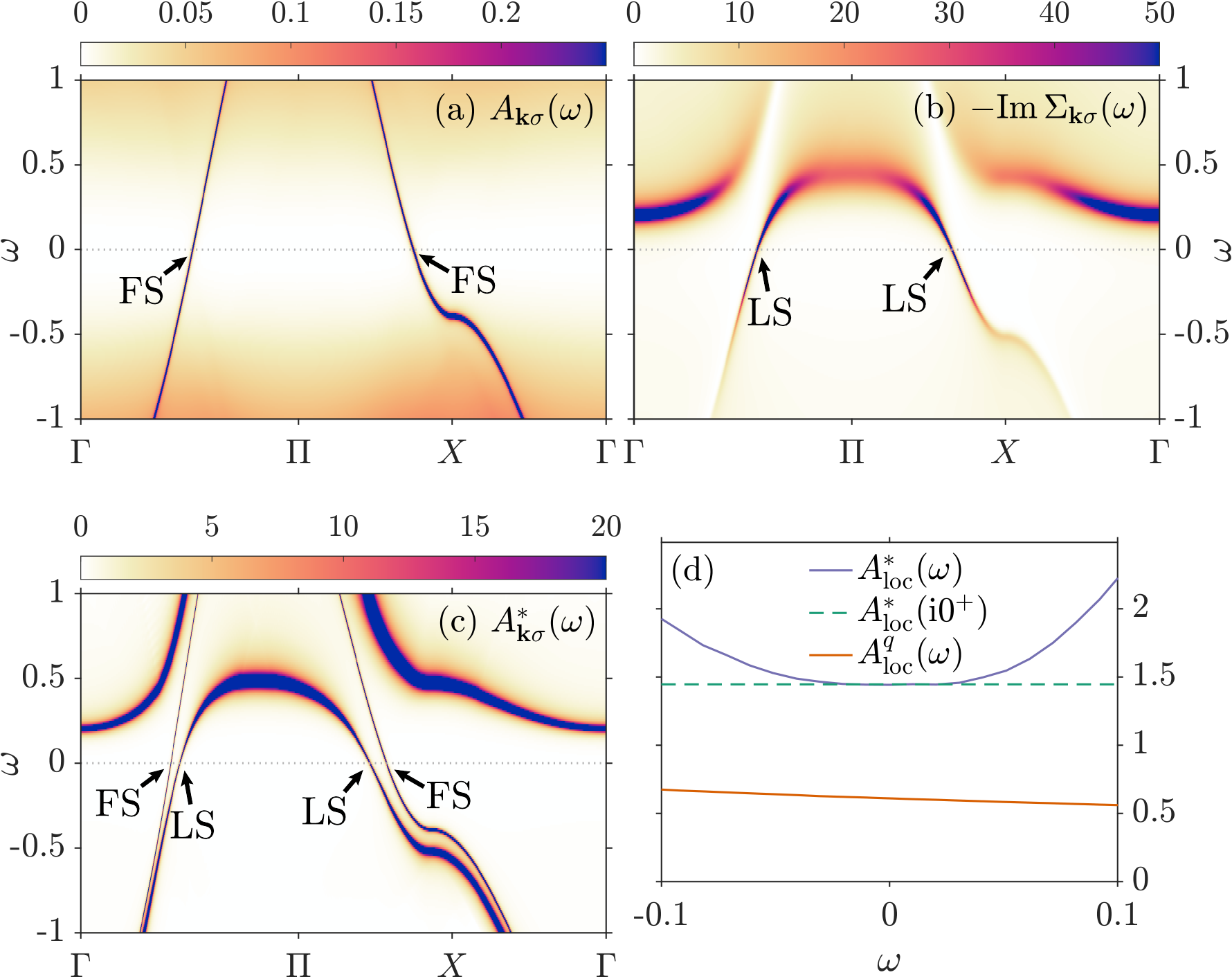}
\caption{(a) $\vec{k}$-resolved spectral function [$\Gamma = (0,0)$, $X = (0,\pi)$, $\Pi = (\pi,\pi)$], (b) corresponding self-energy, (c) $\vec{k}$-resolved QP spectral function, and (d) QP DOS and $q$-particle DOS
for the $t$-$t'$ square lattice model discussed in the main text, at filling $n_{\sigma} = 0.6$. For better visualization, we applied Lorentzian broadening of width $\eta = 5\times 10^{-3}$ in (a-c).
To obtain the data in panel (d), we regularized the integrals in Eqs.~\eqref{eq:Aast_loc} and \eqref{eq:Aq_loc}  with Lorentzian broadening of width $\eta = 10^{-3}$.}
\label{fig:G_SE_Gqp_n06}
\end{figure}

\begin{figure}[t]
\includegraphics[width=\linewidth]{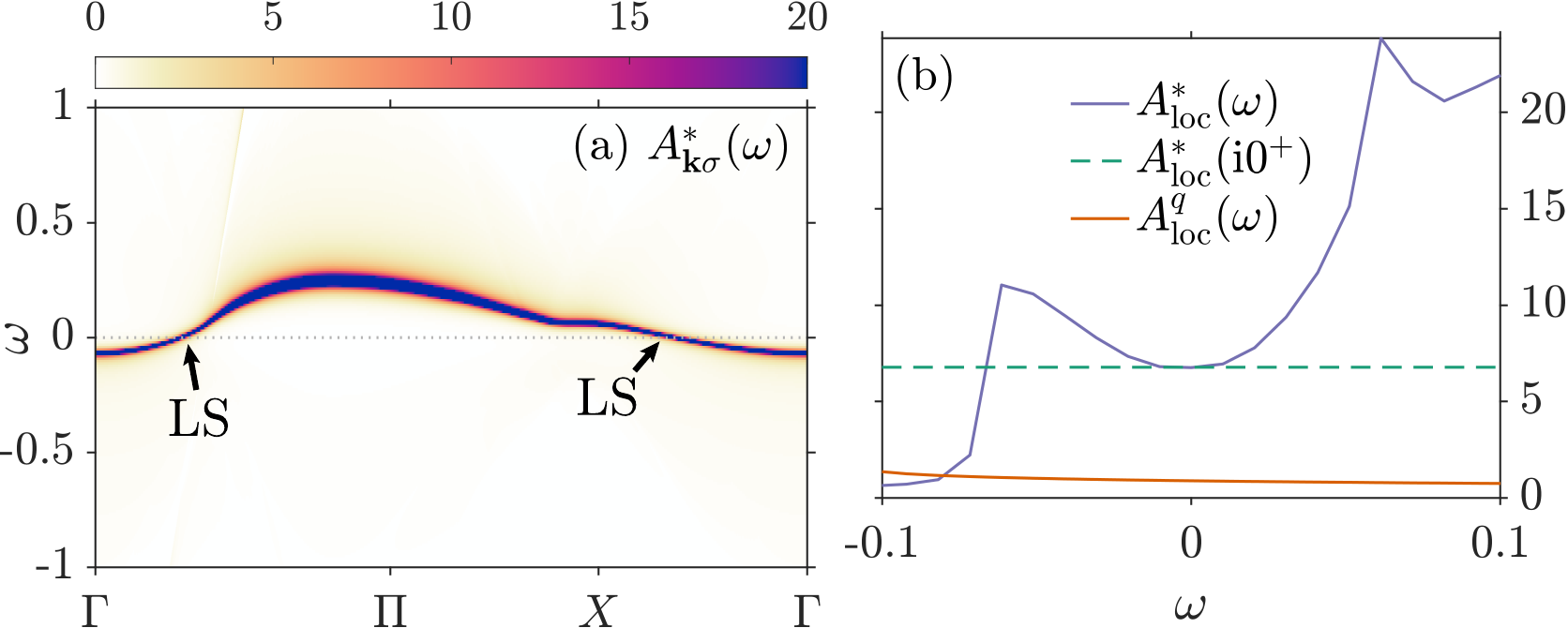}
\caption{(a) $\vec{k}$-resolved QP spectral function corresponding to the spectral function and self-energy shown in Fig.~\ref{fig:Hq_ttp_square}(a,b), and (b) corresponding QP and $q$-particle DOS.
We applied Lorentzian broadening of width $\eta = 5\times 10^{-3}$ in (a) for better 
visualization, and of width $\eta = 10^{-3}$ in (b) to regularize the integrals in Eqs.~\eqref{eq:Aast_loc} and \eqref{eq:Aq_loc}.}
\label{fig:G_SE_Gqp_n05}
\end{figure}

To make this more concrete, we have computed $G_{\vec{k}\sigma}(\omega)$, $\Sigma_{\vec{k}\sigma}(\omega)$, $G^{\ast}_{\vec{k}\sigma}(\omega)$ and the QP-DOS 
\vspace{-3pt}
\begin{align}
\label{eq:Aast_loc}
A^{\ast}_{\mr{loc}}(\omega) = -\tfrac{1}{\pi} \mr{Im} \int_{\vec{k}} \sum_{\sigma} G^{\ast}_{\vec{k}\sigma}(\omega) 
\end{align}
for the square lattice model of the main text, at $n_{\sigma} = 0.6$ [Fig.~\ref{fig:G_SE_Gqp_n06}] and $n_{\sigma} = 0.5$, see Fig.~\ref{fig:Hq_ttp_square} (main text) and Fig.~\ref{fig:G_SE_Gqp_n05}.
We also show the Matsubara result for the QP-DOS at the Fermi level, denoted $A^{\ast}_{\mr{loc}}(\mr{i}0^+)$.

At $n_{\sigma} = 0.6$, both a FS and a LS is present, the former visible as a coherent pole in $A_{\vec{k}\sigma}(\omega) = -\tfrac{1}{\pi} \mr{Im} G_{\vec{k}\sigma}(\omega)$ [see Fig.~\ref{fig:G_SE_Gqp_n06}(a)] crossing the Fermi level, 
and the latter as a coherent pole in $-\mr{Im} \Sigma_{\vec{k}\sigma}(\omega)$ crossing the Fermi level, see Fig.~\ref{fig:G_SE_Gqp_n06}(b).
The resulting QP spectral function $A^{\ast}_{\vec{k}\sigma}(\omega) = -\tfrac{1}{\pi} \mr{Im} G^{\ast}_{\vec{k}\sigma}(\omega)$, shown in Fig.~\ref{fig:G_SE_Gqp_n06}(c), 
exhibits two coherent poles crossing the Fermi level, associated with the Fermi and Luttinger surfaces, respectively.
Figure~\ref{fig:G_SE_Gqp_n06}(d) shows the resulting $A_{\mr{loc}}^{\ast}(\omega)$, together with the $q$-QP DOS
\vspace{-2.5pt}
\begin{align}
\label{eq:Aq_loc}
A^{q}_{\mr{loc}}(\omega) = \int_{\vec{k}} \sum_{\sigma} \delta(\omega - \epsilon_{\vec{k}\sigma}) \, .
\end{align}
Evidently, $A_{\mr{loc}}^{\ast}(0) \neq A^{q}_{\mr{loc}}(0)$, due to the LS contribution.
The specific heat $C^{\ast}_v$ predicted from $G^{\ast}_{\vec{k}\sigma}(\omega)$ in Refs.~\cite{Fabrizio2020,Fabrizio2022,Fabrizio2023}
therefore differs from the exact specific heat $C^{q}_v$, which can be computed form the $q$-QP.
To leading order in $T$, they are given by
\vspace{-3pt}
\begin{align}
C^{q}_v = \frac{\pi^2}{6} T A^{q}_{\mr{loc}}(0) \, , \quad C^{\ast}_v = \frac{\pi^2}{6} T A_{\mr{loc}}^{\ast}(0) \, .
\end{align}

The spectral function and self-energy at $n_{\sigma}$ are shown in Fig.~\ref{fig:Hq_ttp_square} of the main text, and have also been discussed there.
Since the spectral function does not exhibit a QP pole, only the self-energy pole at the LS contributes to $A^{\ast}_{\vec{k}\sigma}(\omega)$, see Fig.~\ref{fig:G_SE_Gqp_n05}(a).
The bandwidth of the self-energy pole and therefore also the QP bandwidth is almost an order of magnitude smaller than the $q$-particle bandwidth $W_q = 4t$ (with $t = 0.5$).
As a result, the $A^{\ast}_{\mr{loc}}(\omega)$ is almost an order of magnitude larger than $A^{q}_{\mr{loc}}(\omega)$, and therefore $C^{\ast}_v/T \gg C^{q}_v/T$.

This suggests that some aspect of the arguments put forward in Refs.~\cite{Fabrizio2020,Fabrizio2022,Fabrizio2023} is incorrect.
The Ward identities used in these arguments should hold in the $q$-particle model.
It is currently not clear to us where exactly the arguments go wrong; a detailed investigation is left to future work.

\vspace{-0.6cm}
\section{Two-impurity Anderson model}
\label{sec:2IAM}
\vspace{-0.3cm}

Impurity models frequently exhibit Fermi liquids featuring a pseudogapped impurity spectral function.
Apart from numerous theoretical accounts~\cite{Jones1987,DeLeo2004,Sakai1992,Zarand2006,Chung2007,Nishikawa2012,Nishikawa2012a,Nishikawa2018,Blesio2018,Zitko2021}, 
there are also experimental realizations~\cite{Minamitani2012,Hiraoka2017,Yang2019,Guo2021} of pseudogapped impurity systems.
An example featuring this phenomenon is the two-impurity Anderson model (2IAM)~\cite{Sakai1992,Zarand2006,Chung2007,Nishikawa2012,Nishikawa2012a,Nishikawa2018},
\vspace{-2pt}
\begin{align}
\nonumber
H_{\mr{2I}} &= K \vec{S}_1 \!\cdot \vec{S}_2 + \! \sum_{k i \sigma} \Bigl[V_{k} \bigl(f^{\dagger}_{i\sigma} a_{k i\sigma} + \mr{h.c.}\bigr) + \epsilon_{k} a^{\dagger}_{k i\sigma} a^{\phantom{\dagger}}_{k i\sigma}\Bigr]  .
\end{align}
Here, $f_{i\sigma}$ annihilates a spin-$\sigma$ electron on impurity site $i \in \{1,2\}$; $n_{i\sigma}$ and $\vec{S}_i$ are the corresponding number and spin operators. 
Both impurity sites hybridize with separate bath channels, with annihilation operator $a^{\dagger}_{k i\sigma}$ describing the corresponding bath mode at energy $\epsilon_k$ and hybridization $V_k$.
We choose the bath parameters such that they give rise to a semi-circular, channel-independent hybridization density of states
\begin{align}
\Gamma(\omega) = \frac{\gamma}{\pi} \sqrt{1 - \omega^2} \, \theta(1-|\omega|) \, .
\end{align}

It is well known that the $K$-tuned phase diagram of this model exhibits two distinct Fermi liquid phases, separated by a quantum critical point~(QCP) at $K_c$. 
For $K < K_c$, the impurity spectral function exhibits a Kondo resonance, while the $K > K_c$ phase features a pseudogapped impurity spectral function. 
The low-energy spectra of both phases exhibit a Fermi liquid form, but differ in their respective scattering phase shifts, as can be verified by NRG.

\begin{figure}
\includegraphics[width=\linewidth]{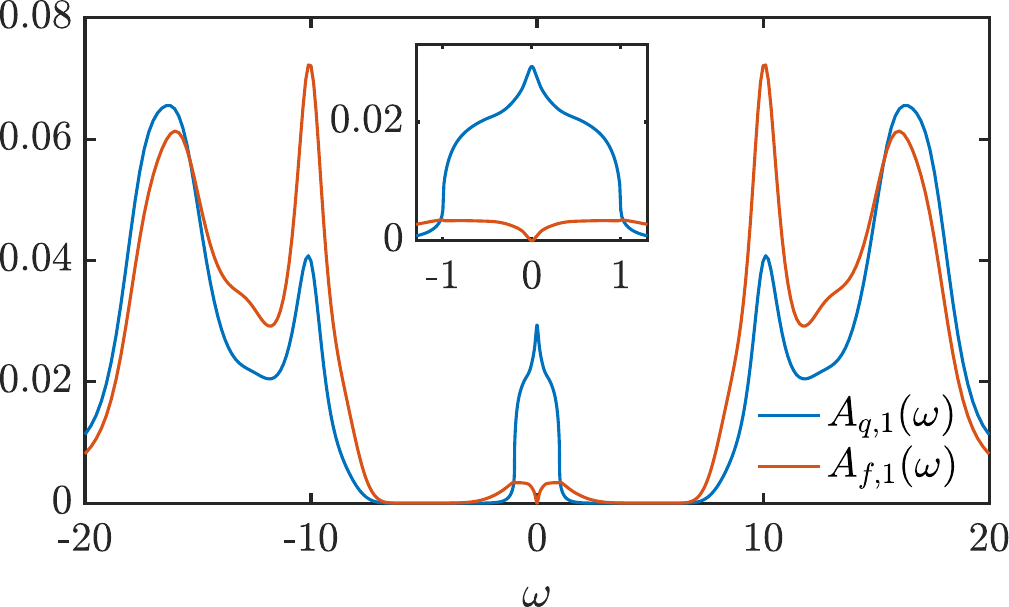}
\caption{Spectral functions $A_{q,1}(\omega)$ and $A_{f1}(\omega)$, corresponding to the fermionic operators $q_{1\sigma} = f_{1\sigma} (2n_{1\bar{\sigma}} - 1)$ and $f_{1\sigma}$, respectively.
The inset shows details of the resonance and PG at the Fermi level.
Parameters are $K = \gamma = 20$.} 
\label{fig:TwoImp_Afq}
\end{figure}

Choosing 
$K = \gamma = 20$, we
have computed the spectral function of both $f_{1\sigma}$ and $q_{1\sigma} = f_{1\sigma}(2 n_{1\bar{\sigma}} - 1)$ in the pseudogapped phase, denoted $A_{f,1}(\omega)$ and $A_{q,1}(\omega)$, respectively.
The single-electron spectral function $A_{f,1}(\omega)$ is pseudogapped, shown in Fig.~\ref{fig:TwoImp_Afq} (red curve).
On the other hand, $A_{q,1}(\omega)$ (blue curve) exhibits a resonance at the Fermi level, consistent with earlier findings~\cite{Zhu2013}.
From the fermionic QP perspective, the PG phase of the 2IAM therefore appears to fit the PG-FL discussed in the main text.
This suggests that this model exhibits emergent CH of the form discussed in the main text. 
A thorough analysis of this, including self-consistent solutions within cluster DMFT~\cite{Gleis2024,Gleis2025}, will be the subject of future work.

\clearpage

\thispagestyle{empty}

\setcounter{equation}{0}
\setcounter{figure}{0}
\setcounter{page}{1}

\renewcommand{\theequation}{S\arabic{equation}}
\renewcommand{\thefigure}{S\arabic{figure}}
\renewcommand{\thepage}{S\arabic{page}}

\setcounter{secnumdepth}{2} 
\renewcommand{\thefigure}{S\arabic{figure}}
\setcounter{figure}{0}
\setcounter{section}{0}
\setcounter{equation}{0}
\renewcommand{\thesection}{S-\Roman{section}}
\renewcommand{\theequation}{S\arabic{equation}}
%

%
\title{Supplemental Material for 
``\maintitle''}

\date{\today}
\maketitle

In this supplemental material, we provide additional information on $q$-particle coherent states and the path integral in Sec.~\ref{sec:coh_PI},
details on the formal calculation of the single-electron Green's function in Sec.~\ref{sec:SP_GF},
details on its numerical evaluation in Sec.~\ref{sec:G3_numerical},
more information on the dynamical quasiparticle weight and quasiparticle propagators in Sec.~\ref{sec:QP_properties},
a discussion of the non-freeness of the PG-FL in Sec.~\ref{sec:nonFreeness},
a proof of the existence of at least one quantum phase transition separating PG-FL from Landau FL for a real, particle-hole symmetric Hamiltonian in Sec.~\ref{sec:ph-trafo},
more information on GDMFT in Sec.~\ref{sec:GDMFT},
and details on impurity model calculations in Sec.~\ref{sec:impurity_model}.

\section{Coherent states and path integral}
\label{sec:coh_PI}

\subsection{Coherent states}

We can write down coherent states for the $q$ operators by introducing Grassmann fields $\chi_{i\sigma}$,
\begin{align}
|\{ \chi_{i\sigma} \} \rangle_q &= \prod_{i\sigma} \mr{e}^{-\chi_{i\sigma} q^{\dagger}_{i\sigma}} |0\rangle
\\ \nonumber
q_{j\tilde{\sigma}} |\{ \chi_{i\sigma} \} \rangle_q &= \chi_{j\tilde{\sigma}} |\{ \chi_{i\sigma} \} \rangle_q \, ,
\end{align}
where the subscript $q$ indicates that it is a right eigenstate of $q$ operators. 
Instead of directly generating the coherent states for the $q$ operators, we can also use coherent states for the single-particle operators $c$ and transform them into the $q$ basis using the 
unitary in Eq.~\eqref{eq:cq_unitary} of the main text,
\begin{align}
|\{ \chi_{i\sigma} \} \rangle_c &= \prod_{i\sigma} \mr{e}^{-\chi_{i\sigma} c^{\dagger}_{i\sigma}} |0\rangle \\
|\{ \chi_{i\sigma} \} \rangle_q &= U |\{ \chi_{i\sigma} \} \rangle_c = \prod_{i\sigma} \mr{e}^{-\chi_{i\sigma} q^{\dagger}_{i\sigma}} |0\rangle \, .
\end{align}

The resolution of the identity has the familiar form
\begin{align}
\mathds{1} = \int \prod_{j,\sigma} d\bar{\chi}_{j\sigma} \, d\chi_{j\sigma} \; 
\mr{e}^{-\sum_{j,\sigma} \bar{\chi}_{j\sigma} \chi_{j\sigma}} 
|\{\chi_{j\sigma}\}\rangle_q {\hspace{-3pt}\phantom{\rangle}_q} \langle \{\chi_{j\sigma}\}| \, .
\end{align}
We use the short notation
\begin{align}
\label{eq:q_identity}
\mathds{1} = \int \mathcal{D}[\bar{\chi}, \chi] \; 
\mr{e}^{-\bar{\chi} \chi} |\chi\rangle_q {\hspace{-3pt}\phantom{\rangle}_q}\langle \chi| \, .
\end{align}

\subsection{Path integral}

We now consider the Hamiltonian
\begin{align}
H &= \sum_{ij\sigma}  t^{c}_{i j\sigma} c^{\dagger}_{i\sigma} c_{j\sigma} - \mu \sum_{i \sigma} n_{i\sigma}
\\ \nonumber
&+ \sum_{i j \sigma} t^{q}_{i j\sigma} (2n_{i\bar{\sigma}}-1) c^{\dagger}_{i\sigma} c_{j\sigma} (2n_{j\bar{\sigma}}-1)
\\ \nonumber
&+ U \sum_i n_{i\uparrow} n_{i\downarrow} + \sum_{i\neq j} V_{ij} n_i n_j \, ,
\\ \nonumber
H &= \sum_{\vec{k}\sigma} \epsilon_{\vec{k}\sigma} q^{\dagger}_{\vec{k}\sigma} q_{\vec{k}\sigma} + \sum_{\vec{k}\sigma}  \tilde{\epsilon}_{\vec{k}\sigma} c^{\dagger}_{\vec{k}\sigma} c_{\vec{k}\sigma} 
\\ \nonumber
&+ U \sum_i n_{i\uparrow} n_{i\downarrow} + \sum_{i\neq j} V_{ij} n_i n_j \, ,
\end{align}
where $n_i = n_{i\uparrow} + n_{i \downarrow}$ is the local density. The local density operator $n_{i\sigma} = c^{\dagger}_{i\sigma} c_{i\sigma} = q^{\dagger}_{i\sigma} q_{i\sigma}$ 
is quadratic in either the $q$ or $c$ operators, see Eq.~\eqref{eq:n_i} of the main text.

By using the resolution of the identity in Eq.~\eqref{eq:q_identity}, we can now represent the corresponding partition function,
\begin{align}
Z = \mr{Tr} \,  \mr{e}^{-\beta H} \, ,
\end{align}
in terms of a path integral over $q$ coherent states. The procedure is essentially identical to the usual path integral representation of
fermionic partition functions using $c$ coherent states, see for example Ref.~\onlinecite{Altland2023}.

The only term that requires some additional attention is the kinetic term $c^{\dagger}_{\vec{k}\sigma} c_{\vec{k}\sigma}$, which we have to normal order.
Using the representation of $c_{\bk\sigma}$ in terms of $q_{\bk\sigma}$, Eq.~\eqref{eq:ck_qk} of the main text, we get
\begin{align}
&c^{\dagger}_{\vec{k}\sigma} c^{\phantom{\dagger}}_{\vec{k}\sigma} 
\\ \nonumber
&= \frac{4}{N^2}\sum_{\vec{k}_1\vec{q}_1} \sum_{\vec{k}_2\vec{q}_2}  
q^{\dagger}_{\vec{k}_1-\vec{q}_1\bar{\sigma}} q_{\vec{k}_1\bar{\sigma}}^{\phantom{\dagger}}  q^{\dagger}_{\vec{k}+\vec{q}_1\sigma} 
q^{\phantom{\dagger}}_{\vec{k}+\vec{q}_2\sigma} q_{\vec{k}_2\bar{\sigma}}^{\dagger} q^{\phantom{\dagger}}_{\vec{k}_2-\vec{q}_2\bar{\sigma}}  
\\ \nonumber
&- \frac{2}{N}\sum_{\vec{k}_1\vec{q}_1} 
q^{\dagger}_{\vec{k}_1-\vec{q}_1\bar{\sigma}} q_{\vec{k}_1\bar{\sigma}}^{\phantom{\dagger}}  q^{\dagger}_{\vec{k}+\vec{q}_1\sigma} q^{\phantom{\dagger}}_{\vec{k}\sigma} 
\\ \nonumber
&- \frac{2}{N} \sum_{\vec{k}_2\vec{q}_2} q^{\dagger}_{\vec{k}\sigma} q^{\phantom{\dagger}}_{\vec{k}+\vec{q}_2\sigma} q_{\vec{k}_2\bar{\sigma}}^{\dagger} q^{\phantom{\dagger}}_{\vec{k}_2-\vec{q}_2\bar{\sigma}}
\\ \nonumber
 &+ q^{\dagger}_{\vec{k}\sigma} q^{\phantom{\dagger}}_{\vec{k}\sigma} \, .
\end{align}
The term with six $q$ operators is not normal-ordered yet. Rearranging it gives
\begin{align}
\phantom{=}& q^{\dagger}_{\vec{k}_1-\vec{q}_1\bar{\sigma}} q_{\vec{k}_1\bar{\sigma}}^{\phantom{\dagger}}  q^{\dagger}_{\vec{k}+\vec{q}_1\sigma} 
q^{\phantom{\dagger}}_{\vec{k}+\vec{q}_2\sigma} q_{\vec{k}_2\bar{\sigma}}^{\dagger} q^{\phantom{\dagger}}_{\vec{k}_2-\vec{q}_2\bar{\sigma}}  
\\ \nonumber
=& q^{\dagger}_{\vec{k}_1-\vec{q}_1\bar{\sigma}}  q^{\dagger}_{\vec{k}+\vec{q}_1\sigma} 
q^{\phantom{\dagger}}_{\vec{k}+\vec{q}_2\sigma} 
q_{\vec{k}_1\bar{\sigma}}^{\phantom{\dagger}} q_{\vec{k}_2\bar{\sigma}}^{\dagger} q^{\phantom{\dagger}}_{\vec{k}_2-\vec{q}_2\bar{\sigma}}  
\\ \nonumber
=& \delta_{\vec{k}_1\vec{k}_2} q^{\dagger}_{\vec{k}_1-\vec{q}_1\bar{\sigma}}  q^{\dagger}_{\vec{k}+\vec{q}_1\sigma} 
q^{\phantom{\dagger}}_{\vec{k}+\vec{q}_2\sigma} 
q^{\phantom{\dagger}}_{\vec{k}_1-\vec{q}_2\bar{\sigma}}  
\\ \nonumber
+& q^{\dagger}_{\vec{k}_1-\vec{q}_1\bar{\sigma}}  q^{\dagger}_{\vec{k}+\vec{q}_1\sigma} 
q_{\vec{k}_2\bar{\sigma}}^{\dagger} q^{\phantom{\dagger}}_{\vec{k}_2-\vec{q}_2\bar{\sigma}}
q^{\phantom{\dagger}}_{\vec{k}+\vec{q}_2\sigma} 
q_{\vec{k}_1\bar{\sigma}}^{\phantom{\dagger}}   \, .
\end{align}
With the definition
\begin{align}
\label{eq:rho_q}
\rho_{\vec{q}\sigma} = \frac{1}{N} \sum_{\vec{k}} q^{\dagger}_{\vec{k}+\vec{q}\sigma} q^{\phantom{\dagger}}_{\vec{k}\sigma} \, ,
\end{align}
we find
\begin{align}
&c^{\dagger}_{\vec{k}\sigma} c^{\phantom{\dagger}}_{\vec{k}\sigma} 
\\ \nonumber
&= \frac{4}{N}\sum_{\vec{k}_1\vec{q}_1} \sum_{\vec{q}_2}  
 q^{\dagger}_{\vec{k}_1-\vec{q}_1\bar{\sigma}}  q^{\dagger}_{\vec{k}+\vec{q}_1\sigma}  \rho_{\vec{q}_2 \bar{\sigma}}
q^{\phantom{\dagger}}_{\vec{k}+\vec{q}_2\sigma} 
q_{\vec{k}_1\bar{\sigma}}^{\phantom{\dagger}}  
\\ \nonumber
&+
\frac{4}{N}\sum_{\vec{q}_1 \vec{q}_2} 
\rho_{\vec{q}_2 - \vec{q}_1 \bar{\sigma}}
 q^{\dagger}_{\vec{k}+\vec{q}_1\sigma} 
q^{\phantom{\dagger}}_{\vec{k}+\vec{q}_2\sigma} 
\\ \nonumber
&- 2\sum_{\vec{q}} q^{\dagger}_{\vec{k}+\vec{q}\sigma} q^{\phantom{\dagger}}_{\vec{k}\sigma} \rho_{-\vec{q}\bar{\sigma}} 
- 2 \sum_{\vec{q}} q^{\dagger}_{\vec{k}\sigma} q^{\phantom{\dagger}}_{\vec{k}+\vec{q}\sigma} \rho_{\vec{q}\bar{\sigma}} 
 + q^{\dagger}_{\vec{k}\sigma} q^{\phantom{\dagger}}_{\vec{k}\sigma} \, .
\end{align}

Using that, the partition function has the form 
\begin{widetext}
\begin{align}
\label{eq:Path_integral_supplement}
Z &= \int \mathcal{D}[\bar{\chi}, \chi] \, \mr{e}^{-S[\bar{\chi},\chi]}
\; , \quad
S[\bar{\chi},\chi] = S_2[\bar{\chi},\chi] + S_4[\bar{\chi},\chi] + S_6[\bar{\chi},\chi]
\\ \nonumber
S_2[\bar{\chi},\chi] &= \int_{0}^{\beta} \mr{d}\tau \, \sum_{\vec{k}\sigma} \bar{\chi}_{\vec{k}\sigma}(\tau) \left( \partial_{\tau} + \epsilon_{\vec{k}\sigma} + \tilde{\epsilon}_{\vec{k}\sigma} \right) \chi_{\vec{k}\sigma}(\tau)
\\ \nonumber
S_4[\bar{\chi},\chi] &= \int_{0}^{\beta} \mr{d}\tau \, U \sum_i n_{i\uparrow}(\tau) n_{i\downarrow}(\tau) 
+ \int_{0}^{\beta} \mr{d}\tau \, \sum_{i \neq j} V_{ij} n_{i}(\tau) n_{j}(\tau)
\\ \nonumber
&+ \int_{0}^{\beta} \mr{d}\tau \, \sum_{\vec{k}\sigma} \tilde{\epsilon}_{\vec{k}\sigma} \bigg\{ \frac{4}{N}\sum_{\vec{q}_1 \vec{q}_2}  \rho_{\vec{q}_2 - \vec{q}_1 \bar{\sigma}}(\tau)
\bar{\chi}_{\vec{k}+\vec{q}_1\sigma}(\tau)
\chi_{\vec{k}+\vec{q}_2\sigma}(\tau)
\\ \nonumber
&- 2\sum_{\vec{q}} \left[ \bar{\chi}_{\vec{k}+\vec{q}\sigma}(\tau) \chi_{\vec{k}\sigma}(\tau) \rho_{-\vec{q}\bar{\sigma}}(\tau) 
+\bar{\chi}_{\vec{k}\sigma}(\tau) \chi_{\vec{k}+\vec{q}\sigma}(\tau) \rho_{\vec{q}\bar{\sigma}}(\tau) 
\right]
\bigg\}
\\ \nonumber
S_6[\bar{\chi},\chi] &= 4 \int_{0}^{\beta} \mr{d}\tau \, \sum_{\vec{k}\sigma} \tilde{\epsilon}_{\vec{k}\sigma} 
\sum_{\vec{q}_1 \vec{q}_2} 
\rho_{-\vec{q}_1\bar{\sigma}}(\tau)
\bar{\chi}_{\vec{k}+\vec{q}_1\sigma}(\tau)  \rho_{\vec{q}_2 \bar{\sigma}}(\tau)
\chi_{\vec{k}+\vec{q}_2\sigma}(\tau) \, .
\end{align}
\end{widetext}

\section{Single-electron Green's function}
\label{sec:SP_GF}

In this section, we derive the single-electron Green's function of the $q$-particle model. 
In this section, we denote analytically continued retarded fermionic propagators involving operators $A$ and $B$ by
\begin{align}
\G{A,B}(z) = \sum_{nm} \frac{\mr{e}^{-\beta E_n} + \mr{e}^{-\beta E_m}}{Z} \frac{\langle n |A | m \rangle \langle m | B | n \rangle}{z - E_n + E_m} \, ,
\end{align}
where $\beta$ is the inverse temperature, $Z$ the partition function, and $|n\rangle$ and $|m\rangle$ are many-body eigenstates with eigenenergies $E_n$ and $E_m$, repsectively.

In a periodic lattice, in any dimension, the $q$-particle propagator is
\begin{align}
\G{q^{\phantom{\dagger}}_{\vec{k}\sigma},q^{\dagger}_{\vec{k}\sigma}}(z) = \frac{1}{z + \mu - \epsilon_{\vec{k}}} \, ,
\end{align}
where $\epsilon_{\vec{k}}$ is the dispersion.
The electron annihilation operator in $\vec{k}$-space is
\begin{align}
c_{\vec{k}\sigma} &= \frac{1}{\sqrt{N}}\sum_{i} \mr{e}^{-\mr{i}\vec{k}\vec{r}_i} c_{i\sigma} = \frac{1}{\sqrt{N}}\sum_{i} \mr{e}^{-\mr{i}\vec{k}\vec{r}_i} q_{i\sigma} (2 q_{i\bar{\sigma}}^{\dagger} q_{i\bar{\sigma}} - 1) 
\nonumber \\
&= \frac{2}{N}\sum_{\vec{k}'\vec{q}}  q^{\phantom{\dagger}}_{\vec{k}+\vec{q}\sigma} q_{\vec{k}'\bar{\sigma}}^{\dagger} q^{\phantom{\dagger}}_{\vec{k}'-\vec{q}\bar{\sigma}}  - q^{\phantom{\dagger}}_{\vec{k}\sigma}
\end{align}
We then have to evaluate the following propagators:
\begin{widetext}
\begin{align}
&\G{q^{\phantom{\dagger}}_{\vec{k}+\vec{q}\sigma} q_{\vec{k}_1\bar{\sigma}}^{\dagger} q^{\phantom{\dagger}}_{\vec{k}_1-\vec{q}\bar{\sigma}},
q^{\dagger}_{\vec{k}_2-\vec{q}\bar{\sigma}} q_{\vec{k}_2\bar{\sigma}}^{\phantom{\dagger}}  q^{\dagger}_{\vec{k}+\vec{q}\sigma}} (z) = 
\frac{\langle\{q^{\phantom{\dagger}}_{\vec{k}+\vec{q}\sigma} q_{\vec{k}_1\bar{\sigma}}^{\dagger} q^{\phantom{\dagger}}_{\vec{k}_1-\vec{q}\bar{\sigma}},
q^{\dagger}_{\vec{k}_2-\vec{q}\bar{\sigma}} q_{\vec{k}_2\bar{\sigma}}^{\phantom{\dagger}}  q^{\dagger}_{\vec{k}+\vec{q}\sigma} \} \rangle}
{z + \mu - \epsilon_{\vec{k}_2 - \vec{q}} + \epsilon_{\vec{k}_2} - \epsilon_{\vec{k}+\vec{q}}} =
\\ \nonumber
&\frac{ \delta_{\vec{k}_1 \vec{k}_2} \left[\langle q_{\vec{k}+\vec{q}\sigma} q_{\vec{k}+\vec{q}\sigma}^{\dagger} \rangle \langle q^{\dagger}_{\vec{k}_2 \bar{\sigma}} q_{\vec{k}_2\bar{\sigma}}\rangle 
+\langle q^{\dagger}_{\vec{k}+\vec{q}\sigma} q_{\vec{k}+\vec{q}\sigma} \rangle \langle q^{\dagger}_{\vec{k}_2 - \vec{q} \bar{\sigma} } q_{\vec{k}_2 - \vec{q}\bar{\sigma}}\rangle
- \langle q^{\dagger}_{\vec{k}_2 - \vec{q}\sigma} q_{\vec{k}_2 - \vec{q}\sigma} \rangle \langle q^{\dagger}_{\vec{k}_2 \bar{\sigma}} q_{\vec{k}_2\bar{\sigma}}\rangle
\right]
}{z + \mu - \epsilon_{\vec{k}_2 - \vec{q}} + \epsilon_{\vec{k}_2} - \epsilon_{\vec{k}+\vec{q}}}
\\ \nonumber
&+ \frac{\delta_{\vec{q}\vec{0}}  \langle q^{\dagger}_{\vec{k}_2 \bar{\sigma}} q_{\vec{k}_2 \bar{\sigma}} \rangle \langle q^{\dagger}_{\vec{k}_1 \bar{\sigma}} q_{\vec{k}_1\bar{\sigma}}\rangle}{z + \mu - \epsilon_{\vec{k}}} 
\\ 
&\G{q_{\vec{k}\sigma} , q^{\dagger}_{\vec{k}_2-\vec{q}\bar{\sigma}} q_{\vec{k}_2\bar{\sigma}} q_{\vec{k}+\vec{q}\sigma}^{\dagger} }(z) = 
\frac{\delta_{\vec{q}\vec{0}}  \langle q^{\dagger}_{\vec{k}_2 \bar{\sigma}} q_{\vec{k}_2 \bar{\sigma}} \rangle}{z + \mu - \epsilon_{\vec{k}}}
\, , \quad 
\G{q^{\phantom{\dagger}}_{\vec{k}+\vec{q}\sigma} q_{\vec{k}_1\bar{\sigma}}^{\dagger} q^{\phantom{\dagger}}_{\vec{k}_1-\vec{q}\bar{\sigma}} ,q_{\vec{k}\sigma}^{\dagger} }(z) = 
\frac{\delta_{\vec{q}\vec{0}}  \langle q^{\dagger}_{\vec{k}_1 \bar{\sigma}} q_{\vec{k}_1 \bar{\sigma}} \rangle}{z + \mu - \epsilon_{\vec{k}}}
\end{align}

Combining everything, we get
\begin{align}
\langle q^{\dagger}_{\vec{k} \sigma} q_{\vec{k} \sigma} \rangle &= \frac{1}{\mr{e}^{\beta (\epsilon_{\vec{k}\sigma} - \mu)} + 1} = f(\epsilon_{\vec{k}\sigma}) \, , \\
\G{c_{\vec{k}\sigma},c^{\dagger}_{\vec{k}\sigma}}(z) &= 
4\int_{\vec{k}'}\int_{\vec{q}}
\frac{\left[1-f(\epsilon_{\vec{k}+\vec{q}\sigma})\right] f(\epsilon_{\vec{k}'\bar{\sigma}}) + f(\epsilon_{\vec{k}+\vec{q}\sigma}) f(\epsilon_{\vec{k}'-\vec{q}\bar{\sigma}}) - f(\epsilon_{\vec{k}'\bar{\sigma}})f(\epsilon_{\vec{k}'-\vec{q}\bar{\sigma}})}{z + \mu - \epsilon_{\vec{k}' - \vec{q}\bar{\sigma}} + \epsilon_{\vec{k}'\bar{\sigma}} - \epsilon_{\vec{k}+\vec{q}\sigma}}
+ \frac{4 n_{\bar{\sigma}}^2 - 4n_{\bar{\sigma}} + 1}{z + \mu - \epsilon_{\vec{k}\sigma}} \, , 
\end{align}
\end{widetext}
where we have used
\begin{align}
n_{\bar{\sigma}} = \int_{\vec{k}} f(\epsilon_{\vec{k}\bar{\sigma}}) \, , \quad \int_{\vec{k}} = \int_{\mr{BZ}} \frac{\mr{d}^{d} k}{\mc{V}_{\mr{BZ}}} \, .
\end{align}


\section{Numerical evaluation of $G^{(3)}_{\vec{k}\sigma}$}
\label{sec:G3_numerical}

In this section, we outline how we evaluate the three-quasiparticle contribution
$G^{(3)}_{\vec{k}\sigma}(z)$ defined in \Eq{eq:Gk} on a large periodic lattice.
A direct evaluation of the double Brillouin-zone integral in \Eq{eq:Gk} is computationally expensive, since it involves two momentum integrations
(for the square lattice in the main text, this is a four-dimensional integral).
Instead, we exploit the convolution structure of the integrand and the analyticity of the retarded Green's function to obtain an efficient formulation based on Fourier transforms~\cite{Rojas1995,Rieger1999,Zlatic2000}.
Finally, we reconstruct the real part from the imaginary part by a Kramers--Kronig (KK) transformation, which substantially improves the numerical stability of
$\re\, G^{(3)}_{\vec{k}\sigma}(\omega)$.

\subsection{Time representation and fast momentum sums}

We evaluate the retarded Green's function at $z=\omega^{+}\equiv \omega+\mi\eta$ with $\eta>0$.
Using the standard integral representation
\begin{align}
\frac{1}{\omega^{+}-E} = -\mi\int_{0}^{\infty}\mr{d}t\;
\mr{e}^{\mi(\omega-E)t}\,\mr{e}^{-\eta t}\,,
\end{align}
we can rewrite \Eq{eq:Gk} as
\begin{align}
\label{eq:G3_time_rep}
G^{(3)}_{\vec{k}\sigma}(\omega^{+})
= -4\mi \int_{0}^{\infty}\mr{d}t\;\mr{e}^{\mi\omega t}\,\mr{e}^{-\eta t}\,
S_{\vec{k}\sigma}(t)\,,
\end{align}
with the time-domain quantity
\begin{align}
\label{eq:Skt_def}
S_{\vec{k}\sigma}(t) \equiv \int_{\vec{k}'}\!\int_{\vec{q}}\;
A^{(3)}_{\vec{k}\vec{k}'\vec{q}\sigma}\;
\mr{e}^{-\mi E^{(3)}_{\vec{k}\vec{k}'\vec{q}\sigma}t}\,.
\end{align}
The advantage of this representation is that $S_{\vec{k}\sigma}(t)$ can be evaluated for \emph{all} $\vec{k}$ at fixed $t$ using only a small number of Fourier transforms.

To make the convolution structure explicit, we introduce the (lattice) Fourier transforms
\begin{align}
F_{\sigma}(\vec{r},t) &\equiv \int_{\vec{k}} \mr{e}^{\mi\vec{k}\cdot\vec{r}}\,
f(\epsilon_{\vec{k}\sigma})\,\mr{e}^{-\mi\epsilon_{\vec{k}\sigma}t}\,,
\\
\bar{F}_{\sigma}(\vec{r},t) &\equiv \int_{\vec{k}} \mr{e}^{\mi\vec{k}\cdot\vec{r}}\,
\bigl[1-f(\epsilon_{\vec{k}\sigma})\bigr]\,\mr{e}^{-\mi\epsilon_{\vec{k}\sigma}t}\,.
\end{align}
These functions are obtained by Fourier transforming momentum-space factors of the form
$f(\epsilon_{\vec{k}\sigma})\mr{e}^{-\mi\epsilon_{\vec{k}\sigma}t}$ and
$[1-f(\epsilon_{\vec{k}\sigma})]\mr{e}^{-\mi\epsilon_{\vec{k}\sigma}t}$.
Using that, the two momentum integrals in \Eq{eq:Skt_def} can be written as products in real space followed by a Fourier transform,
\begin{align}
\label{eq:Skt_realspace}
S_{\vec{k}\sigma}(t)
&= \int_{\vec{r}} \mr{e}^{-\mi\vec{k}\cdot\vec{r}}\,
\Big[
\bar{F}_{\bar{\sigma}}(\vec{r},t)\,F^{\ast}_{\bar{\sigma}}(\vec{r},t)\,\bar{F}_{\sigma}(\vec{r},t)
\\ \nonumber
&+
F_{\bar{\sigma}}(\vec{r},t)\,\bar{F}^{\ast}_{\bar{\sigma}}(\vec{r},t)\,F_{\sigma}(\vec{r},t)
\Big] \, ,
\end{align}
where $F^{\ast}$ is the complex conjugate of $F$. 
Equation~\eqref{eq:Skt_realspace} shows that $S_{\vec{k}\sigma}(t)$ is obtained by
(i) Fourier transforming a small set of momentum-space functions to real space,
(ii) multiplying them pointwise in $\vec{r}$, and (iii) Fourier transforming back to momentum space.
This reduces the computational cost from a direct evaluation of the double momentum integral to a sequence of fast Fourier transforms at each time $t$.
Finally, the frequency dependence of $G^{(3)}_{\vec{k}\sigma}(\omega^{+})$ is obtained from \Eq{eq:G3_time_rep} by a (discrete) Fourier transform in time.

The spectral function is then computed as
\begin{align}
A^{(3)}_{\vec{k}\sigma}(\omega) = -\frac{1}{\pi}\,\im\, G^{(3)}_{\vec{k}\sigma}(\omega^{+})\,.
\end{align}
In practice, we find that the imaginary part, and thus $A^{(3)}_{\vec{k}\sigma}(\omega)$, converges rapidly once the frequency range is sufficiently large to cover the support of the spectrum.

\subsection{Kramers--Kronig reconstruction of $\re\,G^{(3)}$}

While the spectral part $A^{(3)}_{\vec{k}\sigma}(\omega)$ is directly obtained from $\im\,G^{(3)}_{\vec{k}\sigma}(\omega^{+})$,
the real part $\re\,G^{(3)}_{\vec{k}\sigma}(\omega)$ is more sensitive to residual discretization effects in the time-to-frequency transform.
To enforce analyticity and improve numerical stability, we therefore reconstruct $\re\,G^{(3)}_{\vec{k}\sigma}(\omega)$ from the computed imaginary part by the KK relation,
\begin{align}
\label{eq:KK_relation}
\re\,G^{(3)}_{\vec{k}\sigma}(\omega)
=
\frac{1}{\pi}\,\mc{P}\!\!\int_{-\infty}^{\infty}\mr{d}\omega'\;
\frac{\im\,G^{(3)}_{\vec{k}\sigma}(\omega')}{\omega'-\omega}\,,
\end{align}
where $\mc{P}$ denotes the Cauchy principal value.
On a discrete frequency mesh $\{\omega_j\}$, we treat $\im\,G^{(3)}_{\vec{k}\sigma}(\omega)$ as piecewise linear between adjacent mesh points.
With this choice, the principal-value integral can be evaluated analytically on each interval, yielding logarithmic contributions of the generic form
$\ln|(\omega_{j+1}-\omega)/(\omega_j-\omega)|$.

\begin{figure}[t]
\includegraphics[width=\linewidth]{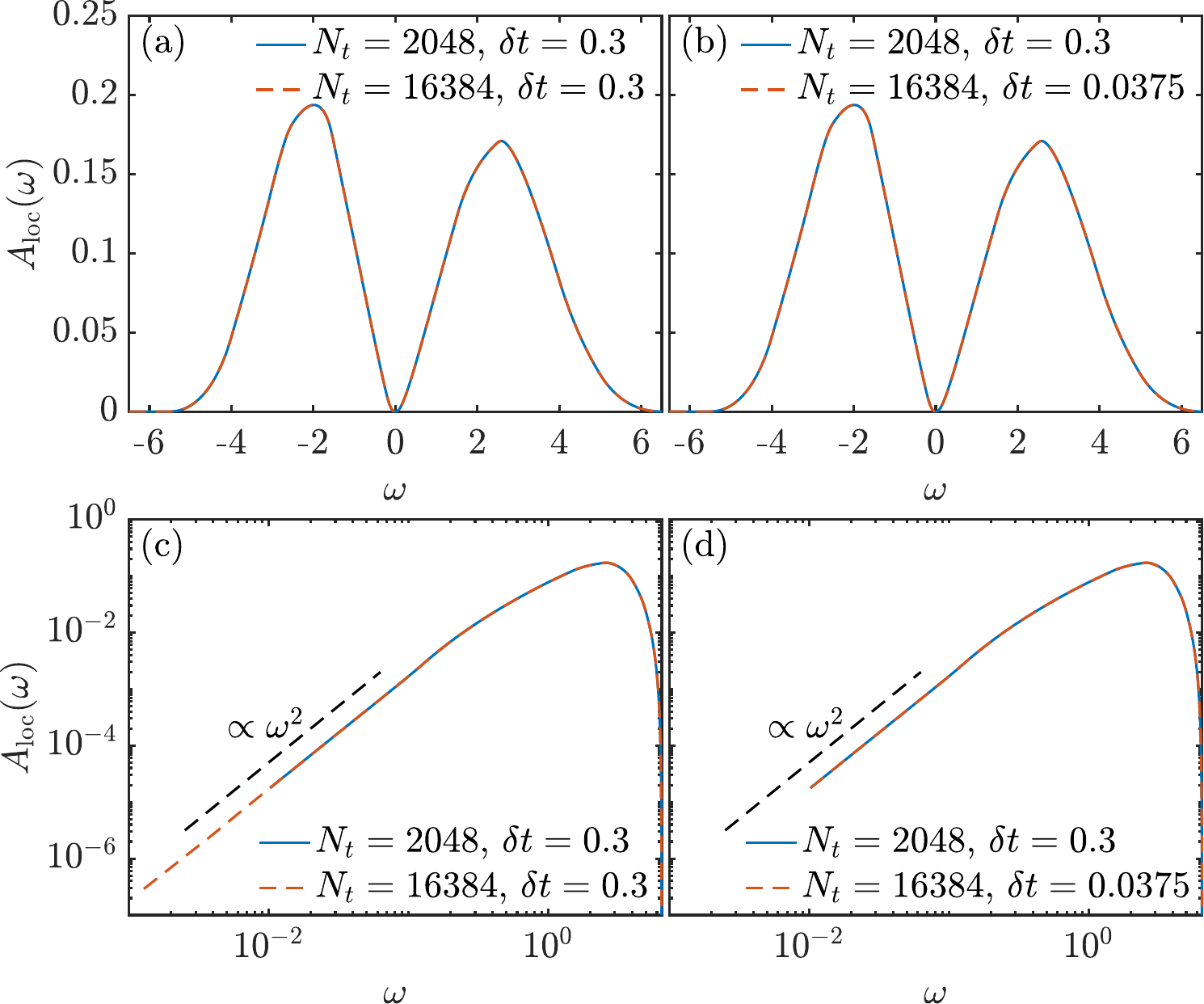}
\caption{Local spectral function of the square lattice model discussed in the main text, obtained with different numbers of time steps $N_t$ and different time step sizes $\delta t$.
We used $(\delta t,N_t) = (0.3,2048)$ for the results in the main text.}
\label{fig:Aloc_FFT_convergence}
\end{figure}

\subsection{Convergence}

We now discuss the convergence of our results with time step size $\delta t$ and the number of time steps $N_t$.
For the results shown in the main text ($L_x \times L_y = 1024\times 1024$), we use $\delta t = 0.15/t = 0.3$ (where $t$ is the nearest-neighbor hopping amplitude) and $N_t = 2048$.
In Fig.~\ref{fig:Aloc_FFT_convergence}(a,c), we compare the local spectral function $A_{\mr{loc}}(\omega)$ obtained with this setting $(\delta t,N_t) = (0.3,2048)$ 
to the $A_{\mr{loc}}(\omega)$ obtained with eight times as many time steps, $(\delta t,N_t) = (0.3,16384)$. 
Since there are essentially no differences between the spectra, $N_t = 2048$ is large enough to obtain reliable results. 
Finally, Figure~\ref{fig:Aloc_FFT_convergence}(b,d) compares $A_{\mr{loc}}(\omega)$ obtained from $(\delta t,N_t) = (0.3,2048)$ to that obtained from $(\delta t,N_t) = (0.0375,16384)$, 
i.e.\ we use an eight times smaller time step while keeping $t_{\mr{max}} = \delta t N_t$ constant in the second setting. Again, there are evidently no differences between the obtained spectra, thus demonstrating convergence in the time step size.


\section{Dynamical quasiparticle weight and quasiparticle propagator}
\label{sec:QP_properties}

In the ``End Matter'', we have discussed the dynamical quasiparticle weight and the quasiparticle propagator as defined in Refs.~\cite{Fabrizio2020,Fabrizio2022,Fabrizio2023}.
This section provides details on the calculations.

\subsection{Low-frequency and temperature behavior of the self-energy at the Luttinger surface}

From an electronic perspective the Hamiltonian $H_q$ can be divided free part $H_0$ and an interaction term $H_{\mr{int}}$,
\begin{subequations}
\begin{align}
H_q &= H_0 + H_{\mr{int}} \, , \;\;
H_0 = \sum_{ij\sigma} (t_{ij\sigma}^{q} - \mu \delta_{ij}) c^{\dag}_{i\sigma} c^{\pdag}_{j\sigma}
\\
H_{\mr{int}} &= \sum_{ij\sigma} t_{ij\sigma}^{q} (4 n_{i\bar{\sigma}} n_{j\bar{\sigma}} - 2 n_{i\bar{\sigma}} - 2 n_{j\bar{\sigma}}) c^{\dag}_{i\sigma} c^{\pdag}_{j\sigma} \, .
\end{align}
\end{subequations}

The interaction gives rise to a self-energy 
\begin{align}
\Sigma_{\vec{k}\sigma}(z) = \Sigma^{\mr{HF}}_{\vec{k}\sigma} + \Sigma^{\mr{dyn}}_{\vec{k}\sigma}(z) \, ,
\end{align}
with a static Hartree-Fock part $\Sigma^{\mr{HF}}_{\vec{k}\sigma}$ and a dynamical part $\Sigma^{\mr{dyn}}_{\vec{k}\sigma}(z)$ which goes as $\sim 1/z$ at large $|z|$.
The Hartree-Fock term is given by a bunch of expectation values,
\begin{subequations}
\begin{align}
\Sigma^{\mr{HF}}_{ij\sigma} &= \langle\{c^{\pdag}_{i\sigma},[H_{\mr{int}},c_{j\sigma}^{\dag}]\}\rangle
\\ \nonumber
&= t_{ij\sigma}^{q} \langle 4 n_{i\bar{\sigma}} n_{j\bar{\sigma}} - 2 n_{i\bar{\sigma}} - 2 n_{j\bar{\sigma}}\rangle
\\ \nonumber
&- t_{ij\bar\sigma}^{q} \langle 4 c^{\dag}_{j\sigma} c^{\pdag}_{i\sigma} c^{\dag}_{i\bar{\sigma}} c^{\pdag}_{j\bar{\sigma}}\rangle 
\\
\label{eq:SigmaHF_q}
&= t_{ij\sigma}^{q} \big[ 4 \langle n_{i\bar{\sigma}} \rangle \langle n_{j\bar{\sigma}} \rangle
- 4 \langle q^{\dag}_{i\bar{\sigma}} q^{\pdag}_{j\bar{\sigma}} \rangle \langle q^{\dag}_{j\bar{\sigma}} q^{\pdag}_{i\bar{\sigma}} \rangle
\\ \nonumber
&- 2 \langle  n_{i\bar{\sigma}} + n_{j\bar{\sigma}} \rangle\big]  - 4 t^{q}_{ij\bar\sigma} \langle q^{\dag}_{i\bar{\sigma}} q^{\pdag}_{j\bar{\sigma}} \rangle \langle q^{\dag}_{j\sigma} q^{\pdag}_{i\sigma} \rangle \, ,
\end{align}
\end{subequations}
where we have used the fact that the density matrix is a Gaussian state in terms of $q$-operators to decouple the quartic expectation values in Eq.~\eqref{eq:SigmaHF_q}.
Using translation symmetry and Fourier transforming $\Sigma_{ij\sigma}^{\mr{HF}}$ gives
\begin{align}
\Sigma_{\vec{k}\sigma}^{\mr{HF}} &= (Z_{\sigma} - 1) (\epsilon_{\vec{k}\sigma} +\mu)
\\ \nonumber
&- 4 \int_{\vec{p}} \int_{\vec{p}'} (\epsilon_{\vec{p}\sigma}  +\mu) f(\epsilon_{\vec{p}'\bar\sigma}) f(\epsilon_{\vec{p}' - \vec{p} + \vec{k}\bar\sigma}) 
\\ \nonumber
&- 4 \int_{\vec{p}} \int_{\vec{p}'} (\epsilon_{\vec{p}\bar\sigma}  +\mu) f(\epsilon_{\vec{p}'\bar\sigma}) f(\epsilon_{\vec{p}' - \vec{p} + \vec{k}\sigma}) \, ,
\end{align}
with $Z_{\sigma} = (2 n_{\bar\sigma} - 1)^2$, and $\mu$ appears because we have defined $\epsilon_{\vec{k}\sigma}$ as the Fourier transform of $t_{ij\sigma}^{q} - \mu \delta_{ij}$.

Without loss of generality, the dynamical part of the self-energy can be written as
\begin{align}
\Sigma^{\mr{dyn}}_{\vec{k}\sigma}(z) = \frac{\widetilde\Delta^2_{\vec{k}\sigma}}{z - \widetilde\epsilon_{\vec{k}\sigma} - \widetilde\Sigma_{\vec{k}\sigma}(z)} \, .
\end{align}
Correspondingly, the electronic Green's function can be written as
\begin{align}
G_{\vec{k}\sigma}(z) &= \frac{1}{z - \epsilon_{\vec{k}\sigma} - \Sigma^{\mr{HF}}_{\vec{k}\sigma} - \Sigma^{\mr{dyn}}_{\vec{k}\sigma}(z)}
\\ \nonumber
&= \frac{z - \widetilde\epsilon_{\vec{k}\sigma} - \widetilde\Sigma_{\vec{k}\sigma}(z)}{(z - \epsilon_{\vec{k}\sigma} - \Sigma^{\mr{HF}}_{\vec{k}\sigma})(z - \widetilde\epsilon_{\vec{k}\sigma} - \widetilde\Sigma_{\vec{k}\sigma}(z)) - \widetilde\Delta^2_{\vec{k}\sigma}} \, .
\end{align}

At a Luttinger surface, $G_{\vec{k}\sigma}(0) = 0$, which means $\widetilde\epsilon_{\vec{k}\sigma} + \widetilde\Sigma_{\vec{k}\sigma}(0) = 0$.
Therefore, in the vicinity of a Luttinger surface, we find
\begin{subequations}
\begin{align}
G_{\vec{k}\sigma}(z) &\simeq \frac{\widetilde{\epsilon}_{\vec{k}\sigma} + \widetilde{\Sigma}_{\vec{k}\sigma}(z) - z}{\widetilde\Delta^2_{\vec{k}\sigma}}  \, ,
\\
A_{\vec{k}\sigma}(\omega) &\simeq -\frac{1}{\pi \widetilde\Delta^2_{\vec{k}\sigma}} \mr{Im} \widetilde{\Sigma}_{\vec{k}\sigma}(\omega) \, .
\end{align}
\end{subequations}
In the close vicinity of a Luttinger surface, 
\begin{align}
-\frac{\mr{Im}\, \widetilde\Sigma_{\vec{k}\sigma}(\omega)}{\pi \widetilde\Delta^2_{\vec{k}\sigma}} = A_{\vec{k}\sigma}(\omega) = A^{(3)}_{\vec{k}\sigma}(\omega) \propto \omega^2 + \pi^2 T^2 \, ,
\end{align}
except in $d = 1$ or if the Fermi level is at a van Hove singularity, c.f.\ Eq.~\eqref{eq:A3_lowfrequency} of the main text.
At low frequencies and $T=0$, we can then expand
\begin{align}
\label{eq:SEtilde_lowfrequency}
\widetilde{\Sigma}_{\vec{k}\sigma}(\omega) = \widetilde{\alpha}_{\vec{k}\sigma} + \widetilde{\beta}_{\vec{k}\sigma} \omega - \mr{i}\widetilde{\gamma}_{\vec{k}\sigma} \omega^2 + \dots \, ,
\end{align}
which leads to the low-frequency form of the self-energy,
\begin{align}
\label{eq:SM_Sigma_lowE}
\Sigma_{\vec{k}\sigma}(\omega) = \Sigma^{\mr{HF}}_{\vec{k}\sigma} + \frac{\Delta^2_{\vec{k}\sigma}}{\omega - \epsilon^{\ast}_{\vec{k}\sigma} + \mr{i} \gamma^{\ast}_{\vec{k}\sigma} \omega^2} \, .
\end{align}
Here, we have defined
\begin{subequations}
\begin{align}
\widetilde Z^{-1}_{\vec{k}\sigma} &= 1 - \left. \frac{\partial \mr{Re} \, \widetilde\Sigma_{\vec{k}\sigma}(\omega)}{\partial \omega} \right|_{\omega = 0} = 1 - \widetilde\beta_{\vec{k}\sigma} \, ,
\\
\Delta^{2}_{\vec{k}\sigma} &= \widetilde Z_{\vec{k}\sigma} \widetilde\Delta^{2}_{\vec{k}\sigma} \, , 
\\
\epsilon^{\ast}_{\vec{k}\sigma} &= \widetilde Z_{\vec{k}\sigma} (\widetilde\epsilon_{\vec{k}\sigma} + \widetilde\Sigma_{\vec{k}\sigma}(0)) = \widetilde Z_{\vec{k}\sigma} (\widetilde\epsilon_{\vec{k}\sigma} + \widetilde\alpha_{\vec{k}\sigma}) \, ,
\\
\gamma^{\ast}_{\vec{k}\sigma} &= \widetilde Z_{\vec{k}\sigma} \widetilde\gamma_{\vec{k}\sigma} \, .
\end{align}
\end{subequations}

A self-energy with a low-frequency form as in Eq.~\eqref{eq:SM_Sigma_lowE} has been studied in Ref.~\cite{Fabrizio2022} on phenomenological grounds.
The low-frequency $\omega^2$ dependence of both $A_{\vec{k}\sigma}(\omega)$ and $\mr{Im} \, \widetilde\Sigma_{\vec{k}\sigma}(\omega)$ at the Luttinger surface
is a direct consequence of the presence of quasiparticles with a sharp Fermi surface (the $q$-QP in our model). 
However, that Fermi surface does not necessarily coincide with the Luttinger surface.

\subsection{Dynamical quasiparticle weight: real-frequency axis}

In Refs.~\cite{Fabrizio2020,Fabrizio2022}, the quasiparticle weight is defined as a dynamical quantity,
\begin{align}
\label{eq:Z_omega}
Z_{\vec{k}\sigma}^{-1}(\omega) &= 1 - \frac{\partial \mr{Re} \, \Sigma_{\vec{k}\sigma}(\omega)}{\partial \omega} = \frac{\partial \, \mr{Re} \, G^{-1}_{\vec{k}\sigma}(\omega)}{\partial \omega} \, .
\end{align}
In the vicinity of a Fermi surface, $Z_{\vec{k}\sigma}(\omega)$ is constant at low frequencies, while at a Luttinger surface, it becomes zero.
For a discussion in terms of a more recent formulation on the Matsubara axis~\cite{Fabrizio2023} is provided in the next section.

To compute the real-frequency dynamical quasiparticle weight close to the Luttinger surface, 
we follow Ref.~\cite{Fabrizio2022} and expand Eq.~\eqref{eq:SM_Sigma_lowE} at low frequencies while retaining the singularity at $\omega = \epsilon^{\ast}_{\vec{k}\sigma}$,
\begin{align}
\Sigma_{\vec{k}\sigma}(\omega) \simeq \Sigma^{\mr{HF}}_{\vec{k}\sigma} + \frac{\Delta^2_{\vec{k}\sigma}}{\omega - \epsilon^{\ast}_{\vec{k}\sigma}} - \mr{i}\frac{\Delta^2_{\vec{k}\sigma} \gamma^{\ast}_{\vec{k}\sigma} \omega^2}{(\omega - \epsilon^{\ast}_{\vec{k}\sigma})^2} \, .
\end{align}
At the Luttinger surface, where $\epsilon^{\ast}_{\vec{k}\sigma} = 0$, the imaginary part is constant to leading order.
The corresponding dynamical quasiparticle weight and quasiparticle propagator in the vicinity of a Luttinger surface is then given by
\begin{align}
\label{eq:SM_dynZ}
&Z^{-1}_{\vec{k}\sigma}(\omega) = \frac{(\omega - \epsilon^{\ast}_{\vec{k}\sigma})^2 + \Delta^2_{\vec{k}\sigma}}{(\omega - \epsilon^{\ast}_{\vec{k}\sigma})^2} \, ,
\\
\label{eq:SM_Gast}
&G^{\ast}_{\vec{k}\sigma}(\omega) =
\\ \nonumber
&\quad \frac{(\omega - \epsilon^{\ast}_{\vec{k}\sigma})^2 + \Delta^2_{\vec{k}\sigma}}{(\omega - E_{\vec{k}\sigma}) (\omega - \epsilon^{\ast}_{\vec{k}\sigma})^2  -  (\omega -\epsilon^{\ast}_{\vec{k}\sigma})\Delta^2_{\vec{k}\sigma} + \mr{i}\gamma_{\vec{k}\sigma} \Delta^2_{\vec{k}\sigma} \omega^2} \, ,
\end{align}
with $E_{\vec{k}\sigma} = \epsilon_{\vec{k}\sigma} + \Sigma^{\mr{HF}}_{\vec{k}\sigma}$. These equations have been derived already in the ``End Matter'' and mirror those presented in Ref.~\cite{Fabrizio2022}.
Close to the Luttinger surface, the corresponding spectral function is 
\begin{align}
A^{\ast}_{\vec{k}\sigma}(\omega) \simeq \delta(\omega - \epsilon_{\vec{k}\sigma}^{\ast}) \, .
\end{align}
The Luttinger surface contribution to the quasiparticle DOS is therefore given by
\begin{align}
\label{eq:LS_QPDOS}
\left. A^{\ast}_{\mr{loc}}(0) \right|_{\rm{LS}} = \int_{\vec{k}} \sum_{\sigma} \delta(\epsilon_{\vec{k}\sigma}^{\ast}) \, .
\end{align}

For the $q$-particle model, the dynamical quasiparticle weight is
\begin{align}
\label{eq:Z_qModel}
Z^{-1}_{\vec{k}\sigma}(\omega) &= \mr{Re} \, \frac{Z_{\sigma} - (\omega - \epsilon_{\vec{k}\sigma})^2 \partial_{\omega} G^{(3)}_{\vec{k}\sigma}(\omega)}{[G^{(3)}_{\vec{k}\sigma}(\omega) (\omega - \epsilon_{\vec{k}\sigma}) + Z_{\sigma}]^2} \, .
\end{align}
Since $G^{(3)}_{\vec{k}\sigma}(\omega)$ is a regular function without singularities, Eq.~\eqref{eq:Z_qModel} can be used everywhere in the Brillouin zone.
For the plots shown in the ``End Matter'', we have omitted the imaginary part of $G^{(3)}_{\vec{k}\sigma}(\omega)$ in Eq.~\eqref{eq:Z_qModel}, which is $\propto \omega^2$ at low frequencies.
This avoids negative dynamical quasiparticle weights at non-zero frequencies but retains the low-frequency singularities in $Z^{-1}_{\vec{k}\sigma}(\omega)$, similar in spirit to Ref.~\cite{Fabrizio2022} and Eq.~\eqref{eq:SM_dynZ}.

When $Z_{\sigma} \neq 0$, we get $\lim_{\omega\to 0} Z_{\vec{k}_{\mr{F}}\sigma}(\omega) = Z_{\sigma}$ at the Fermi surface, as expected.
Then, the quasiparticle propagator close to the Fermi surface is
\begin{align}
G^{\ast}_{\vec{k}\sigma}(\omega) = \frac{G^{(3)}_{\vec{k}\sigma}(\omega) (\omega - \epsilon_{\vec{k}\sigma})/Z_{\sigma} + 1}{\omega - \epsilon_{\vec{k}\sigma}} \simeq \frac{1}{\omega - \epsilon_{\vec{k}\sigma}} \, ,
\end{align}
i.e.\ $G^{\ast}_{\vec{k}_{\mr{F}}\sigma}(\omega)$ is identical to the $q$-particle propagator.

If $Z_{\sigma} = 0$, on the other hand, we get
\begin{align}
Z^{-1}_{\vec{k}\sigma}(\omega) = - \mr{Re} \, \frac{\partial_{\omega} G^{(3)}_{\vec{k}\sigma}(\omega)}{[G^{(3)}_{\vec{k}\sigma}(\omega)]^2} \, ,
\end{align}
which is generically regular at $\vec{k}_{\mr{F}}$, with a well-defined $\omega \to 0$ limit. 
As such, at $\vec{k}_{\mr{F}}$, the $Z_{\sigma} \to 0$ and $\omega \to 0$ limits of $Z^{-1}_{\vec{k}_{\mr{F}}\sigma}(\omega)$ do not commute,
\begin{align}
\label{eq:Zw_limits}
\lim_{Z_{\sigma} \to 0} \lim_{\omega \to 0} Z^{-1}_{\vec{k}_{\mr{F}}\sigma}(\omega) \neq \lim_{\omega \to 0} \lim_{Z_{\sigma} \to 0} Z^{-1}_{\vec{k}_{\mr{F}}\sigma}(\omega) \, .
\end{align}
Note that changing $Z_{\sigma}$ means changing the filling.

\subsection{Dynamical quasiparticle weight: Matsubara axis}

Following Ref.~\cite{Fabrizio2023}, we can also define a dynamical quasiparticle weight on the Matsubara axis,
\begin{align}
\label{eq:Z_Matsubara}
Z_{\vec{k}\sigma}^{-1}(\mr{i}\omega) &= 1 - \frac{\Sigma_{\vec{k}\sigma}(\mr{i}\omega) - \Sigma_{\vec{k}\sigma}(-\mr{i}\omega)}{2\mr{i}\omega} 
\\ \nonumber
&= \frac{G^{-1}_{\vec{k}\sigma}(\mr{i}\omega) - G^{-1}_{\vec{k}\sigma}(-\mr{i}\omega)}{2\mr{i}\omega} \, ,
\end{align}
which is positive for all Matsubara frequencies $\omega$. 

Using
\begin{align}
G_{\vec{k}\sigma}(\mr{i}\omega) &= G^{(3)}_{\vec{k}\sigma}(\mr{i}\omega) + \frac{Z_{\sigma}}{\mr{i}\omega - \epsilon_{\vec{k}\sigma}} 
\\ \nonumber
&= \frac{G^{(3)}_{\vec{k}\sigma}(\mr{i}\omega)(\mr{i}\omega - \epsilon_{\vec{k}\sigma}) + Z_{\sigma}}{\mr{i}\omega - \epsilon_{\vec{k}\sigma}} \, ,
\end{align}
we find
\begin{align}
Z^{-1}_{\vec{k}\sigma}(\mr{i}\omega) = \frac{Z_{\sigma} - |\mr{i}\omega - \epsilon_{\vec{k}\sigma}|^2 \frac{G^{(3)}_{\vec{k}\sigma}(\mr{i}\omega) - G^{(3)}_{\vec{k}\sigma}(-\mr{i}\omega)}{2 \mr{i}\omega}}{|G^{(3)}_{\vec{k}\sigma}(\mr{i}\omega)(\mr{i}\omega - \epsilon_{\vec{k}\sigma}) + Z_{\sigma}|^2} \, .
\end{align}
The dynamical quasiparticle weight on the Matsubara axis is real and positive, since
\begin{align}
\frac{G^{(3)}_{\vec{k}\sigma}(\mr{i}\omega) - G^{(3)}_{\vec{k}\sigma}(-\mr{i}\omega)}{2 \mr{i}\omega}
\end{align} 
is real and negative. If $Z_{\sigma} = 0$, we find
\begin{align}
Z^{-1}_{\vec{k}\sigma}(\mr{i}\omega) = -\frac{G^{(3)}_{\vec{k}\sigma}(\mr{i}\omega) - G^{(3)}_{\vec{k}\sigma}(-\mr{i}\omega)}{2 \mr{i}\omega |G^{(3)}_{\vec{k}\sigma}(\mr{i}\omega)|^2} \, .
\end{align}
Therefore, as was the case on the real-frequency axis [c.f.\ Eq.~\eqref{eq:Zw_limits}], the $Z_{\sigma} \to 0$ and $\omega \to 0$ limits of $Z_{\vec{k}_{\mr{F}}\sigma}(\mr{i}\omega)$ at the Fermi surface do not commute.

The Matsubara axis QP propagator is given by
\begin{subequations}
\begin{align}
G^{\ast}_{\vec{k}\sigma}(\mr{i}\omega) &= Z^{-1}_{\vec{k}\sigma}(\mr{i}\omega) G_{\vec{k}\sigma}(\mr{i}\omega) = \frac{1}{\mr{i}\omega - E^{\ast}_{\vec{k}\sigma}(\mr{i}\omega)} \, ,
\\
E^{\ast}_{\vec{k}\sigma}(\mr{i}\omega) &= \mr{i}\omega \frac{G_{\vec{k}\sigma}(\mr{i}\omega) + G_{\vec{k}\sigma}(-\mr{i}\omega)}{G_{\vec{k}\sigma}(\mr{i}\omega) - G_{\vec{k}\sigma}(-\mr{i}\omega)} 
\\ \nonumber
&= \mr{i}\omega \frac{[G^{(3)}_{\vec{k}\sigma}(\mr{i}\omega) \! +\! G^{(3)}_{\vec{k}\sigma}(-\mr{i}\omega)] |\mr{i}\omega \!-\! \epsilon_{\vec{k}\sigma}|^2 - 2 \epsilon_{\vec{k}\sigma} Z_{\sigma} }{[G^{(3)}_{\vec{k}\sigma}(\mr{i}\omega) \!-\! G^{(3)}_{\vec{k}\sigma}(-\mr{i}\omega)] |\mr{i}\omega \!-\! \epsilon_{\vec{k}\sigma}|^2 - 2 \mr{i}\omega Z_{\sigma} } \, .
\end{align}
\end{subequations}
The corresponding quasiparticle dispersion is 
\begin{align}
E^{\ast}_{\vec{k}\sigma} = \lim_{\omega \to 0} E^{\ast}_{\vec{k}\sigma}(\mr{i}\omega) \, .
\end{align} 

For momenta close to the FS and if $Z_{\sigma} \neq 0$, we find 
\begin{align}
E^{\ast}_{\vec{k}\sigma} = \lim_{\omega \to 0} E^{\ast}_{\vec{k}\sigma}(\mr{i}\omega) \simeq \epsilon_{\vec{k}\sigma} \, .
\end{align}
As expected, the QP propagator close to the FS is
\begin{align}
G^{\ast}_{\vec{k}\sigma}(\mr{i}\omega) \simeq \frac{1}{\mr{i}\omega - \epsilon_{\vec{k}\sigma}} \, ,
\end{align}
provided that $Z_{\sigma} \neq 0$.

Momenta $\vec{k}_{\mr{L}}$ at the Luttinger surface are given by
\begin{subequations}
\begin{align}
G^{(3)}_{\vec{k}_{\mr{L}}\sigma}(0) \epsilon_{\vec{k}_{\mr{L}}\sigma} &= Z_{\sigma} \, , \quad &\mr{if} \; Z_{\sigma} &\neq 0 \, ,
\\
G^{(3)}_{\vec{k}_{\mr{L}}\sigma}(0) &= 0 \, , \quad &\mr{if} \; Z_{\sigma} &= 0 \, .
\end{align}
\end{subequations}
Close to Luttinger surfaces, we find
\begin{subequations}
\begin{align}
E^{\ast}_{\vec{k}\sigma} &= \frac{\epsilon_{\vec{k}\sigma}(G^{(3)}_{\vec{k}\sigma}(0) \epsilon_{\vec{k}\sigma} - Z_{\sigma})}{\epsilon_{\vec{k}\sigma}^2 {\displaystyle \lim_{\omega \to 0}} \frac{G^{(3)}_{\vec{k}\sigma}(\mr{i}\omega) -G^{(3)}_{\vec{k}\sigma}(-\mr{i}\omega)}{2\mr{i}\omega} - Z_{\sigma}} \, , &\mr{if} \; Z_{\sigma} &\neq 0
\\
E^{\ast}_{\vec{k}\sigma} &= \frac{G^{(3)}_{\vec{k}\sigma}(0)}{{\displaystyle \lim_{\omega \to 0}} \frac{G^{(3)}_{\vec{k}\sigma}(\mr{i}\omega) -G^{(3)}_{\vec{k}\sigma}(-\mr{i}\omega)}{2\mr{i}\omega} } \, , &\mr{if} \; Z_{\sigma} &= 0 \, .
\end{align}
\end{subequations}

Inserting the low-frequency form Eq.~\eqref{eq:SM_Sigma_lowE} for the self-energy close to a Luttinger surface, we get
\begin{align}
Z^{-1}_{\vec{k}\sigma}(\mr{i}\omega) = \frac{(\mr{i}\omega - \epsilon^{\ast}_{\vec{k}\sigma}) (\mr{i}\omega + \epsilon^{\ast}_{\vec{k}\sigma}) - \Delta^2_{\vec{k}\sigma}}{(\mr{i}\omega - \epsilon^{\ast}_{\vec{k}\sigma}) (\mr{i}\omega + \epsilon^{\ast}_{\vec{k}\sigma})} \, ,
\end{align}
with the corresponding QP propagator
\begin{align}
&G^{\ast}_{\vec{k}\sigma}(\mr{i}\omega) = 
\\ \nonumber
&\qquad \frac{(\mr{i}\omega - \epsilon^{\ast}_{\vec{k}\sigma}) (\mr{i}\omega + \epsilon^{\ast}_{\vec{k}\sigma}) - \Delta^2_{\vec{k}\sigma}}{(\mr{i}\omega - E_{\vec{k}\sigma})(\mr{i}\omega - \epsilon^{\ast}_{\vec{k}\sigma}) (\mr{i}\omega + \epsilon^{\ast}_{\vec{k}\sigma}) - (\mr{i}\omega + \epsilon^{\ast}_{\vec{k}\sigma}) \Delta^{2}_{\vec{k}\sigma}} \, .
\end{align}
Since $\epsilon^{\ast}_{\vec{k}_{\mr{L}}\sigma} = 0$, this becomes
\begin{align}
G^{\ast}_{\vec{k}\sigma}(\mr{i}\omega) \simeq \frac{1}{\mr{i}\omega + \epsilon^{\ast}_{\vec{k}\sigma}} \, .
\end{align}
in close vicinity to the Luttinger surface. The resulting QP-DOS is consistent with the previously obtained real-frequency result.

In the ``End Matter'', we show the Matsubara axis result for the QP-DOS at the Fermi level,
\begin{align}
A^{\ast}_{\mr{loc}}(\mr{i} 0^{+}) = -\tfrac{1}{\pi} \mr{Im} \int_{\vec{k}}\sum_{\sigma} \lim_{\omega \to 0^{+}}G^{\ast}_{\vec{k}\sigma}(\mr{i}\omega) \, .
\end{align}
For the figures shown in the ``End Matter'', we use $\lim_{\omega \to 0^{+}}G^{\ast}_{\vec{k}\sigma}(\mr{i}\omega) \simeq G^{\ast}_{\vec{k}\sigma}(\mr{i}\times10^{-3})$.

\section{Non-freeness of the PG-FL}
\label{sec:nonFreeness}

The distance of a density matrix $\rho$ from its closest electronic Gaussian state can be quantified using the non-freeness of $\rho$~\cite{Gottlieb2005,Gottlieb2014,AlivertiPiuri2024}, defined as 
\begin{align}
\mc{N}(\rho) = \sum_{\sigma} \left[ S(\gamma^{\sigma}) + S(\mathds{1} - \gamma^{\sigma}) \right] - S(\rho) \, .
\end{align}
Here, $\gamma$ is the one-electron reduced density matrix, 
\begin{align}
\gamma^{\sigma}_{ij} = \mathrm{Tr} \bigl[ \rho \, c^{\dagger}_{i\sigma} c^{\phantom{\dagger}}_{j\sigma} \bigr] \, ,
\end{align}
and $S$ is the von Neumann entropy. 
The nonfreeness $\mc{N}(\rho)$ per spin and lattice site is some number between 0 ($\rho$ is Gaussian) and $\ln 2$, when $S(\rho) = 0$ and all eigenvalues of $\gamma^{\sigma}$ are $1/2$.
In the latter case, $\rho$ is strongly non-Gaussian, i.e.\ tracing out all but one electron leads to a maximally mixed density matrix.

\begin{figure}[t]
\includegraphics[width=\linewidth]{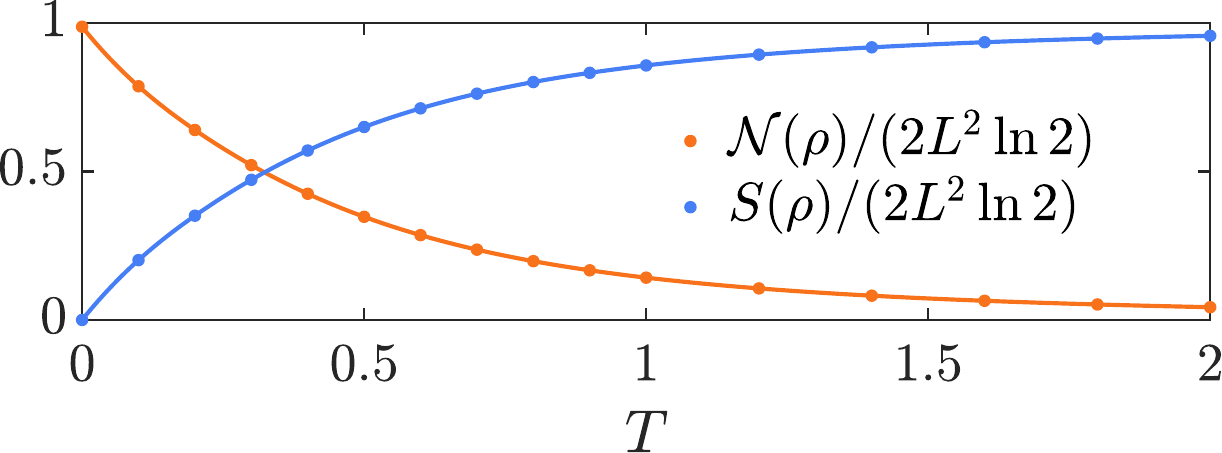}
\caption{Nonfreeness $\mc{N}(\rho)$ and entropy $S(\rho)$ versus $T$ for the square lattice model considered in the main text.}
\label{fig:nonfreeness}
\end{figure}

In Fig.~\ref{fig:nonfreeness}, we show the non-freeness per spin and lattice site for the $t$-$t'$ model discussed in the main text.
At $T=0$, we find a non-freeness per site and spin of $ 0.988 \, \ln 2$, i.e.\ it almost saturates the upper bound, which shows that the ground state is far from an electronic Slater determinant. 
As expected and illustrated in Fig.~\ref{fig:nonfreeness}, the nonfreeness decreases with increasing temperature.

\section{Relation to fermionic Hubbard operators}
\label{sec:Relation_to_HubbardX}

The fermionic Hubbard operators are the most basic local non-canonical operators,
\begin{align}
\label{eq:Hubbard_X}
X^{0\sigma} = |0\rangle \langle \sigma | \, , \quad X^{\sigma 2} = -\sigma |\sigma\rangle \langle 2 | \, ,
\end{align}
where $|0\rangle$, $|\sigma\rangle$, $|2\rangle$ denote empty, singly occupied and doubly occupied states, respectively.
Our definition of $X^{\sigma 2}$ differs by a sign $-\sigma$ ($\sigma = 1$ for $\uparrow$ and $-1$ for $\downarrow$) from the usual convention, which simplifies 
the relation between Hubbard operators and canonical fermionic operators. 
In Eq.~\eqref{eq:Hubbard_X}, we have written all four fermionic Hubbard operators that annihilate electrons. 
The single-particle operators $c_{\sigma}$ can be expressed in terms of fermionic $X$ operators,
\begin{align}
c_{\sigma} = X^{0\sigma} + X^{\bar{\sigma} 2} \, .
\end{align}
The operators $c_{\sigma}$ obey canonical fermionic commutation relations. 
Another obvious choice for canonical fermions are the $q$-operators discussed in the main text,
\begin{align}
q_{\sigma} = X^{0\sigma} - X^{\bar{\sigma} 2} = c_{\sigma}(2n_{\bar{\sigma}}-1) \, .
\end{align}
Of course, any other phase difference between the Hubbard operators would have also provided canonical fermionic operators that are different from 
the single-particle operators.
Similarly, fermionic Hubbard operators can be written in terms of $c_{\sigma}$ and $q_{\sigma}$,
\begin{align}
X^{0\sigma} &= \frac{1}{2}(c_{\sigma} + q_{\sigma}) = c_{\sigma} n_{\bar{\sigma}}
\\
X^{\bar{\sigma} 2} &= \frac{1}{2}(c_{\sigma} - q_{\sigma}) = c_{\sigma} (1- n_{\bar{\sigma}}) \, .
\end{align}

\section{Quantum phase transition: Real-valued, particle-hole symmetric Hamiltonian}
\label{sec:ph-trafo}

In this section, we show that there must be a quantum phase transition~(QPT) between a Landau FL and 
a PG-FL if the Hamiltonian is constrained to be real-valued and particle-hole symmetric.

For that, we consider a spinful one-band model on a bipartite lattice.
The hermitian unitary
\begin{align}
\mc{U}_{\mr{ph}} = \prod_{i\sigma} s_i^{n_{i\sigma}} (c^{\dagger}_{i\sigma} + c^{\phantom{\dagger}}_{i\sigma})
\end{align}
generates particle-hole transformations, with $s_i = 1$ on sublattice A and $s_i = -1$ on sublattice B. 
Since $\mc{U}_{\mr{ph}}$ is hermitian, its eigenvalues are $\pm 1$, i.e.\ particle-hole symmetric states 
can be classified by their particle-hole parity. 

Under a particle-hole transformation, the electron operators transform as
\begin{align}
 c^{\phantom{\dagger}}_{i\sigma} &\to \mc{U}_{\mr{ph}} c^{\phantom{\dagger}}_{i\sigma}\mc{U}_{\mr{ph}} = s_i c^{\dagger}_{i\sigma} \, , 
\\
 c^{\phantom{\dagger}}_{\vec{k}\sigma} &\to  \mc{U}_{\mr{ph}} c^{\phantom{\dagger}}_{\bk\sigma}\mc{U}_{\mr{ph}}  = c^{\dagger}_{\bk + \vec{Q}\sigma} \, , 
\end{align}
where $\vec{Q}$ is the ``antiferromagnetic'' wavevector, satisfying $\vec{Q} \cdot \vec{a}_i = \pm \pi$ for every primitive lattice vector $\vec{a}_i$.
The $q$ operators operators pick up an additional sign due to $(2n_{i\bar{\sigma}}-1)$,
\begin{align}
 q^{\phantom{\dagger}}_{i\sigma} &\to \mc{U}_{\mr{ph}} q^{\phantom{\dagger}}_{i\sigma}\mc{U}_{\mr{ph}} = -s_i q^{\dagger}_{i\sigma} \, , 
\\
 q^{\phantom{\dagger}}_{\vec{k}\sigma} &\to  \mc{U}_{\mr{ph}} q^{\phantom{\dagger}}_{\bk\sigma}\mc{U}_{\mr{ph}}  = -q^{\dagger}_{\bk + \vec{Q}\sigma} \, .
\end{align}

We now assume that the Hamiltonian $H$ is real and particle-hole symmetric, and none of its symmetries are spontaneously broken, i.e.\ we have a unique, symmetric ground state $\ket{\mr{GS}}$.
The ground state is therefore an eigenstate of $\mc{U}_{\mr{ph}}$. 
We further assume that we have a Fermi liquid. If $\vec{k}_F$ is a Fermi wavevector, $\overline{\vec{k}}_F = \vec{k}_F+\vec{Q}$ is also a Fermi wavevector due to particle-hole symmetry.

Consider a quasiparticle excitation $|\Psi_{\vec{k}_F^{+}\sigma}\rangle$ at momentum $\vec{k}_F^{+} = \vec{k}_F + 0^{+}\times \vec{k}_{F}^{\perp}$. Here, $\vec{k}_F$ is a Fermi momentum and $\vec{k}_{F}^{\perp}$ is normal
to the Fermi surface pointing towards its ``outside'' where particle excitations are located.
The corresponding quasiparticle weights are
\begin{subequations}
\label{eq:Z_matrix_elements}
\begin{align}
\alpha_{\vec{k}^{+}_{F}\sigma} &= \langle \Psi_{\vec{k}^{+}_F\sigma} | c^{\dagger}_{\vec{k}^{+}_F\sigma} | \mr{GS} \rangle \, , \quad Z^{c}_{\vec{k}^{+}_F\sigma} = |\alpha_{\vec{k}^{+}_{F}\sigma}|^2
\\
\beta_{\vec{k}^{+}_{F}\sigma} &= \langle \Psi_{\vec{k}^{+}_F\sigma} | q^{\dagger}_{\vec{k}^{+}_F\sigma} | \mr{GS} \rangle \, , \quad Z^{q}_{\vec{k}^{+}_F\sigma} = |\beta_{\vec{k}^{+}_{F}\sigma}|^2 \, .
\end{align}
For quasiholes $|\Psi_{\vec{k}_F^{-}\sigma}\rangle$, we similarly get
\begin{align}
\alpha_{\vec{k}^{-}_{F}\sigma} &= \langle \Psi_{\vec{k}^{-}_F\sigma} | c^{\phantom{\dagger}}_{\vec{k}^{-}_F\sigma} | \mr{GS} \rangle \, , \quad Z^{c}_{\vec{k}^{-}_F\sigma} = |\alpha_{\vec{k}^{-}_{F}\sigma}|^2
\\
\beta_{\vec{k}^{-}_{F}\sigma} &= \langle \Psi_{\vec{k}^{-}_F\sigma} | q^{\phantom{\dagger}}_{\vec{k}^{-}_F\sigma} | \mr{GS} \rangle \, , \quad Z^{q}_{\vec{k}^{-}_F\sigma} = |\beta_{\vec{k}^{-}_{F}\sigma}|^2 \, .
\end{align}
\end{subequations}
Whether we approach the Fermi surface from the outside or inside should not matter, therefore
\begin{align}
\alpha_{\vec{k}_F} = \alpha_{\vec{k}^{-}_{F}\sigma} = \alpha_{\vec{k}^{+}_{F}\sigma} \, , \quad \beta_{\vec{k}_F} = \beta_{\vec{k}^{-}_{F}\sigma} = \beta_{\vec{k}^{+}_{F}\sigma} \, ,
\end{align}
and similarly for $Z^c_{\vec{k}_F}$ and $Z^q_{\vec{k}_F}$. From here on, we write $\vec{k}_F$ whenever the direction of the limit does not matter. 

Since the Hamiltonian is particle-hole symmetric, we can relate quasiparticle excitations at $\vec{k}_F$ to quasihole excitations at $\overline{\vec{k}}_F = \vec{k}_F+\vec{Q}$,
\begin{align}
|\Psi_{\overline{\vec{k}}_F^-\sigma}\rangle = \mc{U}_{\mr{ph}} |\Psi_{\vec{k}_F^+\sigma}\rangle \, ,
\end{align}
and vice versa.
As a result,
\begin{align}
\label{eq:kF_kFbar_relation}
\alpha_{\overline{\vec{k}}_{F}\sigma} = \alpha_{\vec{k}_{F}\sigma} \, , \quad \beta_{\overline{\vec{k}}_{F}\sigma} = \beta_{\vec{k}_{F}\sigma} \, ,
\end{align}
and similar for the quasiparticle weights. 

We now define the operator
\begin{align}
O_{\vec{k}_F\sigma} &= c_{\vec{k}_F^+\sigma} |\Psi_{\vec{k}_F^+\sigma}\rangle \langle\Psi_{\vec{k}_F^+\sigma}| q^{\dagger}_{\vec{k}_F^+\sigma} 
\\ \nonumber
&+ c^{\dagger}_{\overline{\vec{k}}_F^-\sigma} |\Psi_{\overline{\vec{k}}_F^-\sigma}\rangle \langle\Psi_{\overline{\vec{k}}_F^-\sigma}| q^{\phantom{\dagger}}_{\overline{\vec{k}}_F^-\sigma} \, .
\end{align}
This operator is odd under particle-hole transformations,
\begin{align}
\mc{U}_{\mr{ph}} O_{\vec{k}_F\sigma} \mc{U}_{\mr{ph}} &= - O_{\vec{k}_F\sigma} \, .
\end{align}
Its expectation value is
\begin{align}
\bra{\mr{GS}} O_{\vec{k}_F\sigma} \ket{\mr{GS}} &= \alpha_{\vec{k}_{F}\sigma} \beta_{\vec{k}_{F}\sigma} + \beta_{\overline{\vec{k}}_{F}\sigma} \alpha_{\overline{\vec{k}}_{F}\sigma}
\\ \nonumber
&= 2 \alpha_{\vec{k}_{F}\sigma} \beta_{\vec{k}_{F}\sigma} \, ,
\end{align}
where we have used that $\alpha_{\vec{k}_{F}\sigma}$ and $\beta_{\vec{k}_{F}\sigma}$ are real.
However, since $O_{\vec{k}_F\sigma}$ is odd under particle hole transformations, $O_{\vec{k}_F\sigma} \ket{\mr{GS}}$ and $\ket{\mr{GS}}$ have opposite particle-hole parities,
which is why
\begin{align}
\bra{\mr{GS}} O_{\vec{k}_F\sigma} \ket{\mr{GS}} &= 0 \, , \quad \alpha_{\vec{k}_{F}\sigma} \beta_{\vec{k}_{F}\sigma} = 0 \, .
\end{align}
Therefore, 
\begin{align}
Z^{c}_{\vec{k}_F\sigma}Z^{q}_{\vec{k}_F\sigma} = 0 \, ,
\end{align}
i.e.\ $Z^{c}_{\vec{k}_F\sigma}$ and $Z^{q}_{\vec{k}_F\sigma}$ cannot be simultaneously non-zero in the absence of symmetry breaking if the Hamiltonian is particle-hole symmetric and real. 
Therefore, there must be at least one QPT separating the PG-FL and the Landau FL under these constraints.

In the derivation above, we have assumed a periodic lattice, which does not exist, for instance, in the impurity model in the main text.
This ingredient is not essential and just facilitates talking about and relating low-energy quasiparticle and quasihole excitations.
The essential ingredients are that (i) the matrix elements in Eq.~\eqref{eq:Z_matrix_elements} are real and 
(ii) those matrix elements are related at different Fermi points by Eq.~\eqref{eq:kF_kFbar_relation} due to particle-hole symmetry.

\section{Generalized dynamical mean-field theory}
\label{sec:GDMFT}

We resort to the generalized dynamical mean-field theory~(GDMFT) of Ref.~\onlinecite{Stanescu2004} to obtain solutions in $d\to \infty$ with $c$ and $q$ hopping, and a local Hubbard interaction.
GDMFT provides a framework to deal with arbitrary local Hamiltonians, and hopping terms that can be written as bilinears of Hubbard operators at different sites. 
Here, we are interested in the case where these hopping terms only involve fermionic Hubbard operators, 
\begin{align}
\label{eq:GDMFT_impurity_model}
H &= H_0 + H_1 \, , \quad H_0 = \sum_i H_{\mr{loc},i} \, ,
\\
H_1 &= 
\sum_{i\neq j}\sum_{\sigma}
\begin{pmatrix}
X_i^{\sigma 0 } & X_i^{2 \bar{\sigma}}
\end{pmatrix}
\begin{pmatrix}
t^{00}_{ij} & t_{ij}^{02} \\
t_{ij}^{20} & t_{ij}^{22}
\end{pmatrix}
\begin{pmatrix}
X_j^{0 \sigma} \\ 
 X_j^{\bar{\sigma} 2}
\end{pmatrix} 
\\
&= 
\sum_{\vec{k}}\sum_{\sigma}
\begin{pmatrix}
X_\vec{k}^{\sigma 0 } & X_\vec{k}^{2 \bar{\sigma}}
\end{pmatrix}
\underbrace{
\begin{pmatrix}
\epsilon^{00}_{\vec{k}} & \epsilon_{\vec{k}}^{02} \\
\epsilon_{\vec{k}}^{20} & \epsilon_{\vec{k}}^{22}
\end{pmatrix}}_{\mathbf{E}_{\vec{k}}}
\begin{pmatrix}
X_\vec{k}^{0 \sigma} \\ 
 X_\vec{k}^{\bar{\sigma} 2}
\end{pmatrix} 
\end{align}
Here, $H_{\mr{loc},i}$ is an arbitrary local Hamiltonian that only acts non-trivially on site $i$, while $H_1$ describes the hopping. 
The special cases $t^{00}_{ij} = t^{22}_{ij} = t^{20}_{ij} = t^{02}_{ij}$ and $t^{00}_{ij} = t^{22}_{ij} = -t^{20}_{ij} = -t^{02}_{ij}$ correspond to pure single-particle and pure $q$ hopping, respectively.

Below, we will work with the space $\mathbb{W}$ of linear operators on the Hilbert space of our system operators, which has been extensively used in the past by several authors~\cite{Mori1965,Haydock1980a,Viswanath1994,Tiegel2014,julien2008_EROS,Auerbach2018,Auerbach2019,Foley2024_Liouvillian,Pelz2026}.
We will stick closely to the notation used in Ref.~\onlinecite{Pelz2026}, and introduce an inner product for operators,
\begin{align}
(A|B) = \langle \{A^{\dagger},B\} \rangle \, ,
\end{align}
where $\langle \bullet \rangle$ denotes the thermal averave, and we assume $T>0$ such that the thermal density matrix is full rank.
We further introduce a vector notation for operators, $|A) \in \mathbb{W}$.
The Liouvillian $\mc{L} = [H,\bullet]$ acts as a linear map on operators,
\begin{align}
\mc{L} |A) = |[H,A]) \, .
\end{align}
It generates the time evolution of operators,
\begin{align}
\partial_t |A) &= \mr{i} \mc{L} |A) \\
|A(t)) &= \mr{e}^{\mr{i}\mc{L} t} |A) \, .
\end{align}

We are interested in the retarded Green's function for the Hubbard operators,
\begin{align}
\mathbf{G}_{\vec{k}\sigma}(t) = -\mr{i} \theta(t) 
\begin{pmatrix}
\langle\{ X_\vec{k}^{0 \sigma}(t),X_\vec{k}^{\sigma 0}\}\rangle & \langle\{ X_\vec{k}^{0 \sigma}(t),X_\vec{k}^{2 \bar{\sigma}}\}\rangle
\\
\langle\{ X_\vec{k}^{\bar{\sigma} 2}(t),X_\vec{k}^{\sigma 0}\}\rangle & \langle\{ X_\vec{k}^{\bar{\sigma} 2}(t),X_\vec{k}^{2 \bar{\sigma}}\}\rangle
\end{pmatrix} \, .
\end{align}
With the definition $|\mathbf{X}^{\sigma}_{\vec{k}}) = (|X_\vec{k}^{\sigma 0}), |X_\vec{k}^{2 \bar{\sigma}}))$, we can write this Green's function in compact notation,
\begin{align}
\mathbf{G}_{\vec{k}\sigma}(t) = -\mr{i} \theta(t)  (\mathbf{X}^{\sigma}_{\vec{k}}| \mr{e}^{-\mr{i} \mc{L} t} |\mathbf{X}^{\sigma}_{\vec{k}}) \, .
\end{align}
Fourier transform and analytic continuation yield
\begin{align}
\label{eq:Gz_X}
\mathbf{G}_{\vec{k}\sigma}(z) =  (\mathbf{X}^{\sigma}_{\vec{k}}| \frac{1}{z - \mc{L}} |\mathbf{X}^{\sigma}_{\vec{k}}) \, .
\end{align}
We can further obtain a spectral function,
\begin{align}
\mathbf{A}_{\vec{k}\sigma}(\omega) &= \frac{\mr{i}}{2\pi} [\mathbf{G}_{\vec{k}\sigma}(\omega^+) - \mathbf{G}^{\dagger}_{\vec{k}\sigma}(\omega^+)] 
\\
\mathbf{G}_{\vec{k}\sigma}(z) &= \int_{-\infty}^{\infty} \mr{d}\omega \frac{\mathbf{A}_{\vec{k}\sigma}(\omega)}{z - \omega} \, ,
\end{align}
where $\omega^+ = \omega + \mr{i}0^{+}$. The spectral function is not normalized,
\begin{align}
\int_{-\infty}^{\infty} \mr{d}\omega \, \mathbf{A}_{\vec{k}\sigma}(\omega) = \mathbf{N}_0 = \mathbf{t}_0\mathbf{t}_{0}
\\
\hat{\mathbf{A}}_{\vec{k}\sigma}(\omega) = \mathbf{t}_0^{-1} \mathbf{A}_{\vec{k}\sigma}(\omega) \mathbf{t}^{-1}_0 \, ,
\end{align}
where $\mathbf{N}_0 \neq \boldsymbol{1}$ is the norm matrix, $\mathbf{t}_0$ is its principal square root, and $\hat{\mathbf{A}}_{\vec{k}\sigma}(\omega)$ is the normalized spectral function.
Both $\mathbf{N}_0$ and $\mathbf{t}_0$ are $\vec{k}$-independent, since
\begin{align}
(\mathbf{X}^{\sigma}_{i}|\mathbf{X}^{\sigma}_{j}) &= \delta_{ij} 
\begin{pmatrix}
\langle n_{i\bar{\sigma}} \rangle & 0 \\
0 & 1- \langle n_{i\bar{\sigma}} \rangle 
\end{pmatrix}
= \delta_{ij} \mathbf{N}_0 
\\
\mathbf{t}_0 &= 
\begin{pmatrix}
\sqrt{\langle n_{i\bar{\sigma}} \rangle} & 0 \\
0 & \sqrt{1- \langle n_{i\bar{\sigma}} \rangle }
\end{pmatrix}
\, .
\end{align}

It is now useful to define the subspace $\mathbb{W}_{\!X} \subset \mathbb{W}$, which is spanned by the fermionic Hubbard operators with charge $+1$ (raising operators).
The normalized Hubbard operators form an orthonormal basis for $\mathbb{W}_{\!X}$,
\begin{align}
\label{eq:Xk_orthonormal}
|\hat{\mathbf{X}}^{\sigma}_{\vec{k}}) &= |\mathbf{X}^{\sigma}_{\vec{k}})\mathbf{t}^{-1}_0 \, ,
\\
(\hat{\mathbf{X}}^{\sigma}_{\vec{k}}|\hat{\mathbf{X}}^{\sigma'}_{\vec{k}'}) &= \delta_{\vec{k}\vec{k}'} \delta_{\sigma\sigma'} \boldsymbol{1} \, .
\end{align}
Due to spin and momentum conservation, $\mc{L}$ is block-diagonal in this basis,
\begin{align}
\label{eq:L_Xk_block}
(\hat{\mathbf{X}}^{\sigma}_{\vec{k}}|\mc{L}|\hat{\mathbf{X}}^{\sigma'}_{\vec{k}'}) \propto \delta_{\vec{k}\vec{k}'} \delta_{\sigma\sigma'} \, .
\end{align}
We further define
\begin{subequations}
\label{eq:Wx_fullBasis}
\begin{align}
|\mathbf{X}) &= \bigoplus_{\vec{k}\sigma} |\hat{\mathbf{X}}^{\sigma}_{\vec{k}}) \, , \quad \boldsymbol{1}_{x} = (\mathbf{X}|\mathbf{X}) \, ,
\\
\mc{P}_{\!x} &= |\mathbf{X}) (\mathbf{X}| = \sum_{\vec{k}\sigma} |\hat{\mathbf{X}}^{\sigma}_{\vec{k}}) (\hat{\mathbf{X}}^{\sigma}_{\vec{k}}| \, ,
\end{align}
\end{subequations}
where $\boldsymbol{1}_{x}$ is an $M \times M$ identity matrix with $M = \mr{dim} \, \mathbb{W}_{\!X}$, and $\mc{P}_{\!x}$ is the projector on $\mathbb{W}_{\!X}$.
The orthogonal complement of $\mathbb{W}_{\!X}$ is denoted by $\overline{\mathbb{W}}_{\!X}$, with $\mathbb{W} = \mathbb{W}_{\!X} \oplus \overline{\mathbb{W}}_{\!X}$.
Similar to Eq.~\eqref{eq:Wx_fullBasis}, we define a block of orthonormal basis operators $|\overline{\mathbf{X}})$ which span $\overline{\mathbb{W}}_{\!X}$.
In terms of these, the projector on $\overline{\mathbb{W}}_{\!X}$ is given by 
\begin{align}
\mc{P}_{\!\bar{x}} &= |\overline{\mathbf{X}}) (\overline{\mathbf{X}}| \, ,
\end{align}
and $\boldsymbol{1}_{\bar{x}} = (\overline{\mathbf{X}}|\overline{\mathbf{X}})$ is an $\overline{M} \times \overline{M}$ identity with $\overline{M} = \mr{dim} \, \overline{\mathbb{W}}_{\!X}$.

Matrix elements with respect to these bases are represented by subscripts $x$ and $\bar{x}$.
For the Liouvillian, for instance, we get
\begin{align}
\mc{L} &=  |\mathbf{X}) \mc{L}_{xx}(\mathbf{X}| +  |\mathbf{X}) \mc{L}_{x\bar{x}}(\overline{\mathbf{X}}| 
\\ \nonumber
&+  |\overline{\mathbf{X}}) \mc{L}_{\bar{x}x}(\mathbf{X}| + |\overline{\mathbf{X}}) \mc{L}_{\bar{x}\bar{x}}(\overline{\mathbf{X}}| \, ,
\\
\mc{L}_{x'x''} &= (\mathbf{X}'| \mc{L} |\mathbf{X}'') \, , \quad x',x'' \in \{x,\bar{x}\}
\end{align}

We can now formally evaluate the Green's function Eq.~\eqref{eq:Gz_X} by computing the $xx$ block of $(z-\mc{L})^{-1}$,
\begin{align}
[(z-\mc{L})^{-1}]_{xx} = \left[z - \mc{L}_{xx} - \mc{L}_{x\bar{x}}\frac{1}{z - \mc{L}_{\bar{x}\bar{x}}} \mc{L}_{\bar{x}x}\right]^{-1} \, .
\end{align}
Using the basis elements Eq.~\eqref{eq:Xk_orthonormal} and exploiting the block-diagonal structure of $\mc{L}$ [Eq.~\eqref{eq:L_Xk_block}],
we get
\begin{align}
\mathbf{G}_{\vec{k}\sigma}(z) &= \mathbf{t}_0 \frac{\mathbf{1}}{z - (\hat{\mathbf{X}}^{\sigma}_{\vec{k}}|\mc{L}|\hat{\mathbf{X}}^{\sigma}_{\vec{k}}) 
- (\hat{\mathbf{X}}^{\sigma}_{\vec{k}}| \mc{L}_{x\bar{x}}\frac{1}{z - \mc{L}_{\bar{x}\bar{x}}} \mc{L}_{\bar{x}x}|\hat{\mathbf{X}}^{\sigma}_{\vec{k}})} \mathbf{t}_0 \\
&=  \bigl[z \mathbf{t}^{-2}_0 - \mathbf{t}^{-2}_0(\mathbf{X}^{\sigma}_{\vec{k}}|\mc{L}|\mathbf{X}^{\sigma}_{\vec{k}})\mathbf{t}^{-2}_0 
\\ \nonumber
&- \mathbf{t}^{-2}_0(\mathbf{X}^{\sigma}_{\vec{k}}| \mc{L}_{x\bar{x}}\frac{1}{z - \mc{L}_{\bar{x}\bar{x}}} \mc{L}_{\bar{x}x}|\mathbf{X}^{\sigma}_{\vec{k}})\mathbf{t}^{-2}_0 \bigr]^{-1} \, .
\end{align}

To evaluate the non-dynamical part, we divide the Liouvillian into a local part and a kinetic part, $\mc{L}_0 = [H_0,\bullet]$ and $\mc{L}_1 = [H_1,\bullet]$,
\begin{align}
(\mathbf{X}^{\sigma}_{\vec{k}}|\mc{L}|\mathbf{X}^{\sigma}_{\vec{k}}) &= (\mathbf{X}^{\sigma}_{\vec{k}}|\mc{L}_0|\mathbf{X}^{\sigma}_{\vec{k}})  
+ (\mathbf{X}^{\sigma}_{\vec{k}}|\mc{L}_1|\mathbf{X}^{\sigma}_{\vec{k}}) \, .
\end{align}
The expectation value of $\mc{L}_0$ is $\vec{k}$-independent because $H_0$ is local,
\begin{align}
\mathbf{t}^{-2}_0(\mathbf{X}^{\sigma}_{\vec{k}}|\mc{L}_0|\mathbf{X}^{\sigma}_{\vec{k}})\mathbf{t}^{-2}_0 
= \mathbf{t}^{-2}_0(\mathbf{X}^{\sigma}_{i}|\mc{L}_0|\mathbf{X}^{\sigma}_{i})\mathbf{t}^{-2}_0 
= \mathbf{E}_{\mr{loc}} \, .
\end{align}
Within GDMFT, this is computed via local expectation values of the effective impurity model.
To compute the expectation value of $\mc{L}_1$, we compute the (anti)commutators explicitly,
\begin{subequations}
\begin{align}
\{X_{i}^{0\sigma},[X^{\sigma 0}_{\ell} X^{0 \sigma}_{\ell'}, X^{\sigma 0}_j]\} &= \delta_{i\ell} \delta_{j\ell'} n_{i\bar{\sigma}} n_{j\bar{\sigma}}
\\
\{X_{i}^{\bar{\sigma} 2},[X^{2 \bar{\sigma}}_{\ell} X^{0 \sigma}_{\ell'}, X^{\sigma 0}_j]\} &= \delta_{i\ell} \delta_{j\ell'} (1-n_{i\bar{\sigma}}) n_{j\bar{\sigma}}
\\
\{X_{i}^{0\sigma},[X^{\sigma 0}_{\ell} X_{\ell'}^{\bar{\sigma} 2},X^{2 \bar{\sigma}}_{j}]\} &= \delta_{i\ell} \delta_{j\ell'} n_{i\bar{\sigma}} (1-n_{j\bar{\sigma}})
\\
\{X_{i}^{\bar{\sigma} 2},[X^{2 \bar{\sigma}}_{\ell} X_{\ell'}^{\bar{\sigma} 2},X^{2 \bar{\sigma}}_{j}]\} &= \delta_{i\ell} \delta_{j\ell'} (1-n_{i\bar{\sigma}}) (1-n_{j\bar{\sigma}}) \, .
\end{align}
\end{subequations}
All other relevant combinations are zero because fermionic Hubbard operators on different sites anticommute, and because $\{X_{i}^{\bar{\sigma} 2},X^{\sigma 0}_{j}\} = 0$.
There are combinations involving anticommutations of opposite-spin operators, but their expectation value is zero as long as spin and charge are conserved.
When we put everything together, we get
\begin{align}
\mathbf{t}^{-2}_0(\mathbf{X}^{\sigma}_{\vec{k}}|\mc{L}_1|\mathbf{X}^{\sigma}_{\vec{k}})\mathbf{t}^{-2}_0 &= \mathbf{E}_{\vec{k}} + \delta \mathbf{E}_{\vec{k}} \, .
\end{align}
Here, $\delta \mathbf{E}_{\vec{k}\sigma}$ is the Fourier transform of 
\begin{align}
\delta \mathbf{E}_{ij\sigma} = 
\langle \delta n_{i\bar{\sigma}} \delta n_{j\bar{\sigma}} \rangle
\mathbf{N}_0^{-1}
\begin{pmatrix}
t^{00}_{ij} & -t^{02}_{ij} 
\\
-t^{20}_{ij} & t^{22}_{ij} 
\end{pmatrix} \mathbf{N}_0^{-1} \, ,
\end{align}
with $\delta n_{i\sigma} = n_{i\sigma} - \langle n_{i\sigma} \rangle$.
In GDMFT, we approximate $\langle \delta n_{i\bar{\sigma}} \delta n_{j\bar{\sigma}} \rangle = 0$, which is exact in the limit of infinite coordination number~\cite{Stanescu2004}.

Following Ref.~\onlinecite{Stanescu2004}, we can now define the irreducible cumulant
\begin{align}
\mathbf{M}_{\vec{k}\sigma}(z) &= \bigl[z \mathbf{t}_0^{-2} - \mathbf{E}_{\mr{loc}} - \delta \mathbf{E}_{\vec{k}\sigma} - 
\\ \nonumber
&- \mathbf{t}^{-2}_0(\mathbf{X}^{\sigma}_{\vec{k}}| \mc{L}_{x\bar{x}}\frac{1}{z - \mc{L}_{\bar{x}\bar{x}}} \mc{L}_{\bar{x}x}|\mathbf{X}^{\sigma}_{\vec{k}})\mathbf{t}^{-2}_0 \bigr]^{-1}
\\
\mathbf{G}_{\vec{k}\sigma}(z) &= \frac{\boldsymbol{1}}{\mathbf{M}^{-1}_{\vec{k}\sigma}(\omega) - \mathbf{E}_{\vec{k}}} \, .
\end{align}
Within GDMFT (exact in the $d\to \infty$ limit), the irreducible cumulant is approximated as $\vec{k}$-independent. As discussed above, this means $\delta \mathbf{E}_{\vec{k}} = 0$,
and the dynamical part 
\begin{align}
\boldsymbol{\Sigma}^{X}_{\vec{k}\sigma}(z) = (\mathbf{X}^{\sigma}_{\vec{k}}| \mc{L}_{x\bar{x}}\frac{1}{z - \mc{L}_{\bar{x}\bar{x}}} \mc{L}_{\bar{x}x}|\mathbf{X}^{\sigma}_{\vec{k}})
\end{align}
is momentum-independent, ie.\
\begin{align}
\boldsymbol{\Sigma}^{X}_{\vec{k}\sigma}(z) = \boldsymbol{\Sigma}^{X}_{\sigma}(z) = (\mathbf{X}^{\sigma}_{i}| \mc{L}_{x\bar{x}}\frac{1}{z - \mc{L}_{\bar{x}\bar{x}}} \mc{L}_{\bar{x}x}|\mathbf{X}^{\sigma}_{i}) \, .
\end{align}
The resulting irreducible cumulant is
\begin{align}
\mathbf{M}_{\sigma}(z) = \frac{\boldsymbol{1}}{z \mathbf{t}_0^{-2} - \mathbf{E}_{\mr{loc}} - \mathbf{t}_0^{-2} \boldsymbol{\Sigma}^{X}_{\sigma}(z) \mathbf{t}_0^{-2}} \, .
\end{align}

Due to its locality, the irreducible cumulant of the lattice model can be computed from an effective impurity model~\cite{Stanescu2004},
\begin{subequations}
\label{eq:Himp_GDMFT}
\begin{align}
H_{\mr{imp}} &= H_{\mr{loc}} + H_{\mr{hyb}} + H_{\mr{bath}} \, ,
\\
H_{\mr{bath}} &= \sum_{\lambda,\sigma} \sum_{\alpha = 0,2} \epsilon_{\lambda\alpha\sigma} a^{\dagger}_{\lambda\alpha\sigma} a^{\phantom{\dagger}}_{\lambda\alpha\sigma}
\\
H_{\mr{hyb}} &= \sum_{\lambda,\sigma} \bigl[ V^{00}_{\lambda\sigma}  X^{\sigma 0} a^{\phantom{\dagger}}_{\lambda0\sigma} 
+ V^{20}_{\lambda\sigma}  X^{2 \bar{\sigma}} a^{\phantom{\dagger}}_{\lambda0\sigma}
\\ \nonumber
&+ V^{02}_{\lambda\sigma}  X^{\sigma 0} a^{\phantom{\dagger}}_{\lambda2\sigma}
+ V^{22}_{\lambda\sigma}  X^{2 \bar{\sigma}} a^{\phantom{\dagger}}_{\lambda2\sigma} \bigr] + \mr{h.c.}
\end{align} 
\end{subequations}
Here, $H_{\mr{loc}}$ only acts non-trivially on the impurity degrees of freedom, and it has the same form as the local lattice Hamiltonian $H_{\mr{loc},i}$. 
The bath Hamiltonian contains two spinfull fermionic flavors $\alpha \in \{0,2\}$ ($a^{\phantom{\dagger}}_{\lambda\alpha\sigma}$ are canonical fermionic operators). 
The bath parameters $V^{\alpha \beta}_{\lambda\sigma}$ and $\epsilon_{\lambda\alpha\sigma}$ are adjusted such that the local lattice Green's function matches that of the 
impurity model,
\begin{align}
\label{eq:Gimp_M}
\mathbf{G}_{\mr{imp},\sigma}(z) &= \frac{\boldsymbol{1}}{\mathbf{M}^{-1}_{\sigma}(z) - \mathbf{\Delta}_{\sigma}(z)}
\\ 
\label{eq:Gloc}
\mathbf{G}_{\mr{loc},\sigma}(z) &= 
\int_{\vec{k}} \mathbf{G}_{\vec{k}\sigma} = \int_{\vec{k}} \frac{\boldsymbol{1}}{\mathbf{M}^{-1}_{\sigma}(z) - \mathbf{E}_{\vec{k}}} \, ,
\\
\label{eq:GDMFT_update}
\boldsymbol{\Delta}_{\sigma}(z) &= \mathbf{M}^{-1}_{\sigma}(z) - \mathbf{G}^{-1}_{\mr{loc},\sigma}(z) \, .
\end{align}
For a given (not yet self-consistent) $\mathbf{\Delta}_{\sigma}(z)$, we compute the irreducible cumulant form the impurity model, compute the local Green's function via Eq.~\eqref{eq:Gloc}, and finally obtain an updated $\mathbf{\Delta}_{\sigma}(z)$ via Eq.~\eqref{eq:GDMFT_update}. This is iterated until convergence.

Finally, the connection between $\boldsymbol{\Delta}_{\sigma}(z)$ and the bath parameters needs to be known to set up the impurity model Eq.~\eqref{eq:Himp_GDMFT}. First of all, it should be noted that for given $\boldsymbol{\Delta}_{\sigma}(z)$, the impurity model Eq.~\eqref{eq:Himp_GDMFT} is generically interacting, irrespective of $H_{\mr{loc}}$. That is because the bath hybridizes with fermionic Hubbard operators and not with canonical fermion operators.

To separate the bath $\boldsymbol{\Delta}_{\sigma}(z)$ from the irreducible cumulant, we define the operator spaces $\mathbb{W}_{\! X} \subset \mathbb{W}$ and $\mathbb{W}_{a} \subset \mathbb{W}$, which contain the fermionic Hubbard operators on the impurity with charge +1, $|X^{\sigma 0})$ and $|X^{2 \bar{\sigma}})$, and the raising operators $|a^{\dagger}_{\lambda\alpha\sigma})$ of the bath, respectively. The orthogonal complement of their direct sum $ \mathbb{W}_{\! X} \oplus \mathbb{W}_{a}$ is denoted $\overline{\mathbb{W}}_{\! X}$.
We also divide the Liouvillian of the impurity model into separate parts,
\begin{align}
\mc{L}_{\mr{imp}} = \mc{L}_{\mr{loc}} + \mc{L}_{\mr{hyb}} + \mc{L}_{\mr{bath}} \, .
\end{align}

Applying the different parts of the Liouvillian to basis operators in $\mathbb{W}_a$, we get
\begin{subequations}
\begin{align}
\mc{L}_{\mr{loc}} |a^{\dagger}_{\lambda\alpha\sigma}) &= 0
\\
\mc{L}_{\mr{hyb}} |a^{\dagger}_{\lambda\alpha\sigma}) &= V^{0\alpha}_{\lambda\sigma}  |X^{\sigma 0})
+ V^{2\alpha}_{\lambda\sigma}  |X^{2 \bar{\sigma}})
\\
\mc{L}_{\mr{bath}} |a^{\dagger}_{\lambda\alpha\sigma}) &= \epsilon_{\lambda\alpha\sigma} |a^{\dagger}_{\lambda\alpha\sigma}) \, .
\end{align}
\end{subequations}
Thus, by acting with $\mc{L}_{\mr{imp}}$ on any operator in $\mathbb{W}_a$, we end up with some operator in $\mathbb{W}_a \oplus \mathbb{W}_{\! X}$, ie.\ we never end up in the space of irreducible operators $\overline{\mathbb{W}}_{\! X}$. As a result, $\mc{L}_{\mr{imp}}$ has a block-tridiagonal form,
\begin{align}
\label{eq:Limp_tridiag}
\mc{L}_{\mr{imp}} = 
\begin{pmatrix}
[\mc{L}_{\mr{imp}}]_{aa} & [\mc{L}_{\mr{imp}}]_{ax} & \boldsymbol{0} \\
[\mc{L}_{\mr{imp}}]_{xa} & [\mc{L}_{\mr{imp}}]_{xx} & [\mc{L}_{\mr{imp}}]_{x\bar{x}} \\
\boldsymbol{0} & [\mc{L}_{\mr{imp}}]_{\bar{x} x} & [\mc{L}_{\mr{imp}}]_{\bar{x}\bar{x}}
\end{pmatrix} \, ,
\end{align}
where the subscript $a$ refers to the bath block, akin to the subscripts $x$ and $\bar{x}$. 

Then, we formally find
\begin{subequations}
\begin{align}
[(z-\mc{L}_{\mr{imp}})^{-1}]_{xx} &= \biggl[z -  [\mc{L}_{\mr{imp}}]_{xx}
\\ \nonumber
 &- [\mc{L}_{\mr{imp}}]_{xa} \frac{1}{z - [\mc{L}_{\mr{imp}}]_{aa}} [\mc{L}_{\mr{imp}}]_{ax} 
\\ \nonumber
&- [\mc{L}_{\mr{imp}}]_{x\bar{x}} \frac{1}{z - [\mc{L}_{\mr{imp}}]_{\bar{x}\bar{x}}} [\mc{L}_{\mr{imp}}]_{\bar{x}x} \biggr]^{-1} \, .
\end{align}
The impurity Green's function is therefore
\begin{align}
\mathbf{G}_{\mr{imp},\sigma}(z) &= (\mathbf{X}^{\sigma}|\frac{\boldsymbol{1}}{z - \mc{L}_{\mr{imp}}}|\mathbf{X}^{\sigma}) =
\\ \nonumber
 \bigl[ z \mathbf{t}^{-2}_0 &- \mathbf{E}_{\mr{loc}} -  \mathbf{t}^{-2}_0 \boldsymbol{\Delta}^{\! X}(z)  \mathbf{t}^{-2}_0
- \mathbf{t}^{-2}_0 \boldsymbol{\Sigma}^{X}(z) \mathbf{t}^{-2}_0 \bigr]^{-1} 
\\ \nonumber
&= \bigl[\mathbf{M}^{-1}_{\sigma}(z) - \boldsymbol{\Delta}(z)\bigr]^{-1} \, ,
\\
\label{eq:Mimp}
\mathbf{M}_{\sigma}(z) &=  \bigl[ z \mathbf{t}^{-2}_0 - \mathbf{E}_{\mr{loc}} - \mathbf{t}^{-2}_0 \boldsymbol{\Sigma}^{X}(z) \mathbf{t}^{-2}_0 \bigr]^{-1} 
\\
\mathbf{E}_{\mr{loc}} &= \mathbf{t}^{-2}_0(\mathbf{X}^{\sigma}|\mc{L}_{\mr{imp}}|\mathbf{X}^{\sigma})\mathbf{t}^{-2}_0 \, ,
\\ 
\label{eq:Sigmax_dyn}
\boldsymbol{\Sigma}^{X}(z) &= (\mathbf{X}^{\sigma}| [\mc{L}_{\mr{imp}}]_{x\bar{x}} \frac{1}{z - [\mc{L}_{\mr{imp}}]_{\bar{x}\bar{x}}} [\mc{L}_{\mr{imp}}]_{\bar{x}x} |\mathbf{X}^{\sigma}) \, , 
\\
\boldsymbol{\Delta}(z) &= \mathbf{t}^{-2}_0 \boldsymbol{\Delta}^{\! X}(z) \mathbf{t}^{-2}_0 \, ,
\\ 
\boldsymbol{\Delta}^{\! X}(z) &= (\mathbf{X}^{\sigma}| [\mc{L}_{\mr{imp}}]_{xa} \frac{1}{z - [\mc{L}_{\mr{imp}}]_{aa}} [\mc{L}_{\mr{imp}}]_{ax} |\mathbf{X}^{\sigma}) \, , 
\end{align}
\end{subequations}
where $|\mathbf{X}^{\sigma}) = (|X^{\sigma 0}),|X^{2 \bar{\sigma}}))$.

The hybridization part can be evaluated without explicit knowledge of the density matrix, which is otherwise generically required to compute inner products of operators.
Within $\mathbb{W}_a$, we get
\begin{align}
(a^{\dagger}_{\lambda\alpha\sigma}|\mc{L}_{\mr{imp}}|a^{\dagger}_{\lambda'\alpha'\sigma'}) = \delta_{\lambda\lambda'} \delta_{\alpha\alpha'} \delta_{\sigma\sigma'} \epsilon_{\lambda\alpha\sigma} \, ,
\end{align}
which follows from $(a^{\dagger}_{\lambda\alpha\sigma}|a^{\dagger}_{\lambda'\alpha'\sigma'}) = \delta_{\lambda\lambda'} \delta_{\alpha\alpha'} \delta_{\sigma\sigma'}$ and holds independent of the density matrix.
Further, we find
\begin{align}
(a^{\dagger}_{\lambda\alpha\sigma}| \mc{L}_{\mr{imp}} |\mathbf{X}^{\sigma}) &= (V^{\alpha 0}_{\lambda\sigma} \langle n_{\bar{\sigma}} \rangle, V^{\alpha 2}_{\lambda\sigma} \langle 1 - n_{\bar{\sigma}} \rangle ) 
\\ \nonumber
&= (V^{\alpha 0}_{\lambda\sigma} [\mathbf{t}_0^{2}]_{00}, V^{\alpha 2}_{\lambda\sigma} [\mathbf{t}_0^{2}]_{22} ) 
\\ \nonumber
&= (V^{\alpha 0}_{\lambda\sigma}, V^{\alpha 2}_{\lambda\sigma} ) \mathbf{t}_0^{2} \, ,
\end{align}
where we have used $[\mathbf{t}_0^{2}]_{02} = [\mathbf{t}_0^{2}]_{20} = 0$.
Therefore, we get
\begin{align}
\Delta_{\alpha \beta}(z) = \sum_{\lambda\gamma} \frac{V^{\alpha\gamma}_{\lambda\sigma} (V^{\beta\gamma}_{\lambda\sigma})^{\ast}}{z - \epsilon_{\lambda\gamma\sigma}} \, .
\end{align}
Once $\mathbf{\Delta}(z)$ is given, there are well-established methods to (approximately) extract the bath parameters. 
We use the Numerical Renormalization Group to solve the impurity model, which logarithmically discretizes the spectrum of $\mathbf{\Delta}(z)$.
Details on the impurity calculation are explained in Sec.~\ref{sec:impurity_model}.

\section{Impurity model}
\label{sec:impurity_model}

To solve the impurity model of the form in Eq.~\eqref{eq:Himp_GDMFT}, we use the Numerical Renormalization Group~(NRG). 
Within NRG, the spectrum of the hybridization function,
\begin{align}
\boldsymbol{\Gamma}(\omega) = \frac{\mr{i}}{2\pi} \left[\boldsymbol{\Delta}(\omega^+) - \boldsymbol{\Delta}^{\dagger}(\omega^+) \right] \, ,
\end{align}
is discretized employing a logarithmic grid with a discretization parameter $\Lambda > 1$. 
The discretized star geometry is mapped on a Wilson chain with logarithmically decaying hopping amplitudes and on-site energies. 
In case of the impurity Hamiltonian Eq.~\eqref{eq:GDMFT_impurity_model}, every site of the Wilson chain consists of two spinful fermionic orbitals.
The spectrum of the resulting logarithmic Wilson chain Hamiltonian is determined via iterative diagonalization.
We use the full density matrix NRG~(fdmNRG) algorithm to obtain spectral functions. 
Since fdmNRG is a well-established method, we refer the reader to Refs.~\cite{Weichselbaum2007,Bulla2008,Weichselbaum2012,Weichselbaum2012a,Lee2016,Lee2017,Weichselbaum2020,Kugler2022,Weichselbaum2024} for more detailed information on the state-of-the-art NRG approach.

\subsection{Symmetric estimator for irreducible cumulants}

The state-of-the-art method in NRG to obtain a reliable estimate of the self-energy is via equations of motion, pioneered by Bulla \textit{et al.}~\cite{Bulla1998}.
Bulla's estimator has been recently improved by a symmetric version by Fabian Kugler~\cite{Kugler2022}, which generally leads to highly accurate self-energies.
In this section, we extend Kugler's symmetric improved estimator to irreducible cumulants, which have to be computed within GDMFT.

The main challenge when computing the cumulant in Eq.~\eqref{eq:Mimp} is the dynamical part in Eq.~\eqref{eq:Sigmax_dyn}, all other contributions to the cumulant are mere expectation values of local operators. 
To compute the dynamical part, we formally define $\boldsymbol{\mc{G}}(z) = (z - \mc{L}_{\mr{imp}})^{-1}$ and impose the block decomposition used in Eq.~\eqref{eq:Limp_tridiag},
\begin{align}
\boldsymbol{\mc{G}}(z) &= 
\begin{pmatrix}
\boldsymbol{\mc{G}}_{aa}(z) & \boldsymbol{\mc{G}}_{ax}(z) & \boldsymbol{\mc{G}}_{a\bar{x}}(z) \\
\boldsymbol{\mc{G}}_{xa}(z) & \boldsymbol{\mc{G}}_{xx}(z) & \boldsymbol{\mc{G}}_{x\bar{x}}(z) \\
\boldsymbol{\mc{G}}_{\bar{x}a}(z) & \boldsymbol{\mc{G}}_{\bar{x}x}(z) & \boldsymbol{\mc{G}}_{\bar{x}\bar{x}}(z) 
\end{pmatrix} \, .
\end{align}
Due to Eq.~\eqref{eq:Limp_tridiag}, its inverse is block-tridiagonal,
\begin{align}
\boldsymbol{\mc{G}}^{-1}(z) &= (z - \mc{L}_{\mr{imp}}) 
\\ \nonumber
&=
\begin{pmatrix}
z - [\mc{L}_{\mr{imp}}]_{aa} & - [\mc{L}_{\mr{imp}}]_{ax} & \boldsymbol{0} \\
- [\mc{L}_{\mr{imp}}]_{xa} & z - [\mc{L}_{\mr{imp}}]_{xx} & -[\mc{L}_{\mr{imp}}]_{x\bar{x}} \\
\boldsymbol{0} & - [\mc{L}_{\mr{imp}}]_{\bar{x}x} & z - [\mc{L}_{\mr{imp}}]_{\bar{x}\bar{x}}
\end{pmatrix} \, .
\end{align}
This block-tridiagonal structure of $\boldsymbol{\mc{G}}^{-1}(z)$ is crucial to derive symmetric improved estimators. It implies the relations
\begin{subequations}
\begin{align}
- \boldsymbol{\mc{G}}_{xx}(z) [\mc{L}_{\mr{imp}}]_{x\bar{x}} + \boldsymbol{\mc{G}}_{x\bar{x}}(z) (z - [\mc{L}_{\mr{imp}}]_{\bar{x}\bar{x}}) &= \boldsymbol{0}
\\
- \boldsymbol{\mc{G}}_{\bar{x}x}(z) [\mc{L}_{\mr{imp}}]_{x\bar{x}} + \boldsymbol{\mc{G}}_{\bar{x}\bar{x}}(z) (z - [\mc{L}_{\mr{imp}}]_{\bar{x}\bar{x}}) &= \boldsymbol{1} \, .
\end{align}
\end{subequations}
By rearranging these equations, we find
\begin{align}
\label{eq:zLbarxbarx_inv}
\frac{\boldsymbol{1}}{z - [\mc{L}_{\mr{imp}}]_{\bar{x}\bar{x}}} = \boldsymbol{\mc{G}}_{\bar{x}\bar{x}}(z) - \boldsymbol{\mc{G}}_{\bar{x}x}(z) \boldsymbol{\mc{G}}^{-1}_{xx}(z) \boldsymbol{\mc{G}}_{x\bar{x}}(z) \, .
\end{align}

To evaluate Eq.~\eqref{eq:Sigmax_dyn}, we define the operators
\begin{align}
\label{eq:Q_definition}
|\mathbf{Q}^{\sigma}) &= [\mc{L}_{\mr{imp}}]_{\bar{x}x} |\mathbf{X}^{\sigma}) \, ,
\end{align}
which are in are in $\overline{\mathbb{W}}_{\!X}$. For their explicit computation, see Eq.~\eqref{eq:Q_projection} below.
Equation~\eqref{eq:Sigmax_dyn} is the expectation value of Eq.~\eqref{eq:zLbarxbarx_inv} with respect to $|\mathbf{Q}^{\sigma})$,
\begin{subequations}
\label{eq:IFG_cumulant}
\begin{align}
\boldsymbol{\Sigma}^{\! X}(z) &= (\mathbf{Q}^{\sigma}|\frac{\boldsymbol{1}}{z - [\mc{L}_{\mr{imp}}]_{\bar{x}\bar{x}}}|\mathbf{Q}^{\sigma}) 
\\ \nonumber
&= \mathbf{I}(z) - \mathbf{F}^{L}(z) \mathbf{G}^{-1}(z) \mathbf{F}^{R}(z)
\\
\mathbf{G}(z) &= (\mathbf{X}^{\sigma}|\frac{\boldsymbol{1}}{z - \mc{L}_{\mr{imp}}}|\mathbf{X}^{\sigma})
\\
\mathbf{F}^{L}(z) &= (\mathbf{Q}^{\sigma}|\frac{\boldsymbol{1}}{z - \mc{L}_{\mr{imp}}}|\mathbf{X}^{\sigma})
\\
\mathbf{F}^{R}(z) &= (\mathbf{X}^{\sigma}|\frac{\boldsymbol{1}}{z - \mc{L}_{\mr{imp}}}|\mathbf{Q}^{\sigma})
\\
\mathbf{I}(z) &= (\mathbf{Q}^{\sigma}|\frac{\boldsymbol{1}}{z - \mc{L}_{\mr{imp}}}|\mathbf{Q}^{\sigma}) \, .
\end{align}
\end{subequations}
The dynamical correlators in Eq.~\eqref{eq:IFG_cumulant} can be evaluated with impurity solvers such as NRG.

The operator in Eq.~\eqref{eq:Q_definition} can be constructed with the help of simple expectation values,
\begin{align}
\label{eq:Q_projection}
|\mathbf{Q}^{\sigma}) = |[H,\mathbf{X}^{\sigma}]) &- |\mathbf{X}^{\sigma}) \mathbf{t}_0^{-2} (\mathbf{X}^{\sigma}| [H,\mathbf{X}^{\sigma}]) 
\\ \nonumber
&-  \sum_{\alpha} |f^{\dagger}_{0\alpha\sigma}) (f^{\dagger}_{0\alpha\sigma}| [H,\mathbf{X}^{\sigma}]) \, ,
\end{align}
where $f_{0\alpha\sigma}$ are the canonical annihilation operators of the bath orbitals located at the impurity,
\begin{align}
f_{0\alpha\sigma} \propto \sum_{\lambda\beta} V^{\alpha\beta}_{\lambda\sigma} a_{\lambda\beta\sigma} \, .
\end{align}
We note that $|\mathbf{Q}^{\sigma})$ generically acts in a non-trivial manner on the bath orbitals located at the impurity.
In special cases, for instance, when computing the self-energy for a single-impurity Anderson model, the corresponding operator $\mathbf{Q}^{\sigma}$ acts trivially in the bath~\cite{Kugler2022}. 

\begin{figure}
\includegraphics[width=\linewidth]{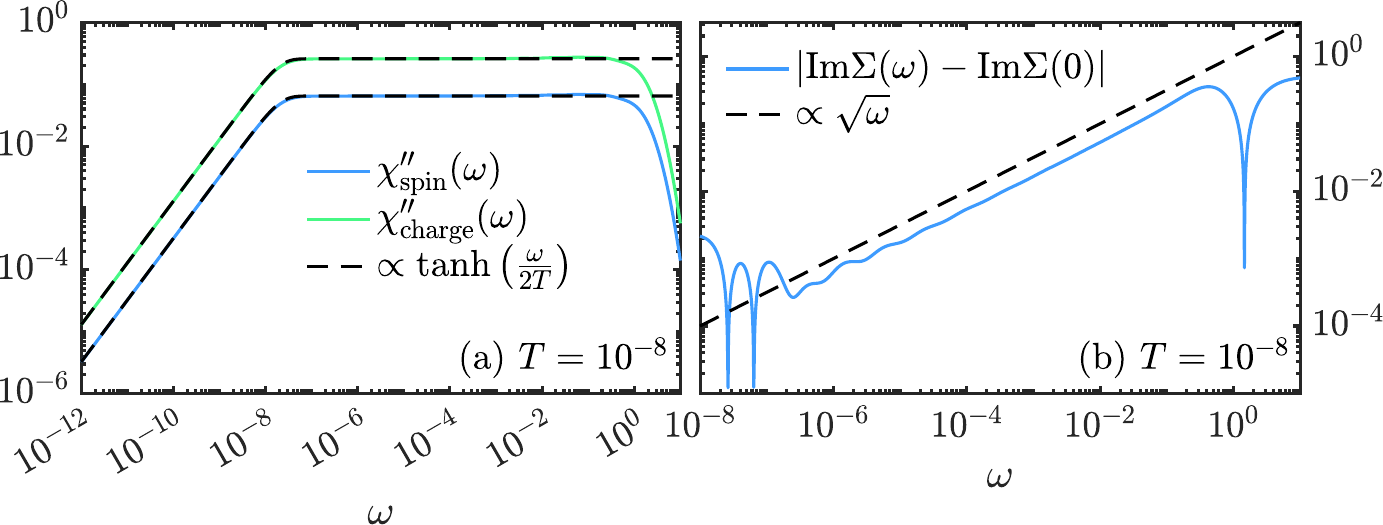}
\caption{(a) spin and charge spectrum and (b) frequency dependence of the imaginary part of the self-energy for the impurity model in the main text at quantum criticality ($V_c = V_q$).}
\label{fig:Q_impurity_susceptibility_SE}
\end{figure}

\begin{figure}
\includegraphics[width=\linewidth]{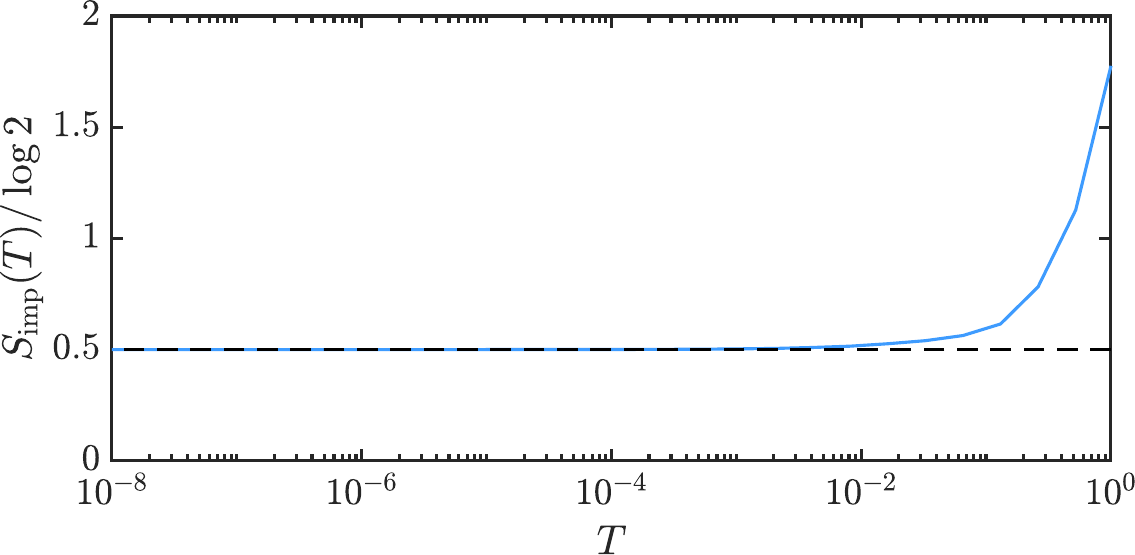}
\caption{Impurity contribution to the entropy versus temperature for the impurity model in the main text, at the quantum critical point at $V_c = V_q$.}
\label{fig:Q_impurity_entropy}
\end{figure}

\subsection{Further results on the impurity model}

In this section, we show that the impurity model in the main text exhibits two-channel Kondo behavior~\cite{Nozieres1980,Andrei1984,Tsvelick1985,Affleck1991,Ludwig1991,Anders2005} at the quantum critical point, where $V_c = V_q$.
For that, we computed both the spin and charge susceptibility at the impurity $\chi_{\mr{spin}}(\omega)$ and $\chi_{\mr{charge}}(\omega)$, respectively. 
Their spectra at $T=10^{-8}$ are shown in Fig.~\ref{fig:Q_impurity_susceptibility_SE}(a). They both fit the phenomenological $\propto \mr{tanh} \frac{\omega}{2T}$ form, typical for two-channel Kondo behavior.
Note that $\chi_{\mr{spin}}(\omega) = \tfrac{1}{4}\chi_{\mr{charge}}(\omega)$, because there is no local Hubbard interaction and the system is both particle-hole and spin symmetric.
As a result, charge and spin can be interchanged.
Figure~\ref{fig:Q_impurity_susceptibility_SE}(b) shows that $-\mr{Im} \Sigma(\omega) = -\mr{Im} \Sigma(0) + b \sqrt{|\omega|}$, which is also typical two-channel Kondo behavior.
Finally, Fig.~\ref{fig:Q_impurity_entropy} shows that the impurity contribution to the entropy approaches $\log \sqrt{2}$ at $T\to 0$, another hallmark signature of the two-channel Kondo model.

If a local Hubbard interaction is included, the charge susceptibility is suppressed for $U>0$. In the $U\to \infty$ limit, while keeping $V_x/U^2$ constant, this impurity model becomes a two-channel Kondo model.

\end{document}